\documentclass[iop]{emulateapj}

\usepackage{natbib}
\usepackage{multirow,graphics,epsfig,times,latexsym,amssymb}

\shorttitle{A relic star cluster in the Sextans dSph -- Implications for near-field cosmology}
\shortauthors{Karlsson, Bland-Hawthorn, Freeman, \& Silk}

\begin{document}

\title{The Chemical Signature of a Relic Star Cluster in \\the Sextans Dwarf Spheroidal Galaxy --- Implications for near-field cosmology}

\author{ }
\affil{ }

\author{Torgny Karlsson\altaffilmark{1}}
\affil{Department of Physics and Astronomy, Uppsala University, Box 516, 751~20 Uppsala, Sweden}
\altaffiltext{1}{Sydney Institute for Astronomy, School of Physics, University of Sydney, NSW 2006, Australia}
\email{torgny.karlsson@physics.uu.se}

\author{Joss Bland-Hawthorn}
\affil{Sydney Institute for Astronomy, School of Physics, University of Sydney, NSW 2006, Australia}

\author{Ken C. Freeman}
\affil{Research School of Astronomy \& Astrophysics, Mount Stromlo Observatory, Cotter Road, Weston ACT 2611, Australia}

\author{Joe Silk}
\affil{Physics Department, University of Oxford, OX1 3RH, UK}

\begin{abstract}
We present tentative evidence for the existence of a dissolved star cluster at $[\mathrm{Fe}/\mathrm{H}]=-2.7$ in the Sextans dwarf spheroidal galaxy. We use the technique of chemical tagging to identify stars that are highly clustered in a multi-dimensional chemical abundance space ($\mathcal{C}$-space). In a sample of six stars, three, possibly four, stars are identified as potential cluster stars. The initial stellar mass of the parent cluster is estimated from two independent observations to $M_{*,\mathrm{init}}=1.9^{+1.5}_{-0.9}~(1.6^{+1.2}_{-0.8})\times 10^5~\mathcal{M_{\odot}}$, assuming a Salpeter (Kroupa) initial mass function (IMF). If corroborated by follow-up spectroscopy, this star cluster is the most metal-poor system identified to date. Chemical signatures of remnant clusters in dwarf galaxies like Sextans provide us with a very powerful probe to the high-redshift Universe.  From available observational data, we argue that the average star cluster mass in the majority of the newly discovered ultra-faint dwarf galaxies was notably lower than it is in the Galaxy today and possibly lower than in the more luminous, classical dwarf spheroidal galaxies. Furthermore, the mean cumulative metallicity function of the dwarf spheroidals falls below that of the ultra-faints, which increases with increasing metallicity as predicted from our stochastic chemical evolution model. These two findings, together with a possible difference in the $\langle [\mathrm{Mg}/\mathrm{Fe}] \rangle$ ratio suggest that the ultra-faint dwarf galaxy population, or a significant fraction thereof, and the dwarf spheroidal population, were formed in different environments and would thus be distinct in origin.

\end{abstract}

%\keywords{Galaxy -- dwarf galaxies -- stellar populations -- star clusters -- elemental abundances}
%\keywords{Galaxies: dwarf Ð Galaxy: abundances Ð Galaxy: evolution Ð Galaxy: formation Ð galaxies: star clusters: general}

\section{Introduction}\label{sect:intro}
\noindent
The formation and long-term evolution of dwarf galaxies are topics of great interest today. A substantial fraction of the dwarf galaxy population of the Local Group is expected to have already merged with the halos of M31 and the Galaxy. This is supported by the evidence that the stellar Galactic Halo has several distinct chemical similarities that it shares with the existing population of dwarf galaxies (Kirby et al. 2008\nocite{kirb08}; Tolstoy et al. 2009\nocite{tols09} and references therein; Starkenburg et al. 2010\nocite{star10}; Frebel et al. 2010a, b\nocite{freb10a,freb10b}; Norris et al. 2010a, c\nocite{norr10a,norr10c}). Despite these similarities, the systems are not chemically identical and exactly how the merging process progressed, how the dwarf galaxies that survived are connected to their disrupted counterparts, and how these still-surviving galaxies relate to each other in a cosmological context, is not fully understood. 

\begin{figure*}[t]
\resizebox{\hsize}{!}{\includegraphics[trim=11.6mm 12mm 34.2mm 32mm,clip]{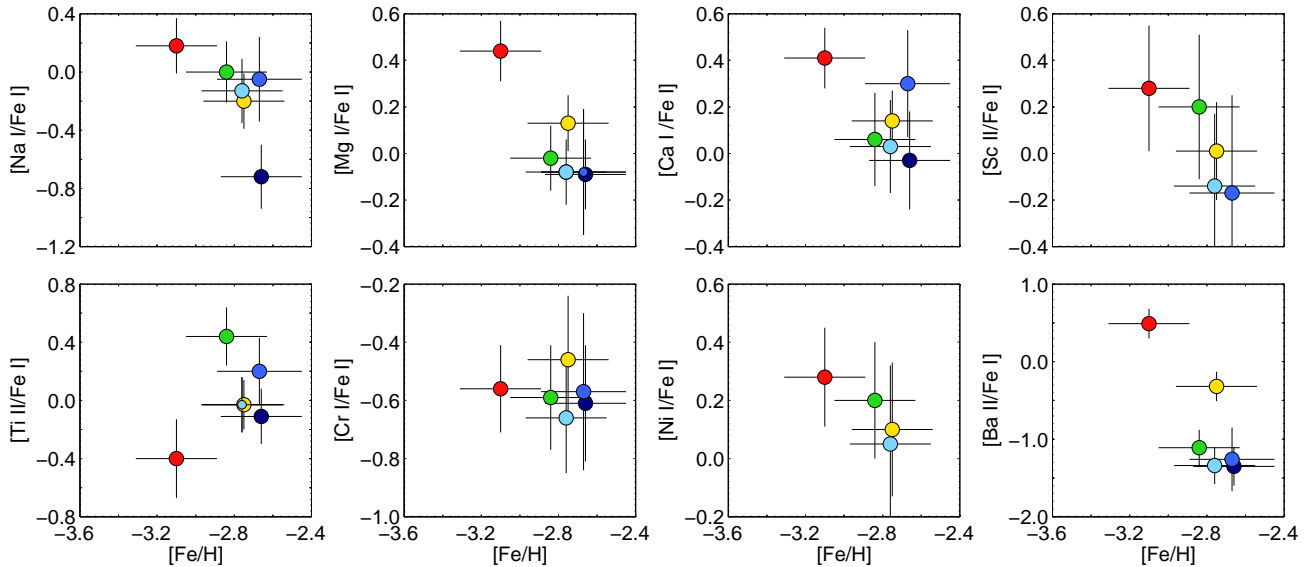}}
\caption{Chemical abundance ratios of six very metal-poor stars in the Sextans dSph. Abundance data are taken from Aoki et al. (2009). Stars are color coded according to their location in the [Mg/Fe] -- [Fe/H] diagram. The three stars with nearly identical [Mg/Fe] (colored in different shades of blue) also have very similar [Fe/H]. These stars all clump together in Ti, Cr, and Ba, as well. One of the ``blue'' stars, S~$10-14$ (dark blue), is substantially deficient in Na. The scandium abundance could not be measured in S~$10-14$, while nickel could not be measured in S~$10-14$ and S~$14-98$. In the [Mg/Fe] (upper, middle left) and [Ti/Fe] (lower, outer left) diagrams, the medium (\mbox{S~$14-98$}) and light blue (S~$11-13$) markers, respectively, are made smaller for visual purposes.}
\label{fig:abnd}
\nocite{aoki09}
\end{figure*}

The study of old stellar populations in these galaxies provides us with an indirect link to the high-redshift Universe. All Local Group dwarf galaxies, including the recently discovered ultra-faint dwarfs, are known to contain ancient, metal-poor stars (Grebel \& Gallagher 2004\nocite{greb04}; Tolstoy et al. 2009\nocite{tols09}). The star formation efficiency in dwarfs is very low which means that there have been relatively few star-formation events over cosmic time. The effects of stochasticity therefore plays a relatively big r\^{o}le (Audouze \& Silk 1995\nocite{audo95}; Karlsson 2005\nocite{karl05}; Karlsson \& Gustafsson 2005\nocite{karl05b}). This aspect must be addressed when seeking to understand the evolution of dwarf galaxies, particularly at early times. 

A key characteristic of star formation is the creation of star clusters. The initial cluster mass function (ICMF), if it can be determined reliably, provides important information on the physical conditions under which the clusters formed (Elmegreen 2010\nocite{elme10}; Larsen 2009\nocite{lars09}). A flat ICMF appears to be indicative of high-pressure environments (e.g. starbursts, mergers), whereas a steep ICMF is typically associated with quiescent environments (e.g. solar neighbourhood). At the present time, the most metal-poor star clusters known have an iron abundance just below $[\mathrm{Fe}/\mathrm{H}]=-2$. One of the globular clusters (Cluster 1) in the Fornax dSph currently holds the record with a metallicity $[\mathrm{Fe}/\mathrm{H}]=-2.5$ (Letarte et al. 2006\nocite{leta06}). But not all clusters have survived as gravitationally bound objects to the present epoch.  Much like the halo, dwarf galaxies have stars with metallicities well below $[\mathrm{Fe}/\mathrm{H}]=-3$ (Kirby et al. 2008\nocite{kirb08}; Starkenburg et al. 2010\nocite{star10}). The relatively simple environments of the low mass dwarfs raises the prospect of identifying disrupted star clusters at much lower metallicity than has been possible before. By extension, this also gives us a unique tool to probe the formation and early evolution of the present dwarf galaxy population and to determine the sequence of events that led to the formation of the stellar Halo of the Milky Way. 

To date, all open clusters have been shown to be extremely homogeneous in elements heavier than Na (e.g. De Silva et al. 2006\nocite{desi06}, 2007\nocite{desi07}). The same holds true for most globular clusters, particularly in [Fe/H] and heavier elements (Sneden et al. 2008\nocite{sned08}). Bland-Hawthorn et al. (2010b\nocite{blan10b}) provide a theoretical framework for determining the mass of a homogeneous star cluster. Clusters with central column densities $\sim0.3$ g cm$^{-2}$ (e.g. open clusters) are homogeneous up to $10^5~\mathcal{M_{\odot}}$. More centrally concentrated clusters, with column densities $\sim3$ g cm$^{-2}$ (e.g. globular clusters), are homogeneous up to at least $10^7 \mathcal{M_{\odot}}$. We can look for signs of clumping in chemical abundance space ($\mathcal{C}$-space), particularly in heavier elements, in order to identify systems that have long since dissolved (Bland-Hawthorn et al. 2010a\nocite{blan10a}). Globular clusters are distinguished by more complex signatures in lighter elements, in particular, an anticorrelation between $\mathrm{O}/\mathrm{Mg}$ and $\mathrm{Na}/\mathrm{Al}$.

In \S 2, we revisit the recent observations of Sextans presented by Aoki et al. (2009). We reconsider their excellent analysis and demonstrate the possible existence of a disrupted stellar cluster at $[\mathrm{Fe}/\mathrm{H}]=-2.7$, well below the metallicity of the most metal poor globular cluster to date. In \S 3, we determine the statistical significance of the cluster in the context of inhomogeneous chemical evolution models. In \S 4, we estimate the mass and discuss the likely nature of this ancient star cluster before discussing implications for near-field cosmology in \S 5. Some final remarks are given in \S \ref{sect:summa}.

\section{Observational evidences}\label{sect:obser}
\noindent
In the following section, we will discuss three key observations suggesting that a significant fraction of the most metal-poor stars in the Sextans dSph were once members of a massive star cluster.

\subsection{Chemical abundance ratios}\label{sect:chemi}
\begin{figure*}[t]
\resizebox{\hsize}{!}{\includegraphics[trim=21.9mm 28mm 13.9mm 30mm,clip]{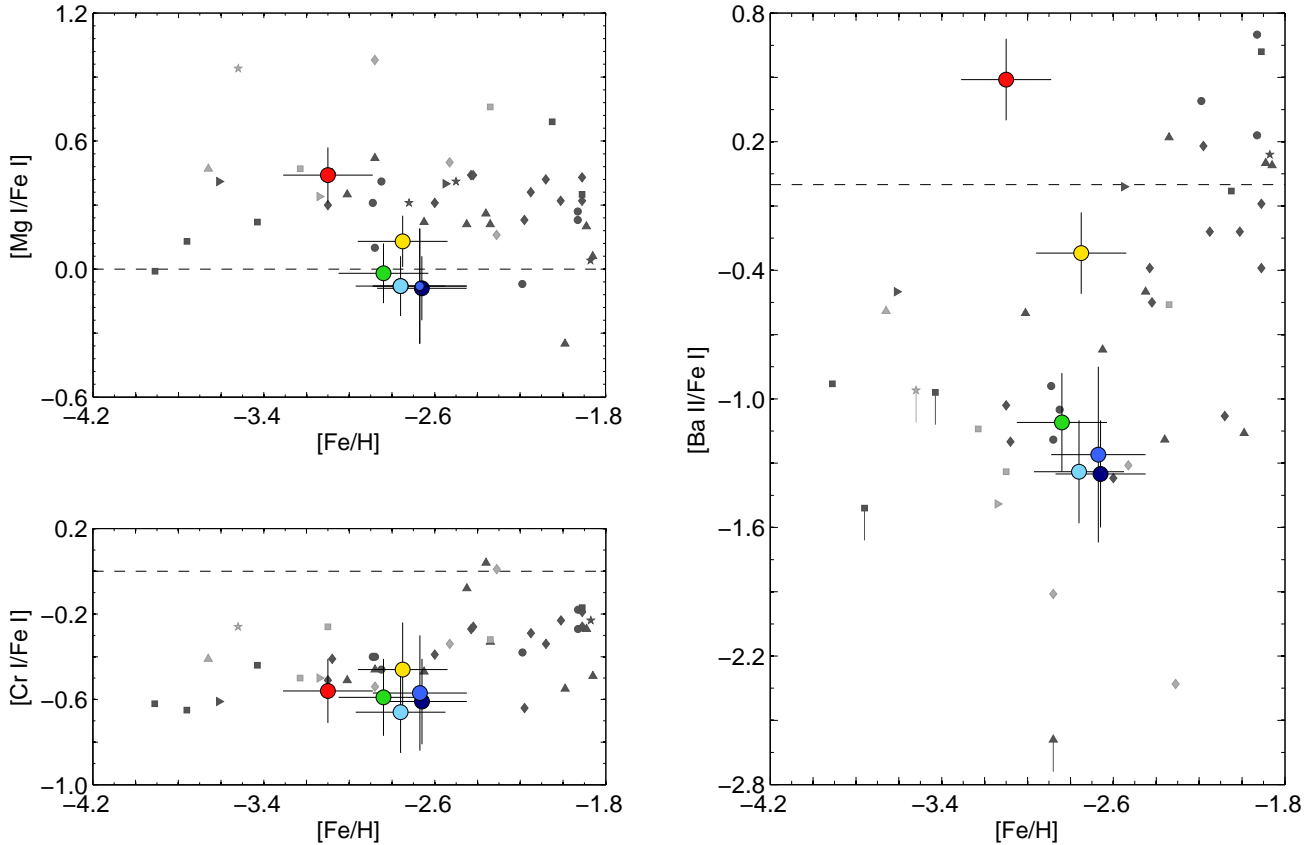}}
\caption{The data of Aoki et al. (2009) for Mg, Cr, and Ba displayed over the background of metal-poor stars of a number of dSphs (dark-grey symbols) and ultra-faint dwarf galaxies (light-grey symbols). The dSphs are: Sextans (circles, Shetrone et al. 2001; Tafelmeyer et al. 2010), Sculptor (squares, Shetrone et al. 2003; Geisler et al. 2005; Frebel et al. 2010a; Tafelmeyer et al. 2010), Ursa Minor (diamonds, Shetrone et al. 2001; Sadakane et al. 2004; Cohen \& Huang 2010), Draco (up-faced triangles, Shetrone et al. 2001; Fulbright et al. 2004; Cohen \& Huang 2009), Fornax (right-faced triangles, Letarte et al. 2010; Tafelmeyer et al. 2010), and Carina (pentagons, Shetrone et al. 2003; Koch et al. 2008). The ultra-faints are: Ursa Major II (squares, Frebel et al. 2010b), Coma Berenices (diamonds, Frebel et al. 2010b), Bo\"{o}tes I (up-faced triangle, Norris et al. 2010c), Leo IV (right-faced triangle, Simon et al. 2010), and Segue 1 (pentagon, Norris et al. 2010a). Note the evident clumping of the ``blue'' Sextans stars in all three diagrams. Note also their anomalous positions, particularly in Mg and to some extent in Cr. To be consistent with Aoki et al. (2009), the background data are converted to the solar abundance scale of Asplund et al. (2005).}
\label{fig:bckg}
\nocite{aoki09}
\nocite{shet01,shet03,tafe10,geis05,freb10a,sada04,cohe10,fulb04,cohe09,leta10,koch08,freb10b,norr10c,simo10,norr10a,aspl05}
\end{figure*}

\noindent
Based on high-resolution ($\mathcal{R}\simeq 40,000$) spectroscopy, Aoki et al. (2009\nocite{aoki09}) determined the chemical abundances of six very metal-poor stars ($[\mathrm{Fe}/\mathrm{H}]<-2.5$) in the Sextans dSph. Four of these stars display a subzero [Mg/Fe] ratio, with a small scatter around the weighted mean $\langle [\mathrm{Mg}/\mathrm{Fe}]\rangle =  -0.06$ (see Fig. \ref{fig:abnd}). Thus, they do not show the Mg-to-Fe enhancement (relative to the Sun) commonly observed in Galactic halo stars of similar metallicity (Cayrel et al. 2004\nocite{cayr04}; Barklem et al. 2005\nocite{bark05}; Lai et al. 2008\nocite{lai08}). This group of stars have very similar metallicity ($[\mathrm{Fe}/\mathrm{H}]\simeq -2.7$) and a closer inspection of Fig. \ref{fig:abnd} shows that they more or less tightly clump together in Cr and Ba as well (see also Fig. \ref{fig:bckg}). The clumping in the [Ba/Fe]--[Fe/H] plane around $[\mathrm{Ba}/\mathrm{Fe}]\simeq -1.3$ (excluding S~$11-37$, see below) is particularly striking. 

At least three of the four stars also group together in Na, Ca, and Ti (i.e., Ti~{\sc ii}). The star S~$14-98$ has a higher Ca abundance as compared to the other three subzero [Mg/Fe] stars, but as Aoki et al. (2009\nocite{aoki09}) points out, this may simply be due to a larger observational uncertainty in Ca for this star. 

In terms of the chemistry, the star S~$11-37$ (color coded green in Fig. \ref{fig:abnd}) may be regarded as a borderline case, and it is unclear whether it actually should be affiliated with the same group as S~$10-14$, S~$11-13$, and S~$14-98$ (color coded blue). It is more than $2\sigma$ away from the weighted mean of the ``blue'' stars in Ti,  $1\sigma$ away in Ba, and slightly more than $1\sigma$ away from the two ``blue'' stars for which Sc abundance could be measured (Sc was not measured in S~$10-14$). In all cases (including Mg), the star S~$11-37$ has an excess abundance as compared to the ``blue'' stars, and is closer to the typical abundances found in Galactic halo stars (e.g. Cayrel et al. 2004\nocite{cayr04}). Such a systematic behavior would not be expected if the abundance differences were due to random errors alone.  In what follows, we assume that S~$11-37$ does not belong to the ``blue'' group of stars. However, higher $S/N$ spectra and a detailed investigation of the abundance response to stellar atmosphere parameters are necessary in order to determine the true origin of this star.

It is clear from Figs. \ref{fig:abnd} and \ref{fig:bckg} that the ``blue'' stars tend to clump together in $\mathcal{C}$-space. This clumping implies a common chemical history, and we will here postulate that these stars were once members of a chemically homogeneous star cluster that is now, at least partly, dissolved. Note that the ``blue'' stars could in principle have been distributed anywhere along the mean trends as defined by the data of a number of dSphs in the Local Group (Fig. \ref{fig:bckg}, dark-grey symbols). The fact that they instead clump together at an anomalous position in $\mathcal{C}$-space (particularly in [Mg/Fe], but also, e.g., in [Cr/Fe], see Fig. \ref{fig:bckg}) argues for a cluster membership in the past. A possible argument against a common origin is that \mbox{S~$10-14$} has an observed [Na/Fe] ratio nearly $3\sigma$ below the other two potential cluster stars. This could, however, be a manifestation of the Na-O anti-correlation observed in Galactic globular clusters. This will be further discussed in \S \ref{sect:type}.  In principle, a comprehensive group-finding analysis (see Sharma \& Johnston 2009\nocite{shar09} for a presentation of the algorithm \textsf{EnLink}) would provide us with the likelihood of this being a true homogeneous clump in $\mathcal{C}$-space (Bland-Hawthorn et al. 2010a\nocite{blan10a}). Unfortunately, the number of stars in the current sample, particularly the number of stars \textit{not} belonging to the potential cluster, is too small and more data below $[\mathrm{Fe}/\mathrm{H}]=-2.5$ are required in order to give a reliable estimate of this probability. 

Finally, we note that the star S~$12-28$ (color coded yellow) has similar abundances to the ``blue'' stars in Ti and Cr. However, the fact that it is over $5\sigma$ away in Ba makes the affiliation with the potential cluster stars improbable.

\subsection{Metallicity distribution function}\label{sect:metal}
\noindent
Starkenburg et al. (2010\nocite{star10}) performed a re-calibration of the Ca {\sc ii} triplet (CaT) lines and determined new metallicities for metal-poor stars in four classical dSphs, including Sextans, that are a part of the Dwarf galaxies Abundances and Radial-velocities Team ({\sf DART}) program. The new data reveal a ``bump'' in the metallicity distribution function (MDF) of Sextans around the inferred $[\mathrm{Fe}/\mathrm{H}] \simeq-2.9$. By applying to the {\sf DART} data the presumed Ca-to-Fe enhancement of $[\mathrm{Ca}/\mathrm{Fe}]=+0.25$ (Starkenburg et al. 2010\nocite{star10}; see also \S \ref{sect:esti}), the ``bump'' appears around $[\mathrm{Ca}/\mathrm{H}]\simeq -2.65$ (see Fig. \ref{fig:mdf}). Interestingly, this is close to the weighted mean abundance of Ca for the three potential cluster (``blue'') stars identified in the data of Aoki et al. (2009\nocite{aoki09}). If the ``bump'' indicates a real excess of stars in the MDF of Sextans, this excess could be  due to the presence of a relic star cluster.

A ``bump''  at $[\mathrm{Fe}/\mathrm{H}]\sim -3$ exists also in the medium-resolution ($\mathcal{R}\simeq 7,000$) data of Sextans\footnote{Kirby et al. (2011a\nocite{kirb11a}) make use of a particular metallicity extraction technique that determines an ``overall metallicity'' of spectrally low-resolved stars, rather than [Fe/H].} presented by Kirby et al. (2011a\nocite{kirb11a}). They noticed that similar ``bumps'' are present in the MDFs of the Draco and Ursa Minor dSphs as well. They speculated that these ``bumps'' could be due to an early burst of star formation in these galaxies, but did not consider the particular possibility that the ``bumps'' may be indirect signs of star clusters.  In fact, if stars were formed in clusters already at the earliest epochs of galaxy formation, irregular and ``bumpy'' MDFs are likely to be expected, i.e., depending on the functional form of the local ICMF (see \S \ref{sect:discu}). The irregularities should appear preferentially in regions of small number statistics, such as in the metal-poor tails of the MDFs of the least massive galaxies. If so, conclusions from a straightforward comparison between MDFs of individual dwarf galaxies should be drawn with some caution since any recognized difference could simply be due to the effects of stochasticity (see discussion in \S \ref{sect:discu}).

\subsection{Kinematical substructure?}\label{sect:kinem}
\noindent
There is evidence for a kinematically cold substructure in the central region of Sextans, possibly originating from a remnant star cluster (Kleyna et al. 2004\nocite{kley04}; Battaglia et al. 2011\nocite{batt11}; see, however, Walker et al. 2006\nocite{walk06}). Battaglia et al. (2011\nocite{batt11}) identify a group of stars with cold kinematics at a projected distance of $R<0^{\circ}.22$ from the optical center. This substructure has an observed mean metallicity of $[\mathrm{Fe}/\mathrm{H}] \simeq-2.6$ (Battaglia et al. 2011\nocite{batt11}), which is close to that of the ``blue'' group of stars discussed in \S \ref{sect:chemi}, and a common origin is not implausible. It should, however, be noted that the potential cluster stars identified in the sample of Aoki et al. (2009\nocite{aoki09}) are located around $2$ core radii, at a projected distance of $0^{\circ}.4 \lesssim R \lesssim 0^{\circ}.6$, well outside $R=0^{\circ}.22$. Evidently, if the ``blue'' stars and the stars associated with the central substructure do originate from the same disrupted (or disrupting) cluster, its former member stars are scattered over a significant fraction of the entire galaxy (Pe\~{n}arrubia et al. 2009\nocite{pena09}). In this framework, it is close at hand to assume that the ``bump'' of stars in the MDF, in fact, is a manifestation of the very same cluster. In what follows, we will assume that this is the case and from the relative size of the ``bump'', we will estimate the total mass of stars that were once members of the cluster. 
 
We note that Sextans is unusual in having the lowest-density dark-matter halo of any of the classical dSph systems (e.g., Gilmore et al. 2007\nocite{gilm07}). The stars in the dSphs are likely to be located well within a constant density dark matter core, which implies that tidal stresses generally are low. For Sextans, they would be especially low because of the low density. This would inhibit the disruption of clusters, leading to longer disruption times (cf. Pe\~{n}arrubia et al. 2009\nocite{pena09}; see also Walker \& Pe\~{n}arrubia 2011\nocite{pena11}). N-body simulations of the disruption process in an environment typical for dSphs like Sextans should be performed in order to shed light on this issue.

\section{Modeling of Sextans}\label{sect:model}
\noindent
To illustrate what effect the presence of a massive star cluster could have on how $\mathcal{C}$-space is populated with stars, and to estimate the mass of the cluster, we will model the chemical evolution of a small dwarf galaxy, comparable in mass to the Sextans dSph.

\subsection{Stochastic chemical evolution model}\label{sect:stoch}
\noindent
The model is based on the assumption that all stars originally are formed as members of chemically homogeneous star clusters. The mass distribution of newly formed clusters is known as the initial cluster mass function (ICMF), and is assumed to have the form

\begin{equation}
\mathrm{d}N/\mathrm{d}M_{*}=\chi(M_{*}) \propto M_{*}^{-\gamma},
\label{eq:dndm}
\end{equation}

\noindent
where $M_{*}$ denotes the cluster mass. Although the particular situation for dwarf galaxies is not clear, especially at high redshifts, observations of the local Universe support a generic slope of $\gamma \simeq 2$ in most environments (e.g. Lada \& Lada 2003\nocite{lada03}; Fall et al. 2005\nocite{fall05}, 2009\nocite{fall09}; Elmegreen 2010\nocite{elme10}), and cluster masses typically within the range $50\lesssim M_{*}/\mathcal{M_{\odot}} \lesssim 2\times 10^5$ (Larsen 2009\nocite{lars09}; see also Bland-Hawthorn et al. 2010b\nocite{blan10b} for a detailed discussion). This mass range and value of $\gamma$ define the fiducial ICMF in the present investigation.  

The vast majority of star clusters in the Milky Way is found to be internally chemically homogeneous to a high degree (De Silva et al. 2006\nocite{desi06}, 2007\nocite{desi07}). This is to be expected from time scale arguments (Bland-Hawthorn et al. 2010b\nocite{blan10b}) and is incorporated in the modeling by forcing each molecular cloud to be chemically homogeneous during its entire star formation period. This implies that a volume $V_{\mathrm{cld}}=M/\bar{\rho}$, associated with each cluster, is made chemically homogeneous before star formation commences. Here, $M=M_{*}/\epsilon$ is the mass of the molecular cloud and $\bar{\rho}$ is the mean density of the interstellar medium (ISM). The star formation efficiency is set to $\epsilon=0.1$.  

Associated with each cluster is also a mixing volume, $V_{\mathrm{mix}}$, which is defined as the volume enriched by the ejecta from the SNe in the cluster. The ejecta are assumed to mix homogeneously with the gas within $V_{\mathrm{mix}}$. The size of the mixing volume can be expressed as (cf. Karlsson 2005\nocite{karl05}; Karlsson et al. 2008\nocite{karl08})

\begin{equation}
V_{\mathrm{mix}} = \frac{4\pi}{3}\left(6D_{\mathrm{trb}}\times t+\left( \frac{3  \sum_{j=1}^{k} M_{\mathrm{r}}}{4\pi \bar{\rho}}\right) ^{2/3}\right)^{3/2},
\label{eq:vmix}
\end{equation}

\noindent
where $D_{\mathrm{trb}}$ (kpc$^2$~Myr$^{-1}$) denotes the turbulent diffusion coefficient of the interstellar gas and $M_{\mathrm{r}}$ denotes the typical mass swept up by a single SN remnant when the gas has cooled enough to form new stars. This occurs roughly when the remnant merges with the ambient medium, which implies that $M_{\mathrm{r}}\sim10^5~\mathcal{M_{\odot}}$ (Ryan et al. 1996\nocite{ryan96}). We have introduced a small amount of randomness in the calculation of $V_{\mathrm{mix}}$ insofar as $\log(M_{\mathrm{r}})$ is drawn from a normal distribution centered on $\log(M_{\mathrm{r}})=5$ with a width of $0.25$. For simplicity, we set $D_{\mathrm{trb}}=0.$

The formation of clusters is assumed to occur randomly in the simulation box. Each cluster, randomly distributed according to Eq. (\ref{eq:dndm}), is formed from a gas cloud with volume $V_{\mathrm{cld}}$ which is made chemically homogeneous. If the cluster is massive enough to form high-mass stars (with masses distributed according to the Salpeter IMF), the combined ejecta from the SNe are then mixed homogeneously within the volume $V_{\mathrm{mix}}$, given by Eq. (\ref{eq:vmix}). With the formation of each cluster, an amount of $f_{\mathrm{out}}\times M_{*}$ of gas is subtracted from the galaxy's remaining gas mass. This includes mass expelled from the galaxy and mass locked up in long-lived stars and stellar remnants. The coefficient $f_{\mathrm{out}}$ is empirically determined such that the peak of the model MDF is matched to the peak of the observed MDF (see Fig. \ref{fig:mdf}). For Sextans, $f_{\mathrm{out}}=37.4$. And so, the chemical enrichment proceeds. Star formation is finally terminated when the gas density reaches $\bar{\rho}=1\times 10^5~\mathcal{M_{\odot}}$, unless there is too little gas in the system for the final cluster to form. Further details of the model can be found in Bland-Hawthorn et al. (2010a\nocite{blan10a}).

\subsection{Stellar mass of Sextans}\label{sect:masso}
\noindent
We can estimate the present stellar mass of the Sextans dSph simply by adding up the light from its stars. The luminosity of Sextans is $4.1\pm 1.9\times 10^5~\mathcal{L_{\odot}}$, which corresponds to an absolute magnitude of $M_V=-9.2\pm 0.5$ (Irwin \& Hatzidimitriou 1995\nocite{irwi95}). Since the majority of the stars in Sextans is believed to be older than $10$ Gyr (e.g. Lee et al. 2009\nocite{lee09}), we can estimate the mass simply by integrating the light from a single stellar population (SSP) of \mbox{$12$ Gyr}, assuming a metallicity of $Z=10^{-2}\mathcal{Z_{\odot}}$.  In this way, we approximate the present stellar mass to $M_{\mathrm{Sxt}}=8.9\pm 4.1\times 10^5~\mathcal{M_{\odot}}$ for a Salpeter IMF.  Here, the symmetric uncertainty in luminosity is directly translated to an uncertainty in mass. 

We can, however, do slightly better and account for the observed spread in stellar age. Lee et al. (2009\nocite{lee09}) determined the star formation history (SFH) of Sextans and noticed a spread in the data of at least $6-8$ Gyr (excluding blue stragglers), even if most stars were recovered in the oldest age bins.  From their inferred star formation rate (SFR; their case B), we adopt an SFH with four age bins; $15-13$ Gyr, $13-11$ Gyr, $11-9$ Gyr, and $9-7$ Gyr, containing $66\%$, $25\%$, $7\%$, and $2\%$ of the stars, respectively. The stars in each bin are assumed to be adequately described by an SSP of $14$ Gyr, $12$ Gyr, $10$ Gyr, and $8$ Gyr, respectively. The metallicity of the SSPs are all set to $10^{-2}\mathcal{Z_{\odot}}$, except for the $8$ Gyr SSP, for which a metallicity of $10^{-1.5}\mathcal{Z_{\odot}}$ is assumed.  For the light integration, the isochrones by Marigo et al. (2008\nocite{mari08}) are adopted.\footnote{See http://stev.oapd.inaf.it/cgi-bin/cmd\_2.2} From this modeling of the stellar content, we determine the present stellar mass of Sextans to $M_{\mathrm{Sxt}}=9.4\pm 4.4\times 10^5~\mathcal{M_{\odot}}$ (Salpeter IMF). This is consistent with the mass $M_{\mathrm{Sxt}}=8.5\times 10^5~\mathcal{M_{\odot}}$, obtained by Woo et al. (2008\nocite{woo08}). The stated uncertainty formally includes, apart from the observed error in the luminosity, an estimate of the uncertainty from the inferred SFR ($\sigma_{\mathrm{Sxt,SFR}}=0.05\times 10^5~\mathcal{M_{\odot}}$) as derived from the data of Lee et al. (2009\nocite{lee09}), as well as the intrinsic uncertainty due to the small number of luminous red giants in galaxies of this size ($\sigma_{\mathrm{Sxt,intr}}=0.4\times 10^5~\mathcal{M_{\odot}}$). Evidently, these latter two uncertainties have a negligible impact on the total uncertainty. 

We note that any SFH inferred from photometric data may be marred with model dependent, systematic errors, e.g. from theoretical isochrone fitting, that are difficult to get a handle on and which add to the random errors in the SFR. The uncertainty $\sigma_{\mathrm{Sxt,SFR}}$ could therefore be larger than the adopted value given above. From a set of completely random SFHs, the uncertainty in the mass determination is estimated to no more than $\sigma_{\mathrm{Sxt,SFR}}=0.3\times 10^5~\mathcal{M_{\odot}}$, assuming that the youngest stars in Sextans are $8$ Gyr. In the case of a Kroupa IMF (Kroupa 2001\nocite{krou01}), we get a stellar mass today of $5.4\pm 2.5\times 10^5~\mathcal{M_{\odot}}$. Note that the total mass (within $2.3$ kpc) of the dark-matter dominated Sextans is on the order of $2-4\times 10^8~\mathcal{M_{\odot}}$ (Battaglia et al. 2011\nocite{batt11}).

In order to model the chemical evolution and reconstruct the galaxy with the correct number of clusters, we need to estimate the initial mass of the stellar populations in Sextans. In a $12$ Gyr population, for instance, only stars below $m\simeq 0.83~\mathcal{M_{\odot}}$ are still present. Roughly $40-60\%$ of the stellar mass is therefore gone (excluding stellar remnants, which have a negligible contribution to the total luminosity), depending on the IMF. This implies that, for a Salpeter (Kroupa) IMF, the birth mass of the stellar component of Sextans was $M_{\mathrm{Sxt,init}}=1.6\pm 0.8~(1.3\pm 0.6)\times 10^6~\mathcal{M_{\odot}}$. This mass will be used below to estimate the age-weighted, initial mass of the star cluster.

In \S \ref{sect:mass}, we will, along with the estimate of the most probable mass of the cluster, attempt to get a handle on the various uncertainties involved. To this end, a detailed distribution describing the uncertainty in the mass estimate of Sextans is desired. The luminosity of Sextans determined by Irwin \& Hatzidimitriou (1995\nocite{irwi95}) has a large uncertainty due to confusion between the dwarf and a background galaxy cluster. Consequently, a gaussian uncertainty distribution, implied by the stated symmetric error in luminosity, will have a non-negligible tail of negative luminosities. The probability of obtaining such false detections can, however, be suppressed by considering integral fluxes only over smaller regions centered on the galactic core. As these fluxes are less affected by background noise, the probability of a non-detection is reduced, keeping in mind that the mass inferred from these central fluxes is lower than the total stellar mass of the galaxy. 

Taking this suppression into account and retaining the standard deviation in the luminosity stated by Irwin \& Hatzidimitriou (1995\nocite{irwi95}), we assume that the mass probability density function of Sextans is better approximated by a generalized Poisson distribution rather than the normal distribution. Its functional form is $e^{-m t}(m t)^{k_{\mathrm{R}}}/\Gamma(k_{\mathrm{R}}+1)$, where $m$ denotes the mass, $t$ is a scale factor, and the parameter $k_{\mathrm{R}}$ is a non-negative real number. Factorial $k_{\mathrm{R}}$ is governed by the gamma function $\Gamma(k_{\mathrm{R}}+1)$. The specific values of  $k_{\mathrm{R}}$ and $t$ are determined from the two relations \mbox{$k_{\mathrm{R}}/t=1.6\times 10^6~\mathcal{M_{\odot}}$} and $\sqrt{(k_{\mathrm{R}}+1)/t^2}=0.8\times 10^6~\mathcal{M_{\odot}}$ (Salpeter IMF), describing the most probable mass (mode) and the standard deviation, respectively. It follows that \mbox{$k_{\mathrm{R}}=5.5$} and \mbox{$t=3.3\times 10^{-6}~\mathcal{M_{\odot}}^{-1}$}. As compared to a gaussian, this distribution has a more extended tail towards high masses, distinct from the suppressed tail towards low masses. It retains a fairly symmetric shape, nonetheless. Only a detailed study of how the background flux varies over the field in the immediate vicinity of Sextans, can give us further information on the actual shape of the mass probability density function.

\section{Nature of cluster}\label{sect:natur}
\noindent
We shall now turn to the relic star cluster itself. In the following sections, we will estimate the initial stellar mass of the cluster and briefly discuss what possible type of cluster it might have been.

\subsection{Mass of cluster}\label{sect:mass}
\noindent
The mass of the dissolved cluster can be estimated independently from the size of the stellar ``bump'' in the MDF inferred from the re-calibrated {\sf DART} data (Starkenburg et al. 2010\nocite{star10}) and from the identification of cluster stars in the data by Aoki et al. (2009\nocite{aoki09}). Let us first estimate the mass from the ``bump'' by comparing the simulated MDF with the observed one and then calculate the combined estimate, including the information on the number of ``blue'' stars identified as potential cluster stars in \S \ref{sect:chemi}.

\subsubsection{Mass estimate from the re-calibrated \textsf{DART} MDF}\label{sect:esti}
\begin{figure}[t]
\resizebox{\hsize}{!}{\includegraphics[trim=1mm 8mm 3mm 15.2mm,clip]{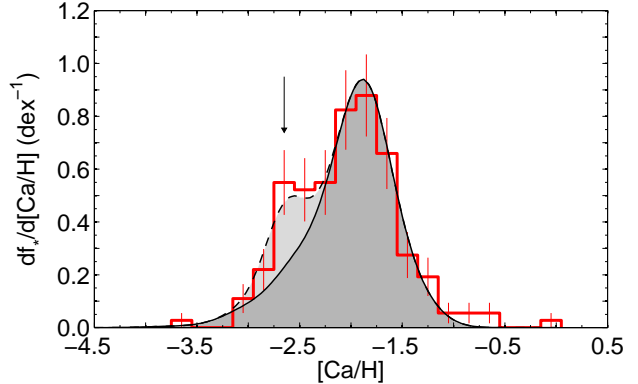}}
\caption{Metallicity distribution function for stars in the Sextans dSph galaxy, as given by $\mathrm{d}f_{*}/\mathrm{d}[\mathrm{Ca}/\mathrm{H}]$. The quantity $f_{*}$ is the fraction of stars that fall below each [Ca/H] bin (1 dex). The black, solid line denotes the fiducial distribution of [Ca/H], predicted from the stochastic chemical evolution model presented in \S \ref{sect:model} while the black, dashed line denotes the corresponding distribution, including an $M_{\star}=2.0\times 10^5~\mathcal{M_{\odot}}$ star cluster at an assumed Ca abundance of $[\mathrm{Ca}/\mathrm{H}] = -2.65$ (arrow). Both MDFs are convolved with a $\sigma = 0.2$ dex gaussian. The red step function (with Poissonian noise) shows the observed distribution of [Ca/H], as inferred from the {\sf DART} data (Starkenburg et al. 2010). Note that the data point at $[\mathrm{Ca}/\mathrm{H}]=-2.65$ lies more than $4\sigma$ above the model of the underlying galaxy.}
\label{fig:mdf}
\nocite{star10}
\end{figure}

\noindent
As suggested by the appearance of  the ``blue'' stars identified in \S \ref{sect:chemi}, we will assume that the ``bump''  predominantly originates from a single, massive star cluster. This will have certain implications for the interpretation of the shape of the MDF. In the {\sf DART} program, the [Fe/H] abundances are inferred from the CaT equivalent widths, which means that they de facto are translated [Ca/H] abundances. From Starkenburg et al. (2010\nocite{star10}, their Appendix A), it is clear that this translation simply reads as an offset very close to $0.25$ dex (i.e., $[\mathrm{Ca}/\mathrm{Fe}]=+0.25^{+0.01}_{-0.00}$ dex) over the entire range $-4\le [\mathrm{Fe}/\mathrm{H}] \le -0.5$, for which the CaT calibration is valid. Any observed deviation from this presumed Ca-to-Fe enhancement (recall that the potential cluster stars have $[\mathrm{Ca}/\mathrm{Fe}]\simeq +0.05$) would lead to an erroneous [Fe/H] determination. Hence, a translation like the one above merely shifts the {\sf DART} MDF to the left in [Ca/H]-space, it does not reshape it into the form it actually should have if displayed in true [Fe/H]-space. In the mass-estimation analysis that follows below, we therefore choose to replace [Fe/H] with [Ca/H] as a measure of metallicity, in order to get a consistent comparison between models and observations (see Fig. \ref{fig:mdf}). By applying a constant offset (sufficient for our purposes), the (re-)conversion to observed Ca abundances reads \mbox{$[\mathrm{Ca}/\mathrm{H}]_{\mathrm{obs}} = [\mathrm{Fe}/\mathrm{H}]_{\mathrm{{\sf DART}}} + 0.25$}.

The simulated MDF is composed of two main components; an underlying MDF of the galaxy and a cluster placed at $[\mathrm{Fe}/\mathrm{H}]=-2.7$ (i.e., $[\mathrm{Ca}/\mathrm{H}]=-2.65$, assuming that \mbox{$[\mathrm{Ca}/\mathrm{Fe}]=0.05$}). A significant fraction of the stars in the sample of Aoki et al. (2009\nocite{aoki09}) are not identified with the remnant cluster, as evident from Fig. \ref{fig:abnd}. The star least likely to be a cluster member (color coded red in Figs. \ref{fig:abnd} and \ref{fig:bckg}) is the most metal-poor star in the sample with $[\mathrm{Fe}/\mathrm{H}]=-3.10$. Moreover, the weighted mean metallicity of the ``non-blue'' stars is $[\mathrm{Fe}/\mathrm{H}]=-2.90\pm0.12$, while that of the ``blue'' stars is $[\mathrm{Fe}/\mathrm{H}]=-2.70\pm0.12$.  This implies that the ``bump'' of cluster stars must lay on top of a broader distribution that extends down towards the extremely metal-poor regime (see Fig. \ref{fig:mdf}). We modeled the underlying MDF using the chemical evolution model described in \S \ref{sect:stoch}, assuming a Salpeter IMF and primordial ($Z=0$), metal-independent SN yields from Nomoto et al. (2006\nocite{nomo06}). The Ca yields were divided by a factor of $1.8$ ($=-0.25$ dex) to match the mean [Ca/Fe] observed in the Galactic halo. Similarly, the Mg yields were divided by a factor of $1.5$ ($=-0.18$ dex). Since we are primarily interested in the chemical evolution below $[\mathrm{Fe}/\mathrm{H}]\simeq -2$, the enrichment due to intermediate-mass stars and SNe type Ia were not included in the modeling. A total mass of $\sim 1.6\times 10^6~\mathcal{M_{\odot}}$ was imposed for the system ``cluster $+$ underlying galaxy''. 

By matching the number of observed stars in the ``cluster region'', arbitrarily defined as the region in the metallicity interval $-3.15\le [\mathrm{Ca}/\mathrm{H}] \le -2.15$, with the model MDF (integrated over the same range), a fiducial cluster mass was determined. The procedure is illustrated in Fig. \ref{fig:mdf}. The integrated number of observed stars within this region is more than $3\sigma$ above the model of the underlying galaxy while the number of stars in the central metallicity bin at $[\mathrm{Ca}/\mathrm{H}]=-2.65$ is more than $4\sigma$ above the underlying galaxy. The ``bump'' (light-gray shaded area) contains $12.3\%$ of the total stellar mass, which implies an initial cluster mass of \mbox{$M_{*,\mathrm{init}}=2.0\times 10^5~\mathcal{M_{\odot}}$}.  For a Kroupa IMF, the initial cluster mass is slightly lower, or $M_{*,\mathrm{init}}=1.6\times 10^5~\mathcal{M_{\odot}}$. These masses translate into present-day cluster masses of $M_{*}=1.2~(0.7)\times 10^5~\mathcal{M_{\odot}}$, for a Salpeter (Kroupa) IMF.

In order to get a proper handle on the uncertainties involved and to estimate the most probable cluster mass, we must account simultaneously for the uncertainties of all parameters that go into the calculation of the cluster mass.  Given the observed and modeled MDF,  the mass of the star cluster is determined from the expression:

\begin{equation}
\hat{M}_{*,\mathrm{init}} = \hat{M}_{\mathrm{Sxt,init}}\frac{\hat{N}_{\mathrm{in}}-\hat{C}_{\mathrm{cnt}}}{\hat{N}_{\mathrm{in}}+\hat{N}_{\mathrm{out}}},
\label{eq:mcl}
\end{equation} 

\noindent
where $\hat{N}_{\mathrm{in}}$ denotes the observed number of stars \textit{within} the ``cluster region'', $\hat{N}_{\mathrm{out}}$ denotes the observed number of stars \textit{outside} this interval, and $\hat{C}_{\mathrm{cnt}}$ denotes the \textit{continuum} level defined as the mean number of stars (normalized to the observations) in the part of the underlying model galaxy that falls within the ``cluster region''. The mass of Sextans is denoted by $\hat{M}_{\mathrm{Sxt,init}}$. Note that all parameters in \mbox{Eq.~(\ref{eq:mcl})} are random variables (here, explicit random variables are indicated by the ``hat'' symbol), each to which a specific distribution must be specified in order to determine the distribution of cluster masses, $\hat{M}_{*,\mathrm{init}}$.  

To evaluate the uncertainty in the cluster mass associated with the observations of the MDF, the random variables $\hat{N}_{\mathrm{in}}$ and $\hat{N}_{\mathrm{out}}$ are assumed to be distributed according to the Poisson distribution $e^{-\mu}\mu^k/k!$, with $\mu_{\mathrm{in}}=71$ and $\mu_{\mathrm{out}}=111$, respectively. These numbers are extracted from the \textsf{DART} data of Sextans (Starkenburg et al. 2010\nocite{star10}). The corresponding dispersion in the cluster mass due to this uncertainty is $\sigma_{*,\mathrm{init}}=0.7\times 10^5~\mathcal{M_{\odot}}$. Moreover, as a result of stars being formed in clusters, the shape of the model MDF of the underlying galaxy should vary somewhat from galaxy to galaxy. We collected $100$ different realizations, assuming the fiducial ICMF discussed in \S \ref{sect:stoch},  to map out this effect. 

With the assumption that the ``bump'' at $[\mathrm{Fe}/\mathrm{H}]=-2.7$ is predominantly due to a single cluster, we accepted only relatively smooth shapes, disregarding any model with obvious ``secondary bumps'' in the metallicity range of interest. The variations in the continuum level were found to be nearly Poisson distributed around the mode (i.e., the continuum level of the fiducial model depicted in Fig. \ref{fig:mdf}) with a standard deviation of $\sigma_{\mathrm{cnt}}=6.2$. We note that with such a $\sigma$, the distribution is fairly close to being normal.  It follows that the uncertainty of the cluster mass due to the underlying model galaxy alone is $\sigma_{*,\mathrm{init}}=0.7\times 10^5~\mathcal{M_{\odot}}$. This uncertainty depends on parameters like the ICMF which have an impact on the shape of the MDF. A larger fraction of smaller clusters would, for example, result in a smaller dispersion of the cluster mass.  Note that, for an ICMF with a slope departing from $\gamma = 2$, the fluctuations in the continuum level may be described by a distribution that deviates from that of the Poisson distribution. A strongly varying star formation rate would also affect the continuum level, but this variation is not included in the present investigation.  Finally, from the uncertainty in the mass of Sextans, governed by the distribution introduced in \S \ref{sect:masso}, the uncertainty in the mass of the cluster is estimated to $\sigma_{*,\mathrm{init}}=0.9\times 10^5~\mathcal{M_{\odot}}$.

When accounting for the uncertainties in the observed and modelled MDF and the mass of Sextans simultaneously, we derive, from the expression in Eq. (\ref{eq:mcl}), a most probable cluster mass of $M_{*,\mathrm{init}}=1.6^{+1.6}_{-1.1}\times 10^5~\mathcal{M_{\odot}}$ (Salpeter IMF). This mass is slightly lower than the fiducial mass deduced above. The asymmetric uncertainties are calculated such that there is a $68.3\%$ probability of having a cluster mass within the stated interval, where the endpoints denote the intersections between the distribution and a horizontal line (cf. Fig. \ref{fig:dst}). Note that the black, dashed line in Fig. \ref{fig:dst} exhibits a non-zero probability having negative cluster masses. This just reflects the small statistical chance that the continuum level of the underlying galaxy may lie above the observed MDF. In these cases, no cluster is detected. If we take this into account,  the mean cluster mass is  $2.3\times 10^5~\mathcal{M_{\odot}}$ while the median is  $2.1\times 10^5~\mathcal{M_{\odot}}$.

\subsubsection{Combined mass estimate}\label{sect:comb} 
\begin{figure}[t]
\resizebox{\hsize}{!}{\includegraphics[trim=22mm 14mm 5mm 21mm,clip]{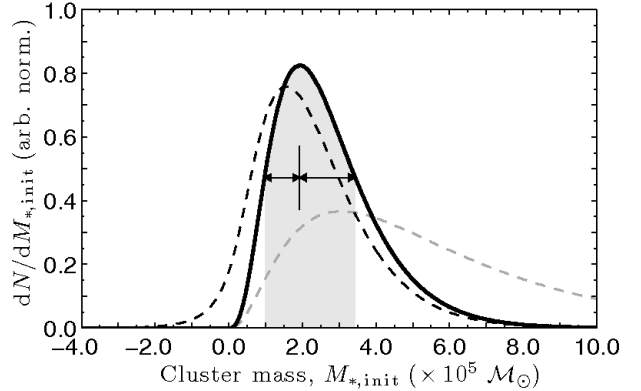}} % trim = l b r t
\caption{The uncertainty in the estimation of the cluster mass. The black, dashed line denotes the distribution derived from Eq. (\ref{eq:mcl}), while the black, solid line denotes the combined, conditional distribution given that three out of six stars below $[\mathrm{Fe}/\mathrm{H}]=-2.5$ are identified as cluster stars in the data of Aoki et al. (2009). The probability for this to occur is indicated by the gray, dashed line. The most probable mass is determined to $1.9^{+1.5}_{-0.9}\times 10^5~\mathcal{M_{\odot}}$ for a Salpeter IMF. The (asymmetric) uncertainty in the estimation is indicated by the horizontal arrows. There is a $68.3\%$ probability that the cluster mass falls within this interval (shaded area).}
\label{fig:dst}
\end{figure}

\noindent
We note that the predicted fraction of cluster stars below $[\mathrm{Fe}/\mathrm{H}]=-2.5$ in the particular model displayed in Fig. \ref{fig:mdf} is $40\%$. This implies a nearly 1-in-2 chance (binomial statistics) of finding three, or more, cluster stars in a random sample of six stars in this metallicity range. In less than $5\%$ of the cases, zero cluster stars are expected in a sample of six stars. Clearly, this is consistent with the recognition of three (or four, including S~$11-37$) potential cluster stars (\S \ref{sect:chemi}) in the data of Aoki et al. (2009\nocite{aoki09}). The likelihood of this independent finding may be utilized to put additional constraints on the mode and mass range of the cluster, as shown in Fig. \ref{fig:dst}. It should be pointed out that the stars in the sample of Aoki et al. (2009\nocite{aoki09}) only probes a fraction of the entire stellar population of Sextans and the local fraction of  ``blue'' stars may very well change with distance from the galactic center. As a zeroth-order approximation, we will, however, assume that the fraction of former member stars of the relic cluster at a distance $0^{\circ}.4 \lesssim R \lesssim 0^{\circ}.6$ is equal to the average fraction taken over the entire galaxy out to $R=1^{\circ}.8$, which is the outer boundary in the \textsf{DART} survey (Battaglia et al. 2011\nocite{batt11}).  

At face value, the identification of the three ``blue'' stars in Fig. \ref{fig:abnd}, if they do belong to a dispersed cluster, point towards a cluster mass on the order of a few hundred thousand solar masses. The probability that exactly three out of six stars below $[\mathrm{Fe}/\mathrm{H}]=-2.5$ should belong to the cluster is higher than for any other choice of $k$ (i.e., assuming a Salpeter IMF and a galaxy mass of $M_{\mathrm{Sxt,init}}=1.6\times 10^6~\mathcal{M_{\odot}}$) in the mass range $2.2\times 10^5\le M_{*,\mathrm{init}}/\mathcal{M_{\odot}}\le 3.6\times 10^5$. That is to say, in this mass range, the binomial distribution \mbox{$\mathrm{Bin}(p_{M_{*}},k,n=6)$}  peaks at $k=3$, where $p_{M_{*}}\equiv M_{*,\mathrm{init}}/M_{\mathrm{Sxt,init\!\!}}\mid_{\,[\mathrm{Fe}/\mathrm{H}]<-2.5}$.  Taking into account the uncertainty in the mass of the galaxy, the most probable cluster mass is $M_{*,\mathrm{init}}=3.1\times 10^5~\mathcal{M_{\odot}}$ for $k=3$. Given this conditional constraint, the combined estimate of the most probable, initial mass of the dissolved cluster (with asymmtric uncertainties, see Fig. \ref{fig:dst}) is determined to $M_{*,\mathrm{init}}=1.9^{+1.5}_{-0.9}~(1.6^{+1.2}_{-0.8})\times 10^5~\mathcal{M_{\odot}}$, for a Salpeter (Kroupa) IMF. This is our most reliable estimate of the mass of the relic star cluster.

\subsection{Type of cluster}\label{sect:type}
\noindent
Besides its relatively large initial mass, can we say something more specific about the possible nature of the relic star cluster?  Most Galactic open clusters are found in the mass range $10^2\lesssim M_{*}/\mathcal{M_{\odot}}\lesssim 10^4$, but more massive, young star clusters do exist, like Westerlund~1. This cluster has a dynamical mass of $1.5\times 10^5~\mathcal{M_{\odot}}$ (Mengel \& Tacconi-Garman 2009\nocite{meng09}). Another, intermediate-age cluster, namely GLIMPSE-CO1, has an estimated mass of $8\times 10^4\mathcal{M_{\odot}}$ (Davies et al. 2011\nocite{davi11}). These clusters are roughly in the same mass range as the remnant cluster in Sextans. Moreover, even if the average present-day mass of the open clusters in the Milky Way is about $7\times 10^2~\mathcal{M_{\odot}}$, the average mass of the ICMF could have been as high as $4.5\times 10^3~\mathcal{M_{\odot}}$, owing to efficient evaporation (Piskunov et al. 2008\nocite{pisk08}). Although rare, open clusters with initial masses on the order of $10^5~\mathcal{M_{\odot}}$ are therefore conceivable, at least in the Galactic Disk. 

The masses of Galactic globular clusters (GCs, hereafter) are, on the other hand, found in the range $1\times 10^4 \lesssim M_{*}/\mathcal{M_{\odot}} \lesssim 3\times 10^6$. The Sextans cluster falls right in the middle of this range. GCs also show specific elemental abundance correlations, such as the Na-O and the Al-Mg anticorrelations. It has recently been established that these anticorrelations are caused by internal pollution of the intracluster gas (see Carretta et al. 2009\nocite{carr09} for a discussion).  Such correlations are not observed in open clusters and could therefore, hypothetically be used to discern the nature of the relic cluster. Unfortunately, Aoki et al. (2009\nocite{aoki09}) were neither able to measure oxygen, nor aluminum in any of the stars in their sample. 

Interestingly, however, the ``blue'' star S~$10-14$ has a significantly lower Na abundance than the other two potential cluster stars, with a measured Na-to-Fe ratio of $[\mathrm{Na}/\mathrm{Fe}]=-0.72$ (see Fig. \ref{fig:abnd}). In comparison, S~$11-13$ and S~$14-98$ both have $[\mathrm{Na}/\mathrm{Fe}]\simeq -0.1$. This may be suggestive of an Na-O anticorrelation present in the data. We note that all the ``blue'' stars are found at the lower end of the [Na/Fe]-distribution (i.e., LTE abundances) usually observed in globular clusters. Aoki et al. (2009\nocite{aoki09}) interpreted the absence of Na enhancements as evidence against a GC origin for the Mg deficient stars. However, if the primordial Na abundance was low in the gas cloud out of which the GC was formed, low Na abundances on average could be expected. Such an effect is also present in the sample of Galactic GCs observed by Carretta et al. (2009\nocite{carr09}). Furthermore, Carretta et al. (2009\nocite{carr09}) conclude that roughly one third of the stars in Galactic GCs exhibit primordial abundances, while two thirds are affected by internal pollution by the GCs' first stellar generation. This is consistent with the data of Aoki et al. (2009\nocite{aoki09}), assuming that S~$10-14$ belongs to the first, primordial generation while S~$11-13$ and S~$14-98$ belong to the second, polluted generation. 

While the vast majority of the Galactic GCs shows no observed variations in the neutron capture elements between individual member stars, two of the most metal-poor GCs, M15 (Sobeck et al. 2011\nocite{sobe11}) and M92 (Roederer \& Sneden 2011\nocite{roed11}), have recently been demonstrated to exhibit a spread in a number of heavy elements, including Ba and Eu. The mechanism behind this star-to-star scatter is not known. We note that at least $80\%$ (excluding upper limits) of the stars in the samples of Sobeck et al. (2011\nocite{sobe11}) and Roederer \& Sneden (2011\nocite{roed11}) are $r$-process enhanced ($r$-I) and all stars have $[\mathrm{Ba}/\mathrm{Fe}] \ge -0.25$. In contrast, the mean ratio of the potential cluster stars is $\langle[\mathrm{Ba}/\mathrm{Fe}]\rangle=-1.33$. Moreover, both M15 and M92 are several times more massive than what the relic cluster would have been, if it had survived to the present epoch. 

It is unclear whether all GCs below $[\mathrm{Fe}/\mathrm{H}] \simeq -2.3$ should exhibit chemical inhomogeneities in the neutron capture elements. Letarte et al. (2006\nocite{leta06}) observed three very metal-poor GCs in the Fornax dSph, of which two were found to have $[\mathrm{Fe}/\mathrm{H}]<-2.3$. Their data show no evidence in support of an intrinsic scatter in any of the GCs, although the statistics is admittedly low. At this point, we can therefore neither rule out the possibility that the relic cluster was a GC based on the observed clumping of stars in [Ba/Fe], nor that additional stars, e.g. $\mathrm{S}~12-28$ (color coded yellow in Fig. \ref{fig:abnd}), should be included as potential members. Further observations, including measurements of O, Al, and neutron capture element abundances are required in order to resolve these issues and determine the true nature of the relic cluster.

\subsection{Possible origins of the atypical $[\alpha/\mathrm{Fe}]$ ratio?}\label{sect:alpha}
\begin{figure*}[t]
\resizebox{\hsize}{!}{\includegraphics[trim=25mm 15mm 47mm 38mm,clip]{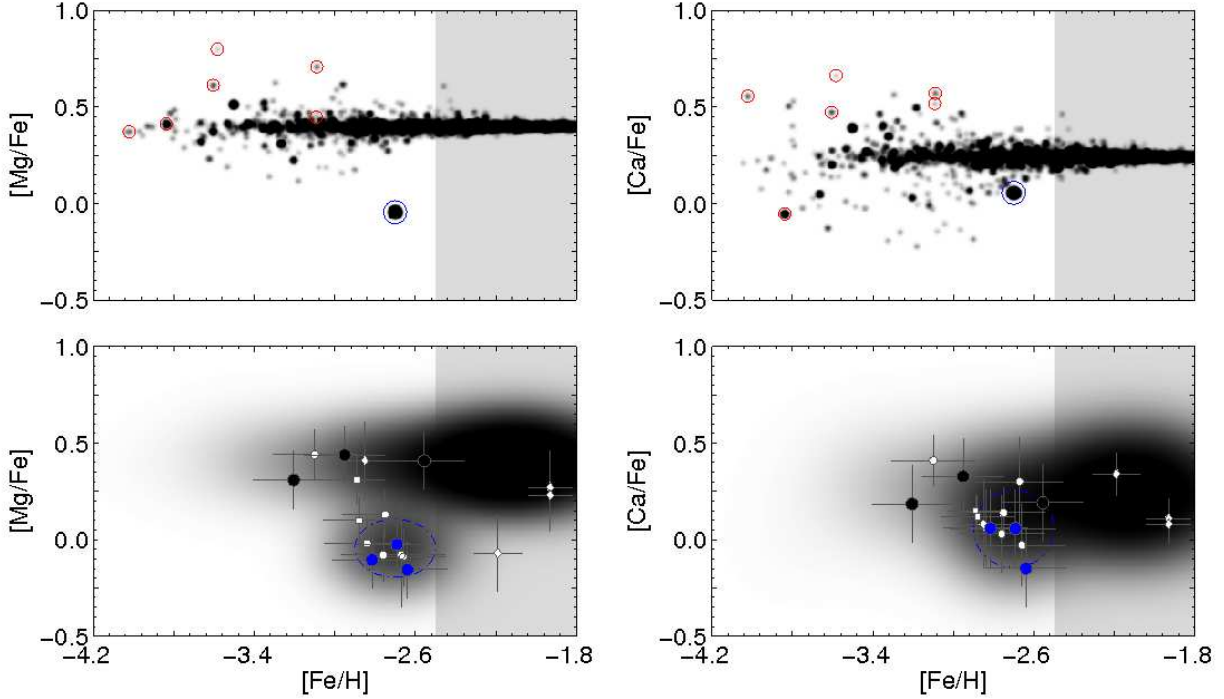}} % [left  bottom  right  top]
\caption{Predicted distributions of stars in $\mathcal{C}$-space for a Sextans-sized model galaxy. The top panels show the probability of finding a star in the [Mg/Fe]--[Fe/H] plane (left) and the [Ca/Fe]--[Fe/H] plane (right). The big, blue circle indicates the location of former member stars of a $M_{*}=1.9\times 10^5~\mathcal{M_{\odot}}$ relic cluster, while the small, red circles indicate the locations of stars which once belonged to clusters that were formed out of gas only enriched by a single, primordial SN, given that the first stars in dSphs were formed from primordial gas. The predicted fraction of these second generation stars in galaxies like Sextans is $1.1\times 10^{-3}$, while the fraction below $[\mathrm{Fe}/\mathrm{H}]=-2.5$ is $3.7\times 10^{-3}$, for the fiducial ICMF. Due to the stochastic nature of the chemical enrichment, second generation stars do not necessarily show up in the most metal-poor regime at $[\mathrm{Fe}/\mathrm{H}]\lesssim -4$. An intrinsic uncertainty of no more than $0.01$ dex is applied to all abundance ratios in order to emphasize the presence of clumps. The bottom panels show the same distributions convolved with a two-dimensional gaussian with $\sigma=0.15,~0.20,~0.20$ dex in [Mg/Fe], [Ca/Fe], and [Fe/H], respectively. Most of the clumping is washed out, except the one originating from the ``blue'' star cluster. The size ($1\sigma$ uncertainty) and location of this clump is indicated by the dash-dotted, blue circles. White circles denote the six stars observed by Aoki et al. (2009), while the squares and diamonds denote observations by Tafelmeyer et al. (2010), and Shetrone et al. (2001), respectively. The six big, colored circles are model stars randomly drawn from the non-shaded part (i.e., $[\mathrm{Fe}/\mathrm{H}]<-2.5$) of the distribution functions in the top panels. The three stars displayed in blue originate from the $M_{*}=1.9\times 10^5~\mathcal{M_{\odot}}$ cluster, while the black stars originate from other clusters. Fiducial uncertainties of $0.15$ dex in [Mg/Fe], $0.20$ dex in [Ca/Fe], and $0.20$ dex in [Fe/H] have been applied, similar to those reported by Aoki et al. (2009).}
\label{fig:sims}
\end{figure*}

\noindent
In Fig. \ref{fig:abnd}, the [Mg/Fe] and [Ca/Fe] ratios of the ``blue'' stars, as well as of \mbox{S~$11-37$}, are unusually low for stars in this metallicity regime (e.g. Tolstoy et al. 2009\nocite{tols09}, their Fig. 16). In the present modeling, we simply appended an $M_{*,\mathrm{init}}=1.9\times 10^{5}~\mathcal{M_{\odot}}$ star cluster at the observed location of the ``blue'' stars in $\mathcal{C}$-space (\S \ref{sect:esti}, see also Fig. \ref{fig:sims}). However, the probability for a cluster of this mass to be found in this part of  \mbox{$\mathcal{C}$-space} (i.e., in [Ca/Fe]) purely by chance is estimated to $<1\%$, given the fiducial model and that normal core-collapse SNe were the only nucleosynthesis sites enriching the early ISM.  We note in passing that relatively low [Ca/Fe] ratios are found in five Galactic Halo GCs (see compilation by Pritzl et al. 2005\nocite{prit05}). Of these, only the two most metal-poor clusters; \mbox{NGC 2419} ($[\mathrm{Fe}/\mathrm{H]}=-2.32$) and \mbox{NGC 5466} ($[\mathrm{Fe}/\mathrm{H]}=-2.05$, both have $[\mathrm{Ca}/\mathrm{Fe}]=+0.1$) are believed not to have been captured from a satellite system with a slower chemical evolution. The Mg-to-Fe ratio of the two clusters is determined to be $+0.30$ and $+0.06$, respectively. Unfortunately, the abundance analysis in both cases is based on one star only.

A decrease of the $[\alpha/\mathrm{Fe}]$ ratio with increasing metallicity is commonly believed to be a result of the delayed enrichment by SNe Ia synthesizing iron and Fe-peak elements (Tinsley 1980\nocite{tins80}; Matteucci \& Greggio 1986\nocite{matt86}). Due to the slower star formation timescale in dSphs, a subsolar $[\alpha/\mathrm{Fe}]$ is reached at a lower $[\mathrm{Fe}/\mathrm{H}]$, as compared to the Galactic Disk. However, in general, the classical dSphs only exhibit stars with a subsolar [$\alpha$/Fe] ratio above $[\mathrm{Fe}/\mathrm{H}]\simeq -1.5$ (Tolstoy et al. 2009\nocite{tols09}; Kirby et al. 2011a\nocite{kirb11a}). At $[\mathrm{Fe}/\mathrm{H}]\sim -3$, typical $[\alpha/\mathrm{Fe}]$ enhancements are observed, also in the dSphs (see Fig. \ref{fig:bckg}). In most ultra-faint dwarf galaxies, we have a similar situation with the bulk of stars being $[\alpha/\mathrm{Fe}]$-enhanced below $[\mathrm{Fe}/\mathrm{H}]\simeq -2$ (Feltzing et al. 2009\nocite{felt09}; Frebel et al. 2010b\nocite{freb10b}; Norris et al. 2010a, c\nocite{norr10a,norr10c}; see, however, Ad\'{e}n et al. 2011\nocite{aden11}). 

\nocite{aoki09}
\nocite{tafe10}
\nocite{shet01}

In a dwarf galaxy with a very low SFR, it is conceivable that SNe Ia start contributing to the metal budget below $[\mathrm{Fe}/\mathrm{H}] =-2$ (see Revaz et al. 2009\nocite{reva09}). Alternatively, infall of metal-free gas could in principle lower the [Fe/H] of an SN Ia-dominated ISM to the desired metallicity regime. Ivans et al. (2003\nocite{ivan03}) reported on three Galactic halo stars  at a metallicity $[\mathrm{Fe}/\mathrm{H}]\simeq -2$ with $[\alpha/\mathrm{Fe}]$ ratios similar to those observed in the ``blue'' stars. They concluded that the abundance patterns of these stars are consistent with an early enrichment by SNe Ia.  But on comparable timescales as the SNe Ia, the asymptotic giant branch (AGB) stars would also contribute to the metal budget. These stars enrich the ISM mostly in C, N, and s-process elements like Ba. Aoki et al. (2009\nocite{aoki09}) therefore reject the SNe Ia as being responsible for the low $[\alpha/\mathrm{Fe}]$ ratios due to the low [Ba/Fe] ratio observed in the ``blue'' stars. Furthermore, in a medium-resolution spectroscopic study, Kirby et al. (2011a\nocite{kirb11a}) mapped out the chemical evolution of $8$ dSphs, including Sextans. In their data for Sextans, [Mg/Fe] reaches subsolar values at $[\mathrm{Fe}/\mathrm{H}]\simeq -1.6$, while $[\mathrm{Ca}/\mathrm{Fe}]\sim 0$ is reached at somewhat lower [Fe/H]. They point out, however, that the $[\alpha/\mathrm{Fe}]$ ratio appears to be affected by SNe Ia already at $[\mathrm{Fe}/\mathrm{H}]\sim -2.5$. At such a low metallicity, small number statistics plays a considerably more prominent r\^{o}le than it does at \mbox{$[\mathrm{Fe}/\mathrm{H}]\gtrsim -1.5$}. The parent molecular cloud from which the cluster was born could theoretically have been enriched in SN Ia material, but not much in AGB material, as implied by Ivans et al. (2003\nocite{ivan03}) for their Galactic halo stars. Such a scenario must be checked against detailed modeling of the early chemical evolution of Sextans. We plan to do so in a forthcoming paper.  

As suggested by Aoki et al. (2009\nocite{aoki09}), a truncated IMF at the highest masses would favor lower $[\alpha/\mathrm{Fe}]$ ratios in the star-forming gas. Weidner \& Kroupa (2005\nocite{weid05}; see also Pflamm-Altenburg et al. 2007\nocite{pfla07}) argue that systems with slow star formation will effectively acquire a truncated IMF as a result of the relation between the SFR and the maximum cluster mass, given that the mass of the most massive star formed in a cluster is limited by the mass of the cluster.  Since $[\alpha/\mathrm{Fe}]$-enhanced stars do exist in the sample (S~$15-19$), some high-mass SNe must, nonetheless, have contributed to the metal budget in Sextans. A similar result is obtained for an IMF with a steeper than usual slope at the high-mass end. 

Could the presence of hypernovae (HNe) contribute to a larger fraction of solar-$[\alpha/\mathrm{Fe}]$ stars? As higher temperatures are realized in models of HNe (Umeda \& Nomoto 2002\nocite{umed02}; Nomoto et al. 2006\nocite{nomo06}), the complete Si-burning region extends outwards in mass, which leads to larger amounts of Fe ($^{56}$Ni) in the ejecta, while the light Fe-peak elements Cr and Mn are underproduced. Lower $[\alpha/\mathrm{Fe}]$ ratios can therefore be achieved for the high-mass SNe, even though the lowest ratios are still maintained in the low-mass SNe. The larger explosion energies associated with the HNe also result in the ejecta being diluted with a larger mass of ambient gas before merging with the ISM.  On a local scale, the presence of HNe thus induce more efficient mixing which decreases the possibility of having star-forming regions with $[\alpha/\mathrm{Fe}]\sim 0$ at any given [Fe/H]. This implies that even fewer solar-$[\alpha/\mathrm{Fe}]$ stars should be able to form around $[\mathrm{Fe}/\mathrm{H}]=-2.7$, despite lower $[\alpha/\mathrm{Fe}]$ ratios in the ejecta of these highly explosive SNe. Alternatively, if the HN ejecta are lost from the system as a result of the higher explosions energies (along with a shallow potential well, see \S \ref{sect:kinem}), a lower $[\alpha/\mathrm{Fe}]$ ratio would also be realized in the remaining, mixed gas, although the overall effect would be smaller.  Consistent with the HN scenario, a relatively low [Cr/Fe]  ratio is measured in the ``blue'' stars (Fig. \ref{fig:bckg}), as compared to the bulk of dSph stars. Why HN enrichment then appears not to have occurred in all dSphs should be further investigated but could be explained by local, chemical inhomogeneities in the star-forming gas.

There is an interesting alternative explanation to the observed [Mg/Fe] and [Ca/Fe] ratios in the ``blue'' stars. These stars could have formed out of gas that was partly enriched by a very massive Pop III star exploding as a pair-instability SN (Heger \& Woosley 2002\nocite{hege02}; Umeda \& Nomoto 2002\nocite{umed02}). Yields for the most massive stars in the pair-instability mass range (i.e., $240\lesssim m/\mathcal{M_{\odot}}\lesssim 260$) produce roughly solar and subsolar [Ca/Fe] and [Mg/Fe] ratios, respectively (Heger \& Woosley 2002\nocite{hege02}). Karlsson et al. (2008\nocite{karl08}) predict that stars with a predominant contribution from primordial pair-instability SNe should have relatively high Ca abundances roughly in the range $-3\lesssim [\mathrm{Ca}/\mathrm{H}]\lesssim -2$, in contrast to what may be expected for stars enriched by a first stellar generation. This result is also supported by hydrodynamics simulations which obtain an overall metallicity of $Z\sim 10^{-3}~\mathcal{Z_{\odot}}$ in proto-galaxies enriched by pair-instability SNe (Wise \& Abel 2008\nocite{wise08}; Greif et al. 2010\nocite{grei10}). Evidently, the Ca abundance of the ``blue'' stars is found within the predicted range. Furthermore, as a result of a very small neutron excess in the interior layers, low [Na/Fe], [Sc/Fe], and [Ba/Fe] ratios (see Fig. \ref{fig:abnd}) are to be expected the ejecta of pair-instability SNe. The pronounced odd-even effect is particularly interesting for Na since it could explain the very low Sodium abundance in S~$10-14$. 

Further observations including measurements of C, O, Si, and Zn abundances are of crucial importance in order to discriminate between the various scenarios presented here. Additional information on stellar ages would also help us to determine whether gas infall has undermined a closed-box-like chemical evolution of Sextans (see, e.g. the discussion on SNe Ia). As a result of the very high explosion energies in pair-instability SNe ($\sim10^{52}-10^{53}$ erg), the oxygen-burning region in the models is significantly extended as compared to the corresponding region in core-collapse SNe, and even HNe. Oxygen-burning products like Si and S should therefore be abundant in the ejecta relative to C and O and high ratios of, e.g. [Si/O] and [S/C] would be expected. As no fall-back onto a remnant occurs (the stellar core is complete disrupted in the explosion), the [O/Fe] ratio is predicted to be subsolar in stars predominantly enriched by pair-instability SNe while it is supersolar in stars enriched by core-collapse SNe and HNe, in line with with the bulk of the stars in the metal-poor Galactic Halo. Zn may be another discriminator, which is predicted to be produced in abundance in HNe, but not in pair-instability SNe (Umeda \& Nomoto 2002\nocite{umed02}).

\subsection{Remark on the cluster signature in $\mathcal{C}$-space}\label{sect:cspac}
Figure \ref{fig:sims} illustrates the expected imprint of a $1.9\times 10^5~\mathcal{M_{\odot}}$ cluster in $\mathcal{C}$-space, assuming that the early star formation in Sextans was clustered. The upper panels indicate the rich spectrum of clumps that is predicted to exist. This is in contrast to predictions in which clusters either are not accounted for (e.g. Karlsson \& Gustafsson 2005\nocite{karl05b}; Cescutti 2008\nocite{cesc08}) or the mass distribution of the clusters is not accounted for (e.g. Revaz et al. 2009\nocite{reva09}). These models produce smoother distributions in abundance space, although the simulations by Revaz et al. (2009\nocite{reva09}) exhibit some clumping.  However, due to the observational uncertainties typical for present-day studies of extra-galactic stars, the wealth of structure in the upper panels of Fig. \ref{fig:sims} will almost entirely be washed out, except for the $1.9\times10^5~\mathcal{M_{\odot}}$ cluster (lower panels). Uncertainties of $\lesssim 0.1$ dex must be achieved in order to reach the level of accuracy required for multiple clumping to appear in $\alpha$-element abundance space (Bland-Hawhorn et al. 2010\nocite{blan10a}; Karlsson et al. 2011\nocite{karl11}). The prediction that underlying clumping exists below the current detection limit is a key point. 

Note that the three blue model stars in the lower panels of Fig. \ref{fig:sims} clump rather tightly together in Mg but slightly less so in Ca. A situation like this may occur purely by chance due to the observational uncertainties and illustrates the need for studies, such as the one by Aoki et al. (2009\nocite{aoki09}), in which accurate stellar abundances of \textit{many} elements are measured. Having access to a multidimensional $\mathcal{C}$-space in combination with a large stellar sample also minimizes the risk of detecting ``false clumps'' in individual elements, which otherwise could lead to erroneous conclusions about the stars' origins. 

But if we do not have access to accurately determined abundances of many elements, can the cluster signature be detected in one dimensional $\mathcal{C}$-space? We shall claim that this is the case, at least in a statistical sense, and argue that the cluster signature may already be prevalent in the distributions of stellar metallicities of dSphs (see \S \ref{sect:discu} below). 

%Finally, it is worth noting that although a number of stars in Sextans (Fig. \ref{fig:sims}, white symbols) with $[\mathrm{Fe}/\mathrm{H}]< -2.5$ are found to have $[\alpha/\mathrm{Fe}]\sim 0$, there exists stars above this limit with supersolar $[\alpha/\mathrm{Fe}]$. 

\section{Discussion and implications}\label{sect:discu}
\noindent
The probable identification of stellar debris from a dissolved, ancient cluster at $[\mathrm{Fe}/\mathrm{H}]=-2.7$ in the Sextans dSph have several interesting implications for the understanding of early star and galaxy formation. In the following section, we shall discuss the interpretation of MDFs and cumulative metallicity functions of dwarf galaxies from a statistical point of view, particularly in connection to the formation of star clusters. We will also briefly discuss what implications this has for the understanding of star formation in metal-poor systems, for the origins of dwarf galaxies, and for the build-up of the stellar Halo around the Milky Way.

\subsection{Implications for early star and cluster formation}\label{sect:star}
\noindent
We know little about the star formation process at the earliest cosmic times. Inefficient fragmentation of collapsing, metal-free gas clouds was believed to prevent the formation of primordial clusters (Yoshida et al. 2008\nocite{yosh08}). However, in more recent hydrodynamics simulations, primordial stars are found to be formed in binary or small multiple systems (Turk et al. 2009\nocite{turk09}; Stacy et al. 2010\nocite{stac10}), although these simulations have not yet reached their asymptotic end state. Assuming turbulent initial conditions, Clark et al. (2011\nocite{clar11}) showed that primordial stars might even have been able to form in small, dense clusters. In their simulations, the resulting stellar IMF is also significantly broader than what previously have been found (e.g. Bromm et al. 2009\nocite{brom09}; see, however, Nakamura \& Umemura 2001\nocite{naka01}), with a noticeable fraction of stars below $1~\mathcal{M_{\odot}}$. See Karlsson et al. (2011\nocite{karl11}) for a recent review on the topic. 

\begin{figure*}[t]
\resizebox{\hsize}{!}{\includegraphics[trim=24.6mm 20mm 40.2mm 35.7mm,clip]{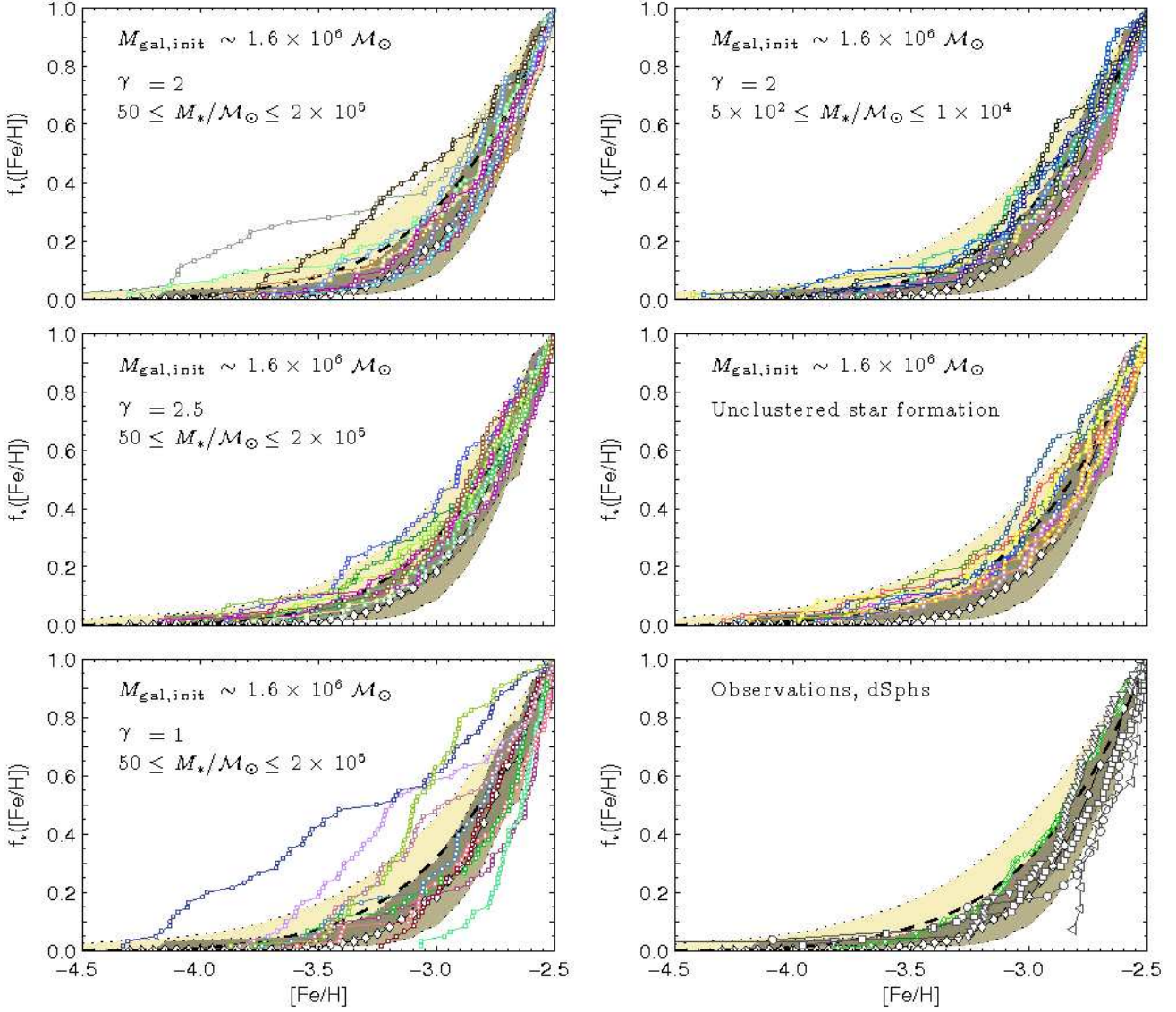}}   % trim = left, bottom, right, top
\caption{Predicted cumulative metallicity functions, $f_{*}$ (normalized to unity at $[\mathrm{Fe}/\mathrm{H}]=-2.5$), for Sextans-sized dSph galaxies. Each of the left-hand-side panels show $10$ model $f_{*}$ (colored squares) for different values of the slope of the ICMF, for which $\gamma=2$ is the fiducial value.  The top right panel shows the resulting cumulative functions assuming the narrower cluster mass range inferred from the simulations by Wise et al. (2012b), while the middle right panel shows the $f_{*}$ in the case of isolated, unclustered star formation.  Each model $f_{*}$ is convolved with a $\sigma=0.2$ dex gaussian and sampled with $60$ stars. The bottom right panel displays the observed $f_{*}$ (Starkenburg et al. 2010) for Carina (gray circles), Sculptor (squares), Sextans (down-faced triangles), and Fornax (left-faced triangles). The combined $f_{*}$ for the $8$ ultra-faint dwarf galaxies (small, green diamonds) discussed by Kirby et al. (2008, 2010, 2011b) is plotted for reference. In all panels, the dashed line denotes the predicted mean, $\bar{f_{*}}$, for a large number of model dSphs, while the black diamonds denote the corrected, cumulative function, $f_{*,\mathrm{Halo}}$ (binned), observed for the Galactic halo (Sch\"{o}rck et al. 2009). The offset between $\bar{f_{*}}$ and $f_{*,\mathrm{Halo}}$ is predominantly due to the lack of stars below $[\mathrm{Fe}/\mathrm{H}]\simeq -3.5$ in the medium-resolution ($\mathcal{R}\sim 2,000$) HES data (see text). The sand-colored areas (bounded by dotted lines) are measures of the dispersion expected from the finite sampling (i.e., $60$ stars) of the cumulative functions and denote the regions within which $80\%$ of the $f_{*}$ are confined, given that they are drawn directly from $\bar{f_{*}}$ (light-shaded) or $f_{*,\mathrm{Halo}}$ (medium-shaded, displayed for reference). The overlap between these two areas is shaded dark. In cases of unclustered star formation (middle right panel) or when the clusters are mostly ``unresolved'' (middle left and top right panels), the vast majority of the model $f_{*}$ is expected to be found within or in close connection to the light-colored region.}
\label{fig:dfdsph}
\nocite{wise12b}
\nocite{scho09}
\nocite{kirb08}
\nocite{star10}
\end{figure*}

Our knowledge of clustered star formation in the metallicity regime \mbox{$-5\lesssim [\mathrm{Fe}/\mathrm{H}]\lesssim -2$} is perhaps even more limited; here only a few of the most metal-poor Galactic globular clusters are found. Wise et al. (2012b\nocite{wise12b}) introduce Population II star clusters in their simulations of the birth of the first galaxies, without actually discussing the cluster formation process itself. From their data (J. Wise, priv. comm.), we find that most clusters below $[\mathrm{Z}/\mathrm{H}]=-2$ were formed with masses in the range $7.8\times 10^2\lesssim M_{*}/\mathcal{M_{\odot}}\lesssim 1.4\times 10^4$. The clusters below the low mass limit were not properly resolved and were rejected. The slope of the numerical ICMF in the above mass range is close to $\gamma=2$, and compatible with the slope of the present-day ICMF. This is, indeed, an interesting result and predicts that $\gamma$ may be universal all the way down to $[\mathrm{Fe}/\mathrm{H}]\sim -3.5$, at least in certain environments. By analyzing the metal content of the stellar populations in old, metal-poor dwarf galaxies, we should be able to put constraints on these numerical results and push our knowledge of the star formation process into the most metal-poor regime. In the analysis that follows, we will, for convenience, adopt a mass range of $5\times 10^2\le M_{*}/\mathcal{M_{\odot}}\le 1\times 10^4$ and a slope of $\gamma=2$ for this ICMF.

If the interpretation of the chemical abundance data presented is \S \ref{sect:chemi} indeed is correct, the ``blue'' stars in Fig. \ref{fig:abnd} were once members of the most metal-poor star cluster known to have formed in the Universe. This clearly suggests that stars could form in massive clusters already around $[\mathrm{Fe}/\mathrm{H}]\sim -3$. However, the question remains whether clustered star formation was common already at these metallicities and whether the mass range and relative frequency of clusters resembled those of the present-day ICMF.

Consider the MDF, or the cumulative metallicity function, $f_{*}$. Suppose that star formation in the early epochs was highly unclustered, i.e., stars were predominantly forming in isolation. If so, the MDF, as well as $f_{*}$, would appear the same for all galaxies with identical (same SFR, infall/outflow rate etc.) evolution. However, the situation would be different if stars were formed in chemically homogeneous clusters. In this case, each MDF would exhibit a unique, irregular and ``bumpy'' shape as a result of stars grouping together in random clumps in metallicity-space. As an effect of small number statistics, these irregularities should be most notable for the least massive galaxies, and at the lowest metallicities. 

We should emphasize that not all ``bumps'' must necessarily be signatures of disrupted clusters. Star bursts, gas infall, the onset of SNe Ia (see Karlsson 2005\nocite{karl05}, their Fig. 10), and even statistical uncertainties in the observations could in principle introduce irregularities. However, ``bumps'' originating from variations in the global SFR or infall of gas may have notably larger intrinsic widths in metallicity space since they do not originate from a single, chemically homogeneous stellar population. ``SN Ia bumps'' in [Fe/H] are formed when the first stars enriched by SNe Ia are pushed towards higher [Fe/H], relative to stars only enriched by core collapse SNe. This actually generates a through in the MDF, whose metal-poor wall appears as a ``bump''. Features like this will only show up in elements predominantly synthesized in the SNe Ia. In elements like Mg or Ca, the ``SN Ia  bump'' is significantly diminished, in contrast to ``bumps'' originating from clusters. We will therefore argue that MDFs and the corresponding $f_{*}$ can be used to probe the early formation of clusters in a statistical sense.

\subsubsection{Star formation in dwarf spheroidal galaxies}\label{sect:sfdsph}
\noindent
We ran a series of simulations of Sextans-sized dSphs, assuming different ICMFs, to map out the effect. For simplicity, Sextans was chosen as a typical dSph in terms of luminosity and stellar mass. In the sample of Starkenburg et al. (2010\nocite{star10}), Carina, Sextans, and Sculptor all have similar masses (within a factor of 2, see Woo et al. 2008\nocite{woo08}). Only Fornax is significantly more massive, with a stellar mass roughly $20\times$ that of Sextans. These four dSphs will be used in the comparison to the simulations. Here, we are only interested in irregularities and ``bumps'' of the $f_{*}$ (and the corresponding MDFs) in a statistical sense. We will therefore take the observational data as given by Starkenburg et al. (2010\nocite{star10}).  No translation to [Ca/H] abundances is performed (cf. \S \ref{sect:esti}). The same goes for the ultra-faint dwarf galaxy data in \S \ref{sect:sfufd}. The results are summarized in Fig. \ref{fig:dfdsph}. Each cumulative metallicity function, $f_{*}$, is convolved with a $\sigma=0.2$ dex gaussian and sampled with $60$ stars. The finite sampling is chosen to match the sampling of $f_{*}$ for Sextans (Starkenburg et al. 2010\nocite{star10}) and introduces an observational scatter indicated by the sand-colored areas (light-colored, see caption of Fig. \ref{fig:dfdsph} for details). The dispersion seen in the middle right panel of Fig. \ref{fig:dfdsph} (unclustered star formation), is solely due to the finite sampling of the cumulative functions.  

At face value, the ICMF with a shallow slope of $\gamma=1$ (Fig. \ref{fig:dfdsph}, bottom left) overproduces the number of model galaxies with a highly irregular $f_{*}$ as compared to observations (bottom right). In contrast, the ICMF with $\gamma=2.5$ (middle left) and the ICMF inferred from the simulation data of Wise et al. (2012b\nocite{wise12b}, top right) produce very few $f_{*}$ that significantly deviate from what is expected from the finite sampling alone, and the plots are virtually indistinguishable from that of the unclustered star formation (middle right). Higher sampling, i.e., more stars, and more precise metallicities are necessary in order to distinguish between these latter two ICMFs. 

The fiducial ICMF (top left) is an intermediate case. It leads to slightly pronounced fluctuations as compared to the case of unclustered star formation. In fact, about $75\%$ of the $f_{*}$ in the top left panel is predicted to fall outside the boundaries (in this case defined by the light-shaded, sand-colored area) at some point in metallicity. For unclustered star formation (i.e., the scatter is solely due to statistical fluctuations) the corresponding fraction is $20\%$. Note that the sand-colored areas in Fig. \ref{fig:dfdsph} are defined such that $80\%$ of all $f_{*}$, drawn from a single parent distribution and sampled with a given number of stars ($60$ in the case of Sextans), should be confined within the limits over the entire metallicity range $[\mathrm{Fe}/\mathrm{H}]\le -2.5$. The scatter generated by the fiducial ICMF is in qualitative agreement with the observations of the four classical dSphs (Fig. \ref{fig:dfdsph}, bottom right panel). At first glance, the scatter in the observations may look similar to the unclustered case. But a careful inspection of the dSph cumulative metallicity functions reveals that all four $f_{*}$ at some point are found outside both the light- and dark-shaded, sand-colored areas, taking into account the poorer sampling of the $f_{*}$ of Sculptor, Carina and Fornax. Assuming unclustered star formation in the dSphs, the chance is less than $0.2\%$ that four out of four (identical) $f_{*}$ should be found outside these limits. The corresponding probability assuming a present-day ICMF in the dSphs is $\sim 30\%$. It thus appears that the four dSphs as a group tend to overproduce the number of galaxies with $f_{*}$ distributions that significantly deviate from the mean, suggesting that the observed cumulative metallicity functions are not identical. This can be interpreted as an indication of clustered star formation and that a significant fraction of the observed stars (i.e., at $[\mathrm{Fe}/\mathrm{H}]<-2.5$) in these galaxies were once formed in a cluster environment similar to that of the present-day Galactic Disk. It is, however, important to identify star clusters in multi-dimensional $\mathcal{C}$-space, as with the cluster in Sextans, in order to firmly establish the existence of a clustered star formation signature. 

We note that below $[\mathrm{Fe}/\mathrm{H}]\simeq-3$ or so, the observations tend to show a smaller scatter than what is expected from the models. Although the observed sample of galaxies is admittedly very small, this could suggest that the formation process of star clusters in the lowest metallicity regime was different from that in the Milky Way today, given that most present-day clusters are formed in the designated mass range. A smaller dispersion between the cumulative functions is achieved, e.g. for a larger fraction of smaller clusters. However, due to the small number of stars and relatively poorly determined metallicities, Starkenburg et al. (2010\nocite{star10}) noted that the exact shape of the observed $f_{*}$ should be considered with caution, in particular below $[\mathrm{Fe}/\mathrm{H}]=-3$. Deeper observations and larger samples of galaxies are certainly required in order to successfully probe the cluster formation process in this metallicity regime.   

\begin{figure*}[t]
\resizebox{\hsize}{!}{\includegraphics[trim=24.5mm 20mm 40mm 35.4mm,clip]{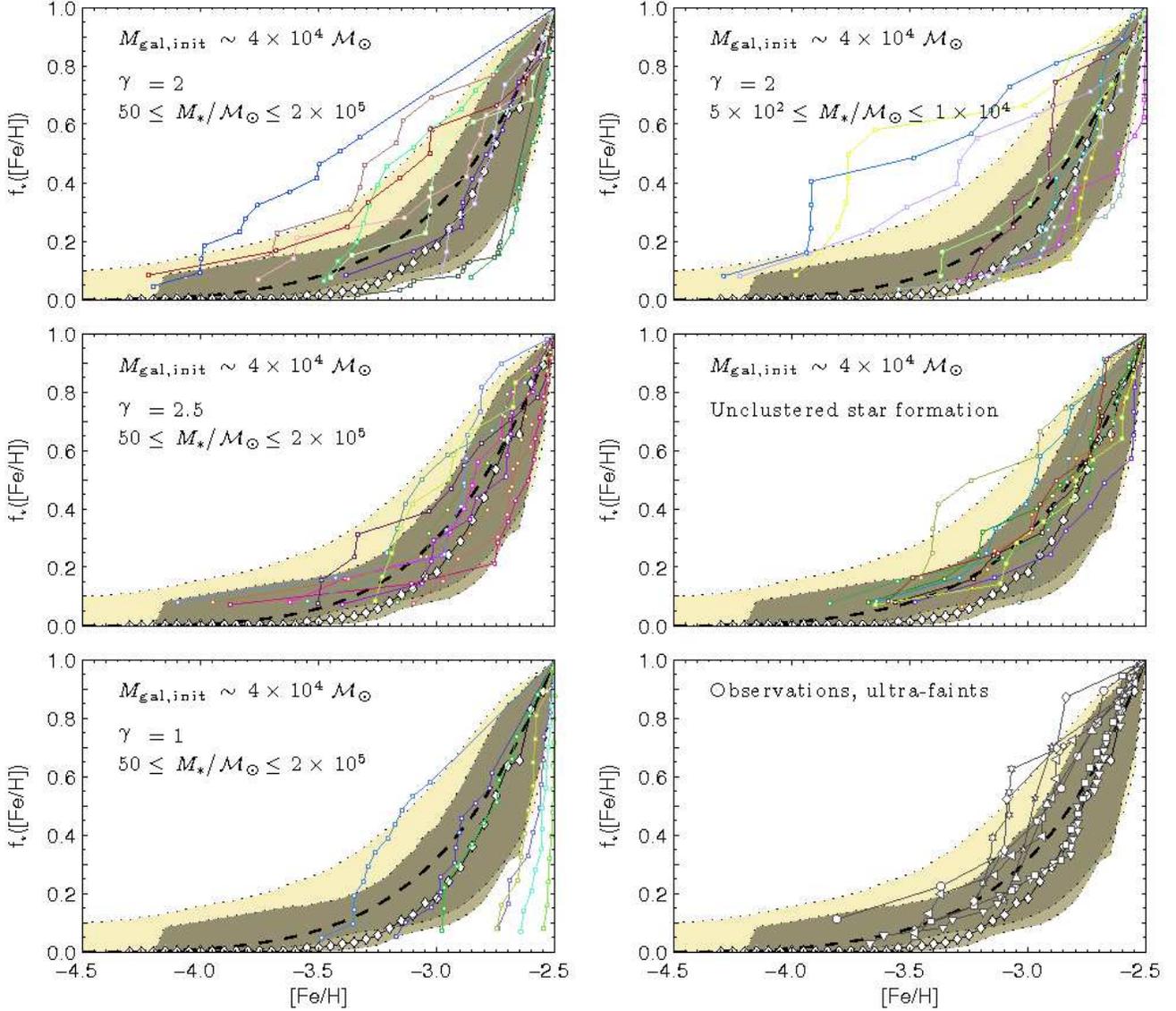}}   % trim = left, bottom, right, top
\caption{Predicted cumulative metallicity functions, $f_{*}$ (normalized to unity at $[\mathrm{Fe}/\mathrm{H}]=-2.5$), for typical ultra-faint dwarf-sized galaxies (cf. Fig. \ref{fig:dfdsph}). Each galaxy is assumed to have a present-day luminosity of $L_{\mathrm{gal}}\sim 10^4~\mathcal{L_{\odot}}$, implying an initial stellar mass of $M_{\mathrm{gal},\mathrm{init}}\sim4\times 10^4~\mathcal{M_{\odot}}$. Here, each $f_{*}$ is sampled with $12$ stars, reflecting the average number of stars in the survey of ultra-faint dwarf galaxies by Kirby et al. (2008, 2010, 2011b). The sand-shaded regions, indicating the typical Poisson noise due to the finite sampling, are changed accordingly. In the bottom right panel, $8$ ultra-faint dwarf galaxies observed by Kirby et al. (2008, 2010, 2011b) are displayed. These are: CVn~I (down-faced triangles), CVn~II (circles), UMa~I (squares), UMa~II (up-faced triangles), Leo~IV (diamonds), ComB (left-faced triangles), Herc (hexagons), and Leo T (pentagons). Otherwise, symbols are the same as in Fig. \ref{fig:dfdsph}. }
\label{fig:dfufd}
\nocite{kirb08,kirb10,kirb11b}
\end{figure*}

\begin{figure*}[t]
\resizebox{\hsize}{!}{\includegraphics[trim=29.7mm 6mm 48mm 24mm,clip]{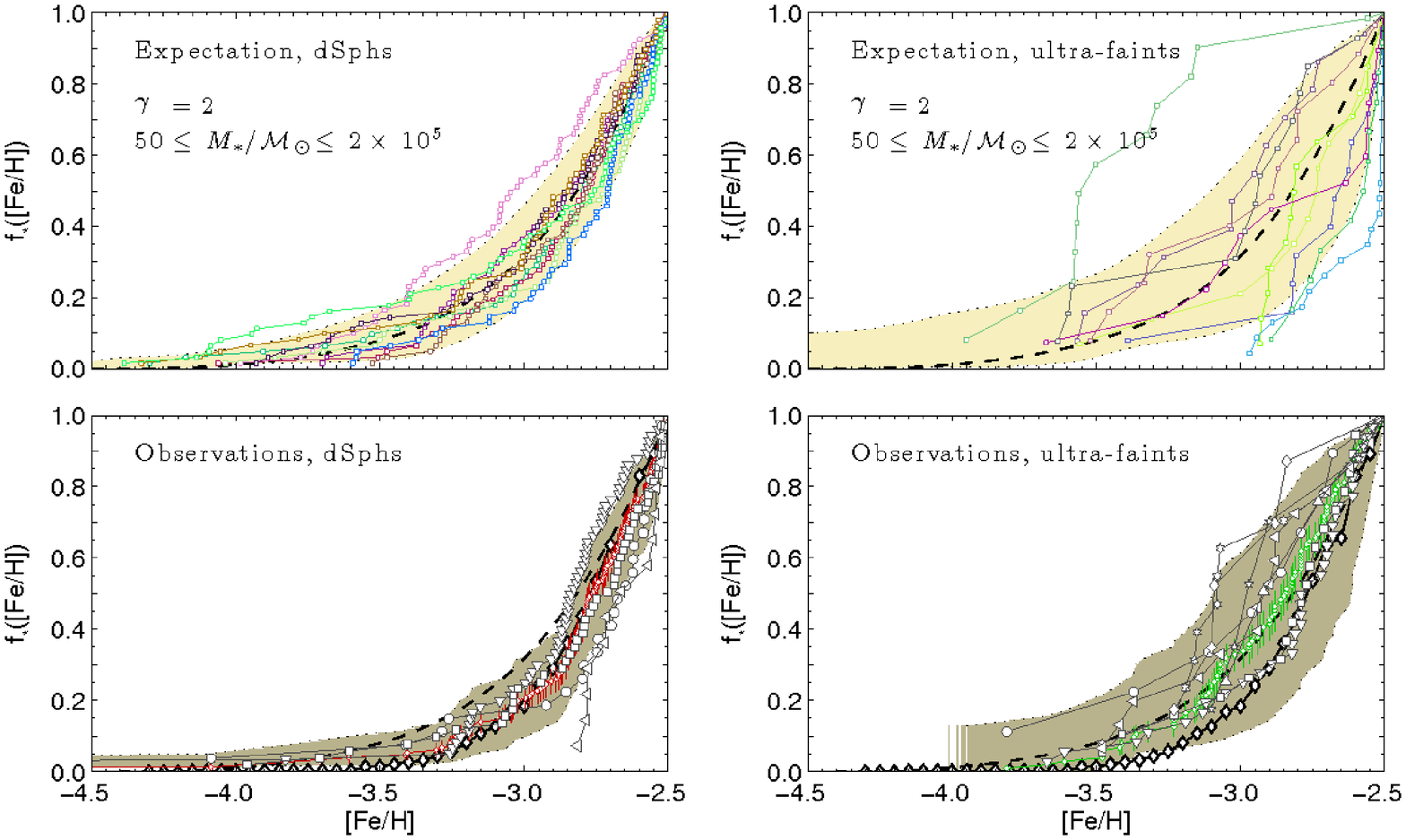}}  
\caption{Predicted and observed cumulative metallicity functions for the dSphs and the ultra-faints. The upper panels display the typical scatter in a sample of $10$ model $f_{*}$ expected from the fiducial, present-day ICMF for dSphs (left) and for ultra-faints (right). The $f_{*}$ of the dSphs are sampled with $60$ stars while the $f_{*}$ of the ultra-faints are sampled with $12$ stars. The lower left and right panels display observations of the dSphs and ultra-faints, respectively (see also Figs. \ref{fig:dfdsph} and \ref{fig:dfufd}, bottom right panels). The mean cumulative metallicity function $\bar{f}_{*,\mathrm{dSph}}$ (left) is denoted by connected, red diamonds while $\bar{f}_{*,\mathrm{ultra-faint}}$ (right) is denoted by connected, green diamonds. The vertical bars denote the $1\sigma$ uncertainty. Here, the light- and dark-shaded areas measure the Poisson noise due to the finite sampling relative to the model mean (dashed line, upper panels), $\bar{f}_{*,\mathrm{dSph}}$ (lower left) and $\bar{f}_{*,\mathrm{ultra-faint}}$ (lower right). Otherwise, symbols are the same as in Figs. \ref{fig:dfdsph} and \ref{fig:dfufd}.}
\label{fig:vs}
\nocite{kirb08,kirb10,kirb11b}
\end{figure*}

\subsubsection{Star formation in ultra-faint dwarf galaxies}\label{sect:sfufd}
\noindent
We also ran a series of simulations of ultra-faint dwarf galaxies with the same setup as for the dSph-sized galaxies. We assume a fiducial luminosity of $L_{\mathrm{gal}}= 1\times 10^4~\mathcal{L_{\odot}}$ for all model galaxies. This translates to an initial mass of $M_{\mathrm{gal,init}}\sim 4\times 10^4~\mathcal{M_{\odot}}$, given that the galaxies only consist of old stars as in Sextans (cf. \S \ref{sect:masso}). In general, a bigger scatter between the $f_{*}$ is therefore to be expected as a result of the much lower galaxy-to-cluster mass ratio relative to the dSphs. To allow a more direct comparison with the models of the dSphs, we assume that each ultra-faint MDF peaks at $[\mathrm{Fe}/\mathrm{H}]\sim-2$ (cf. Fig. \ref{fig:mdf}). This is probably at the high end for most ultra-faint dwarf galaxies in the Milky Way system, which appear to peak in the range $-2.5\lesssim[\mathrm{Fe}/\mathrm{H}]\lesssim -2$ (see Kirby et al. 2008\nocite{kirb08}). However, this should not affect the conclusions presented here. 

Several of the trends for the ultra-faints are the same as in the corresponding dSph case. As shown in Fig. \ref{fig:dfufd}, the ICMF with $\gamma=1$ (bottom left panel) produces cumulative metallicity functions that are substantially different from each other. Since the model galaxies have a very low stellar mass ($M_{\mathrm{gal,init}}\sim 4\times 10^4~\mathcal{M_{\odot}}$), they contain at most a few clusters below $[\mathrm{Fe}/\mathrm{H}]=-2.5$ (i.e., for $\gamma = 1$). In such an extreme case, the assumptions made for generating the $f_{*}$ are on the brink of breaking down, wherefore the exact shape of the $f_{*}$ should be regarded with some caution. Having said that, the cumulative metallicity function of Fornax (Fig. \ref{fig:dfdsph}) appears qualitatively similar to many of the model $f_{*}$ in the bottom left panel of Fig. \ref{fig:dfufd}, with a lack of stars below $[\mathrm{Fe}/\mathrm{H}]\sim-3$. This could suggest that Fornax was, to a high degree, pre-enriched in metals before forming the first generation of low-mass stars, possibly by a supermassive cluster. However, further studies of its metal-poor tail in combination with accurate modeling of the chemical evolution is called for in order to determine whether pre-enrichment occurred to a larger extent in Fornax than in other dwarf galaxies, and, if so, what would be the reason for this pre-enrichment. 

As for the dSph case, the ICMF with $\gamma=2.5$ (Fig. \ref{fig:dfufd}, middle left panel) show no obvious signs of clustering, despite the low mass of the simulated dwarf galaxies. For $\gamma=2$ (top left panel), there is a clear signature of clumping with a major fraction of the $f_{*}$ considerably deviating from the mean (cf. Fig. \ref{fig:dfufd}, middle right panel). The same trend is present in the dSph case, only weaker. In contrast to the dSph case, however, the ICMF inferred from the simulations by Wise et al. (2012b\nocite{wise12b}) appears to generate an increased number of $f_{*}$ that deviate from what is expected from the finite sampling of $12$ stars (Fig. \ref{fig:dfufd}, top right panel). Interestingly, the observed $f_{*}$ (Fig. \ref{fig:dfufd}, bottom right panel) of $8$ ultra-faint dwarf galaxies (Kirby et al. 2008, 2010, 2011b\nocite{kirb08,kirb10,kirb11b}) show no detected dispersion that can readily be attributed to the formation of clusters. Only three out of eight $f_{*}$ are at some point found outside the shaded areas, again taking into account the difference in the sampling of the $f_{*}$. Statistically, this should occur $1.6$ out of eight times, if the cumulative metallicity functions were drawn from a single parent distribution (i.e., no clustered star formation). Hence, it appears that clumping is underproduced in the ultra-faints, also as compared to the case in which the ICMF of Wise et al. (2012b\nocite{wise12b}) is assumed. This is at odds with what naively would be expected. In a set of simulations, Wise et al. (2012a\nocite{wise12a}) investigated the impact of momentum transfer from ionizing radiation. The ICMF in these simulations was found to have a slope of $\gamma\simeq 3.5-4$ (J. Wise, priv. comm.). Such a steep slope would significantly reduce the number of more massive clusters, which are mainly responsible for the excess clumping in the $f_{*}$ (cf. Fig. \ref{fig:dfufd}, middle left panel), in accordance with observations. However, why the radiation pressure in that case would not inhibit the formation of massive clusters in the dSphs remains to be understood. 

One explanation to the apparent overproduction of irregularities in the top panels of Fig. \ref{fig:dfufd} as compared to observations is that we have assumed too low stellar masses in the modeling of these galaxies. The ultra-faints could have been stripped off most of their stellar mass due to tidal interactions with the Milky Way. If so, the number of star clusters that constituted the original galaxies would have been higher, and each cluster would therefore have been sampled with fewer stars, leading to smoother $f_{*}$. However, from N-body simulations of the evolution of dSphs driven by galactic tides, Pe\~{n}arrubia et al. (2008)\nocite{pena08} conclude that ultra-faints are unlikely to be rest products of tidally stripped, more massive dwarf galaxies. Tidal stripping predicts to steep a change in the velocity dispersion as compared to what is observed. This finding is in line with the existence of metallicity gradients in several of the dSphs (de Boer et al. 2011\nocite{debo11}), which indicate that stripping due to tidal interactions at later stages could not have been overly severe. Furthermore, the observed  $Z-L$ relation for dwarf galaxies (e.g. Kirby et al. 2008\nocite{kirb08}), which states that the mean metallicity of the stellar population decreases with decreasing luminosity (i.e., stellar mass), argues against such a scenario.  Another alternative is that our assumption on the chemical homogeneity of star clusters may be incorrect. However, we have no {\it a priori} reason to believe that the physical processes responsible for the homogenization were significantly different in the early universe. Besides, the identification of the three potential cluster stars in Sextans (Fig. \ref{fig:abnd}) argues for the formation of chemically homogeneous clusters, also at early times. 

We conclude that the lack of pronounced irregularities in the cumulative metallicity functions of the ultra-faint dwarf galaxies is a signature of reduced clustering. Evidently, the bulk of the metal-poor clusters that formed in ultra-faints had masses (i.e., the fraction of mass locked-up in low-mass stars) below the current detection limit. This seems to be in contrast to the dSphs, which, despite their significantly higher (stellar) masses, show some evidence of clustering (see \S \ref{sect:sfdsph}).  We will further discuss the possible implications of this finding in the next section.

\subsection{Implications for galaxy formation}\label{sect:galax}
\noindent
Before we immerse into the intricacies of galaxy formation, let us have another look at the data of the dSphs and the ultra-faints.  As an additional check of our findings in \S \ref{sect:star}, we computed the mean $\bar{f}_{*,\mathrm{dSph}}$ and $\bar{f}_{*,\mathrm{ultra-faint}}$ from the observational data of the dSphs and ultra-faint galaxies, respectively (Fig. \ref{fig:vs}, lower panels). These averages have the advantage of being model independent. However, due to the relatively low sampling, they tend to follow the irregularities of the individual galaxies. As a consequence, the chance is higher that a given irregularity of an $f_{*}$ is found inside the region defined by the scatter due to the finite sampling (shaded areas). The probability of detecting pronounced clumping is therefore slightly decreased. Certainly, larger samples of galaxies would be desirable, especially a larger sample of dSphs. With this in mind, we note that two out of four dSphs are found outside the shaded area in Fig. \ref{fig:vs} (lower left panel). Although this is fewer than what is found in \S \ref{sect:sfdsph}, it is more than what is expected from the finite sampling alone and still in agreement with the excess scatter predicted for the fiducial ICMF (upper left panel). To summarize, the fraction of dSphs with detected excess scatter is $50-100\%$, depending on how the boundaries are defined. Recall that the corresponding prediction for the fiducial ICMF is $75\%$ (see \S \ref{sect:sfdsph}). In contrast, only two out of eight ultra-faint dwarf galaxies (and one border-line case) are found outside the shaded area in the lower right panel of Fig. \ref{fig:vs}, in line with the result in \S \ref{sect:sfufd}. Hence, given that cluster masses were distributed according to the present-day ICMF also in the ultra-faints (Fig. \ref{fig:vs}, upper panels), the expected scatter is noticeably larger than what is observed. Evidently, the fiducial ICMF overproduces the amount of scatter and does not seem to adequately describe the birth distribution of star clusters in the ultra-faint dwarf galaxies, unlike in the dSphs. 
     
Figure \ref{fig:vs} also reveals another distinction between the two dwarf galaxy populations. The average cumulative metallicity function of the ultra-faints (connected, green diamonds in the lower right panel) rises above the corresponding cumulative metallicity function, $\bar{f}_{*,\mathrm{Halo}}$, observed in the Galactic Halo (Sch\"{o}rck et al. 2009\nocite{scho09}; connected, black diamonds). In contrast, $\bar{f}_{*,\mathrm{dSph}}$ (connected, red diamonds in the lower left panel) closely follows $\bar{f}_{*,\mathrm{Halo}}$. The difference is quite noticeable. The horizontal error bars on the dwarf galaxy data indicate the $1\sigma$ uncertainty. At face value, it appears that the fraction of extremely metal-poor stars in the classical dSphs, on average, is lower than in the ultra-faint galaxies, while similar to that of the Galactic Halo. Note that if, in fact, $\bar{f}_{*,\mathrm{dSph}}\equiv \bar{f}_{*,\mathrm{Halo}}$, all dSphs in the current sample will at some point be found outside the shaded area as defined by $\bar{f}_{*,\mathrm{Halo}}$ (see \S \ref{sect:sfdsph}). 

Is the difference between $\bar{f}_{*,\mathrm{dSph}}$ and  $\bar{f}_{*,\mathrm{ultra-faint}}$ real? We note that Fe abundances of the stars in the dSphs are estimated from the strength of the CaT, while stellar Fe abundances of the ultra-faints are determined directly from a large number of metallic lines, including Fe. A suppression of the metal-poor tail of the MDF may arise, e.g. if the CaT method is poorly calibrated, a trend in the [Ca/Fe] ratio with [Fe/H] exists that is not properly accounted for, or if there exists a pristine, extremely metal-poor population of stars in the dSphs that is less centrally concentrated and which therefore is not fully sampled. The CaT method have recently been re-calibrated to hold down to $[\mathrm{Fe}/\mathrm{H}]\simeq -4$ (Starkenburg et al. 2010\nocite{star10}). Furthermore, no trend in [Ca/Fe] is observed below $[\mathrm{Fe}/\mathrm{H}]=-2.5$, neither in the Galactic halo (Cayrel et al. 2004\nocite{cayr04}), nor in the dSphs (e.g. Koch et al. 2008\nocite{koch08}; Cohen \& Huang 2010\nocite{cohe10}; Tafelmeyer et al. 2010\nocite{tafe10}; Kirby et al. 2011a\nocite{kirb11a}), although the present data in this metallicity regime are admittedly very sparse for the dSphs. Kirby et al. (2011b\nocite{kirb11b}) have also observed MDFs for a number of dSphs, including Sextans, Sculptor, and Fornax. Examining their data, we see no obvious deviation from the observations by Starkenburg et al. (2010\nocite{star10}). Metallicity and age gradients could be a potential problem, where the oldest and most metal-poor populations usually are found in the outskirts of dwarf galaxies. Battaglia et al. (2011\nocite{batt11}) also detect a metallicity gradient in Sextans. However, judging from their Fig. 8, it is seems to flatten out towards $R=2^{\circ}$. Besides, a less centrally concentrated, pristine stellar population in the dSphs must have a peak metallicity below $[\mathrm{Fe}/\mathrm{H}]\simeq -3$ (and be accompanied by the absence of such a bias in the ultra-faints) in order to compensate for the lack of extremely metal-poor stars. This is not clearly the case, at least not for Sextans. Furthermore, as regards $\bar{f}_{*,\mathrm{Halo}}$, Sch\"{o}rck et al. (2009\nocite{scho09}) note that the rejection of stars with strong G-bands in the Hamburg/ESO data might lead to a bias, in particular against stars with $[\mathrm{Fe}/\mathrm{H}]\lesssim -3.5$. They conclude, however, that a possible effect on the inferred MDF would only be minor.  At this point, we will take the cumulative metallicity functions of the dSphs, ultra-faints, and of the Galactic Halo as given.

\subsubsection{Different origin of the dSphs and the ultra-faints?}
\noindent
Based on the observed difference in the degree of clumping in [Fe/H] between dSphs and ultra-faints and on the offset between $\bar{f}_{*,\mathrm{dSph}}$ and $\bar{f}_{*,\mathrm{ultra-faint}}$, we argue that the two populations of galaxies have formed in different environments, in which the conditions for star formation were not the same.  In their on cold dark matter (CDM) cosmological hydrodynamics simulations, Bovill \& Ricotti (2011a, b\nocite{bovi11a,bovi11b}; see also Bovill \& Ricotti 2009\nocite{bovi09}; Ricotti 2010\nocite{rico10}) identified an early population of faint dwarf galaxies that were able to form stars only prior to reionization (except a fraction of galaxies that may have experienced late-time gas infall and subsequent star formation). These ``pre-reionization fossils'' are defined as having a maximum circular velocity remaining below $v_{\mathrm{c}}=20$~km~s$^{-1}$ at all times and a luminosity $<10^5~\mathcal{L_{\odot}}$. The authors argued that at least a fraction of the ultra-faints are consistent with being ``pre-reionization fossils''. Salvadori \& Ferrara (2009\nocite{salv09}) found in their hierarchical merger tree simulations that the simulated galaxies with similar luminosities and mean metallicities as the ultra-faints were formed within $\mathrm{H}_{2}$-cooling minihaloes at $z>8.5$, i.e., before reionization. This result is in accord with that of Bovill \& Ricotti (2011a, b\nocite{bovi11a,bovi11b}). Star formation in these minihaloes is supposed to be critically affected by the inefficient cooling, including a higher cosmic microwave background (CMB) temperature, and be susceptible to feedback effects like H$_2$-dissociating Lyman-Werner radiation. The characteristic cluster mass could therefore have been much lower than in the present-day Galaxy (cf. Stacy et al. 2010\nocite{stac10}; Clark et al. 2011\nocite{clar11}), and still lower than in the dSphs, if this population of galaxies predominantly were assembled after reionization. 

The reason for the offset between $\bar{f}_{*,\mathrm{dSph}}$ and $\bar{f}_{*,\mathrm{ultra-faint}}$ is not clear. Seemingly, the dSphs exhibit an effect similar to the classical ``G-dwarf problem'' in the Solar neighborhood (van den Bergh 1962\nocite{berg62}). In our present stochastic chemical evolution model (see also Karlsson 2006\nocite{karl06}), we make use of the $D_{\mathrm{trans}}$ criterion for low-mass stars formation (Frebel et al. 2007\nocite{freb07}). This means that low-mass stars can only be formed in carbon and/or oxygen enhanced gas for which the criterion: $D_{\mathrm{trans}}=\log_{10}(10^{[\mathrm{C}/\mathrm{H}]} + 0.3\times 10^{[\mathrm{O}/\mathrm{H}]})\ge-3.5$ is fulfilled. This effectively suppresses the number of stars below $[\mathrm{Fe}/\mathrm{H}]\simeq-3.5$ in the model, mimicking an IMF with a higher characteristic mass. Interestingly, the prediction is in close agreement with the data of the ultra-faints (Fig. \ref{fig:vs}, lower right panel). The suppression is, however, not enough to explain the relative lack of extremely metal-poor stars in the dSphs, why some kind of additional suppression must have occurred. Following the argument above, an offset between the cumulative metallicity functions could be induced if the enrichment of carbon, relative to Fe, was lower in the dSphs than in the ultra-faints. The fraction of carbon-enhanced metal-poor (CEMP) stars (i.e., those with no s-process enhancements) may give us a hint whether different levels of C enrichment were a reality at early stages in the two types of galaxies. It is worth noting that several CEMP stars have recently been discovered in the ultra-faints (Norris et al. 2010a, b\nocite{norr10a,norr10b}; Lai et al. 2011\nocite{lai11}). An alternative explanation to the offset is that the dark-matter haloes in which the more luminous dSphs reside, were accreting gas from the intergalactic medium that was already slightly pre-enriched in metals. This scenario is in line with the results of Salvadori \& Ferrara (2009\nocite{salv09}). A continuous infall of sufficiently metal-poor gas would further enhance the effect (cf. Kirby et al. 2011b\nocite{kirb11b}).

From simulations of primordial stars (Bromm et al. 2009\nocite{brom09}; Karlsson et al. 2011\nocite{karl11} and references therein), it is widely accepted that the special conditions in the Universe at the end of the cosmological ``Dark Ages'' led to a top-heavy primordial/early IMF. If these conditions prevailed in the ultra-faints more so than in the dSphs, we would expect this to be reflected in the stellar chemical abundance data. A close inspection of the [Mg/Fe] ratio in Fig. \ref{fig:bckg} may, indeed, hint towards such a difference. The mean $[\mathrm{Mg}/\mathrm{Fe}]$ ratio below $[\mathrm{Fe}/\mathrm{H}]=-2.5$ for the compiled dSph data (dark-grey symbols), excluding the data of Aoki et al. (2009), is $\langle[\mathrm{Mg}/\mathrm{Fe}]\rangle=0.30\pm0.14$. The corresponding ratio including all the stars of the ultra-faints (light-grey symbols) is $\langle[\mathrm{Mg}/\mathrm{Fe}]\rangle=0.56\pm0.27$. The $1\sigma$ deviation denote the total deviation, including both observational uncertainties and cosmic scatter. This is a fairly noticeable difference. At face value, a higher mean $\alpha/\mathrm{Fe}$ ratio in the ultra-faints could be suggestive of a top-heavy IMF favoring the formation of more massive stars and would be in line with our conclusion above that ultra-faints and dSphs are distinct. However, the mean [Mg/Fe] ratio for the ultra-faints is based on $9$ stars only. It is important for a comparison like this to be internally consistent to minimize systematic uncertainties, to avoid observational biases, and to have control over contamination effects such as SNe Ia enrichment, galactic winds, etc. Before more data are available, this particular finding should therefore be regarded merely as tentative.

\subsubsection{Remarks on the origin of the Galactic Halo}
\noindent
Based upon their data, Aoki et al. (2009\nocite{aoki09}) conclude that dwarf galaxies like Sextans could not be a major contributor to the Galactic stellar Halo. In contrast to the bulk of metal-poor Galactic Halo stars (as well as to dSph stars, see Fig. \ref{fig:bckg}), the majority of stars in their very metal-poor sample is not enhanced in $\alpha$ elements as compared to Fe. However, in the light of the results presented above, we would like to add a small caveat to this conclusion. If we accept the scenario that stars could form in clusters already at $[\mathrm{Fe}/\mathrm{H}]\simeq -3$, and that the three ``blue'' stars once belonged to such a cluster, the existence of subsolar [Mg/Fe] stars below $[\mathrm{Fe}/\mathrm{H}]=-2.5$ in Sextans could naturally be explained as a result of chemical inhomogeneities in the star-forming gas (see Fig. \ref{fig:sims}), without invoking changes in global parameters like the IMF. For that reason alone, dwarf galaxies like Sextans should not automatically be rejected as ``progenitors'' of the Galactic Halo; they may just be statistical outliers. Having that said, we note that the low [Mg/Fe] ratio of the potential cluster still needs to be explained (see \S \ref{sect:alpha}). More data are needed, particularly around $-3 \le [\mathrm{Fe}/\mathrm{H}] \le -2$ in order to determine whether a low $[\alpha/\mathrm{Fe}]$ ratio is a global property of the galaxy already at these metallicities.     

Current data from the Sloan Digital Sky Survey (SDSS) suggest that the stellar Galactic Halo is composed of (at least) two components; an inner Halo with a peak metallicity at $[\mathrm{Fe}/\mathrm{H}]\simeq-1.6$ and a more metal-poor, net retrograde-rotating outer Halo with a peak metallicity at $[\mathrm{Fe}/\mathrm{H}]\simeq-2.2$ (Carollo et al. 2007, 2010\nocite{caro07,caro10}; see also, e.g.  Schuster et al. 2012\nocite{schu12} for evidence of a multi-component Halo at $[\mathrm{Fe}/\mathrm{H}]\gtrsim-1$). Carollo et al. (2007\nocite{caro07}) speculated that the outer Halo was formed through dissipationless, chaotic merging of small subsystems, such as faint dwarf galaxies. Recently, Sch\"{o}nrich et al. (2011\nocite{scho11}) claimed to have identified significant biases in the distance estimates of the stars in the Carollo et al. (2007, 2010\nocite{caro07,caro10}) sample, in particular of the lowest metallicity stars.  They conclude that, as a result of these distance overestimates, the evidence for the existence of an outer and inner Halo is weakened (see, however, Beers et al. 2012\nocite{beer12}).  If ultra-faints were a major contributor, exclusively to an outer stellar Halo, we expect the cumulative metallicity function of the outer Halo to differ from that of the inner Halo. Moreover, if our findings hold, the outer Halo may also exhibit less clumping in $\mathcal{C}$-space than the inner Halo (Karlsson et al. 2011\nocite{karl11}). An extensive, multi-elemental abundance survey of the Galactic Halo would be able to disentangle potential components and unveil the progenitor systems from which the Halo was once assembled.

\section{Final remarks}\label{sect:summa}
\noindent
In the published abundance data for six very metal-poor stars in the Sextans dSph by Aoki et al. (2009\nocite{aoki09}), we have identified three stars around $[\mathrm{Fe}/\mathrm{H}]=-2.7$, that potentially were once members of a now disrupted star cluster. If so, this is currently the most metal-poor cluster known to have formed in the Universe. We estimate the initial mass of the cluster to $M_{*,\mathrm{init}}=1.9^{+1.5}_{-0.9}~(1.6^{+1.2}_{-0.8})\times 10^5~\mathcal{M_{\odot}}$, assuming a Salpeter (Kroupa) IMF. Aoki et al. (2009) infer a very low [Na/Fe] ratio for one of the three potential cluster stars. This may be suggestive of an Na-O anticorrelation present in the data, implying that the relic cluster was a GC. One out of three stars with low [Na/Fe] is typical for GCs (Carretta et al. 2009\nocite{carr09}) and the existence of a relic GC at $[\mathrm{Fe}/\mathrm{H}]=-2.7$ in Sextans would also resonate with the presence of metal-poor GCs in other dSphs (Letarte et al. 2006\nocite{leta06}).  We argue that the metal-poor environment of dwarf galaxies like dSphs provides an ideal base for searching for ancient clusters. We note that  ``bumps'' or irregularities in the metal-poor tails of the MDFs of dSphs could be a signature of clusters. In the data of Kirby et al. (2011a\nocite{kirb11a}), several MDFs exhibit ``bumpy'' tails, such as  Leo~II, Draco and Ursa Minor. Interestingly, the Ursa Minor dSph is known to possess a second concentration of stars roughly $14'$ from the center. Kleyna et al. (2003\nocite{kley03}) claimed that these stars belong to a cold substructure, consistent with the remnants of a disrupted star cluster. Whether the stars in the ``bump'' at $[\mathrm{Fe}/\mathrm{H}]\sim-3$ (Kirby et al. 2011b\nocite{kirb11b}) have the same origin as the stars in the cold substructure remains to be seen. In their sample of ten Ursa Minor stars, Cohen \& Huang (2010\nocite{cohe10}) detect two stars around $[\mathrm{Fe}/\mathrm{H}]=-3.1$. The abundance data is, however, inconclusive regarding their origin.  

Inspired by the possible detection of a metal-poor star cluster in Sextans, we present a new way of interpreting cumulative metallicity functions of dwarf galaxies to study the star formation process at high redshifts and in the most metal-poor regimes. A comparison between models and available metallicity data for dSphs and ultra-faint dwarf galaxies suggests that the formation of clusters in the ultra-faints differed from that of the dSphs. Together with a detected offset between the mean cumulative metallicity functions of the ultra-faints and the dSphs, and a putative offset in the $\langle[\mathrm{Mg}/\mathrm{Fe}]\rangle$ ratio, we speculate that these two populations of galaxies probably were formed in different environments and would therefore be distinct in origin. A possible explanation to the apparent absence of clumping in the ultra-faints below $[\mathrm{Fe}/\mathrm{H}]= -2.5$ is that these galaxies were formed predominantly before the local Universe was reionized, while the dSphs formed mostly after reionization. However, whether there is a dichotomy, as suggested by the current observational data, or a continuous distribution of galaxies exhibiting a mix of properties should be investigated further. Recently, Lai et al. (2011\nocite{lai11}) observed $25$ stars in the Bo\"{o}tes I ultra-faint dwarf galaxy, whereof $12$ stars were found below $[\mathrm{Fe}/\mathrm{H}]=-2.5$. The cumulative metallicity function inferred from their data falls right on top of the mean (see Fig. \ref{fig:vs}, lower right panel) as defined by the data of Kirby et al. (2008, 2010, 2011b\nocite{kirb08,kirb10,kirb11b}), with no features that can be attributed to clustering. This is entirely in line with our conclusion that the environment in the ultra-faints seemed to have inhibited the formation of massive clusters, at least in the regime $[\mathrm{Fe}/\mathrm{H}]<-2.5$. A search for clumping in the ultra-faints above $[\mathrm{Fe}/\mathrm{H}]=-2.5$ should be performed in order to determine the impact of the metallicity on the ICMF. Finally, we note that if more massive clusters were extremely rare in the ultra-faints, this could potentially have a significant impact on the escape fraction of ionizing radiation from these galaxies, thereby influencing on their capacity as sources of reionization.

Our new work has important implications for near-field cosmology, in particular, in identifying unique chemical signatures in the local universe that we can attribute to the first stellar generations. In principle, it is possible to probe and to reconstruct ancient star clusters that have long since
dissolved into the dwarf population. The reconstructed systems can tell us about the onset of stellar clustering and the build-up of the chemical elements with cosmic time.

We believe that reliable detections of stellar clustering within old stellar populations will be commonplace in an era of extremely large telescopes. At the present time, while several multi-object facilities are under discussion, the only concrete proposal is the MANIFEST project that will provide high resolution spectroscopy over the 20$^{\prime}$ field of the Giant Magellan Telescope (Saunders et al. 2010\nocite{saun10}). This field of view is well matched to the size of largest dwarf galaxies in the Galactic halo. Bland-Hawthorn et al. (2010a\nocite{blan10a}, their Figs. 7 and 8) present simulations of ${\cal C}$-space clustering in low mass dwarfs and show that even a modest multi-object spectrograph will provide a tremendous advance on the capabilities we have today. It will be possible to obtain high quality abundances for stars down to $r=22$ mag. This will allow us to extend our reach and observe dwarf giants to beyond 100 kpc in Galactocentric radius.

Many more dwarf spheroidals and ultra-faint dwarf galaxies are likely to be identified by future all sky survey telescopes, including the Large Survey Synoptic Telescope. With enough resolved clusters in individual dwarfs, or treated collectively, we can learn about the star formation process before, during and after the epoch of reionization. These new data will require a more rigorous theoretical framework than exists at the present time. Towards this end, it will be necessary to resolve star clusters in cosmological simulations from the onset of the first stars through reionization to the present epoch.

\bigskip
\bigskip

\acknowledgements
We would especially like to thank Else Starkenburg and Evan Kirby for kindly providing us with stellar metallicity data for, respectively, the dwarf spheroidal galaxies (\textsf{DART}) and the ultra-faint dwarf galaxies that were discussed in this study. We would also like to thank John Wise for kindly providing us with cluster data from their radiative hydrodynamics simulations. We acknowledge stimulating conversations with Kathryn Johnston, Evan Kirby, Duane Lee, Luis Vargas, and John Wise.  JBH is supported by a Federation Fellowship from the Australian Research Council.
%\end{acknowledgements}

\bibliographystyle{apsrmp}
\bibliography{refs}

\begin{thebibliography}{105}
\expandafter\ifx\csname natexlab\endcsname\relax\def\natexlab#1{#1}\fi
\expandafter\ifx\csname bibnamefont\endcsname\relax
  \def\bibnamefont#1{#1}\fi
\expandafter\ifx\csname bibfnamefont\endcsname\relax
  \def\bibfnamefont#1{#1}\fi
\expandafter\ifx\csname citenamefont\endcsname\relax
  \def\citenamefont#1{#1}\fi
\expandafter\ifx\csname url\endcsname\relax
  \def\url#1{\texttt{#1}}\fi
\expandafter\ifx\csname urlprefix\endcsname\relax\def\urlprefix{URL }\fi
\providecommand{\bibinfo}[2]{#2}
\providecommand{\eprint}[2][]{\url{#2}}

\bibitem[{\citenamefont{{Ad{\'e}n}}
  \emph{et~al.}(2011)\citenamefont{{Ad{\'e}n}, {Eriksson}, {Feltzing},
  {Grebel}, {Koch}, and {Wilkinson}}}]{aden11}
\bibinfo{author}{\bibnamefont{{Ad{\'e}n}}, \bibfnamefont{D.}},
  \bibinfo{author}{\bibfnamefont{K.}~\bibnamefont{{Eriksson}}},
  \bibinfo{author}{\bibfnamefont{S.}~\bibnamefont{{Feltzing}}},
  \bibinfo{author}{\bibfnamefont{E.~K.} \bibnamefont{{Grebel}}},
  \bibinfo{author}{\bibfnamefont{A.}~\bibnamefont{{Koch}}}, and
  \bibinfo{author}{\bibfnamefont{M.~I.} \bibnamefont{{Wilkinson}}},
  \bibinfo{year}{2011}, \bibinfo{journal}{A\&A} \textbf{\bibinfo{volume}{525}},
  \bibinfo{pages}{A153}.

\bibitem[{\citenamefont{{Aoki}} \emph{et~al.}(2009)\citenamefont{{Aoki},
  {Arimoto}, {Sadakane}, {Tolstoy}, {Battaglia}, {Jablonka}, {Shetrone},
  {Letarte}, {Irwin}, {Hill}, {Francois}, {Venn}} \emph{et~al.}}]{aoki09}
\bibinfo{author}{\bibnamefont{{Aoki}}, \bibfnamefont{W.}},
  \bibinfo{author}{\bibfnamefont{N.}~\bibnamefont{{Arimoto}}},
  \bibinfo{author}{\bibfnamefont{K.}~\bibnamefont{{Sadakane}}},
  \bibinfo{author}{\bibfnamefont{E.}~\bibnamefont{{Tolstoy}}},
  \bibinfo{author}{\bibfnamefont{G.}~\bibnamefont{{Battaglia}}},
  \bibinfo{author}{\bibfnamefont{P.}~\bibnamefont{{Jablonka}}},
  \bibinfo{author}{\bibfnamefont{M.}~\bibnamefont{{Shetrone}}},
  \bibinfo{author}{\bibfnamefont{B.}~\bibnamefont{{Letarte}}},
  \bibinfo{author}{\bibfnamefont{M.}~\bibnamefont{{Irwin}}},
  \bibinfo{author}{\bibfnamefont{V.}~\bibnamefont{{Hill}}},
  \bibinfo{author}{\bibfnamefont{P.}~\bibnamefont{{Francois}}},
  \bibinfo{author}{\bibfnamefont{K.}~\bibnamefont{{Venn}}}, \emph{et~al.},
  \bibinfo{year}{2009}, \bibinfo{journal}{A\&A} \textbf{\bibinfo{volume}{502}},
  \bibinfo{pages}{569}.

\bibitem[{\citenamefont{{Asplund}} \emph{et~al.}(2005)\citenamefont{{Asplund},
  {Grevesse}, and {Sauval}}}]{aspl05}
\bibinfo{author}{\bibnamefont{{Asplund}}, \bibfnamefont{M.}},
  \bibinfo{author}{\bibfnamefont{N.}~\bibnamefont{{Grevesse}}}, and
  \bibinfo{author}{\bibfnamefont{A.~J.} \bibnamefont{{Sauval}}},
  \bibinfo{year}{2005}, in \emph{\bibinfo{booktitle}{Cosmic Abundances as
  Records of Stellar Evolution and Nucleosynthesis}}, edited by
  \bibinfo{editor}{\bibnamefont{{T.~G.~Barnes III \& F.~N.~Bash}}}, volume
  \bibinfo{volume}{336} of \emph{\bibinfo{series}{Astronomical Society of the
  Pacific Conference Series}}, p.~\bibinfo{pages}{25}.

\bibitem[{\citenamefont{{Audouze} and {Silk}}(1995)}]{audo95}
\bibinfo{author}{\bibnamefont{{Audouze}}, \bibfnamefont{J.}}, and
  \bibinfo{author}{\bibfnamefont{J.}~\bibnamefont{{Silk}}},
  \bibinfo{year}{1995}, \bibinfo{journal}{ApJ} \textbf{\bibinfo{volume}{451}},
  \bibinfo{pages}{L49}.

\bibitem[{\citenamefont{{Barklem}} \emph{et~al.}(2005)\citenamefont{{Barklem},
  {Christlieb}, {Beers}, {Hill}, {Bessell}, {Holmberg}, {Marsteller}, {Rossi},
  {Zickgraf}, and {Reimers}}}]{bark05}
\bibinfo{author}{\bibnamefont{{Barklem}}, \bibfnamefont{P.~S.}},
  \bibinfo{author}{\bibfnamefont{N.}~\bibnamefont{{Christlieb}}},
  \bibinfo{author}{\bibfnamefont{T.~C.} \bibnamefont{{Beers}}},
  \bibinfo{author}{\bibfnamefont{V.}~\bibnamefont{{Hill}}},
  \bibinfo{author}{\bibfnamefont{M.~S.} \bibnamefont{{Bessell}}},
  \bibinfo{author}{\bibfnamefont{J.}~\bibnamefont{{Holmberg}}},
  \bibinfo{author}{\bibfnamefont{B.}~\bibnamefont{{Marsteller}}},
  \bibinfo{author}{\bibfnamefont{S.}~\bibnamefont{{Rossi}}},
  \bibinfo{author}{\bibfnamefont{F.}~\bibnamefont{{Zickgraf}}}, and
  \bibinfo{author}{\bibfnamefont{D.}~\bibnamefont{{Reimers}}},
  \bibinfo{year}{2005}, \bibinfo{journal}{A\&A} \textbf{\bibinfo{volume}{439}},
  \bibinfo{pages}{129}.

\bibitem[{\citenamefont{{Battaglia}}
  \emph{et~al.}(2011)\citenamefont{{Battaglia}, {Tolstoy}, {Helmi}, {Irwin},
  {Parisi}, {Hill}, and {Jablonka}}}]{batt11}
\bibinfo{author}{\bibnamefont{{Battaglia}}, \bibfnamefont{G.}},
  \bibinfo{author}{\bibfnamefont{E.}~\bibnamefont{{Tolstoy}}},
  \bibinfo{author}{\bibfnamefont{A.}~\bibnamefont{{Helmi}}},
  \bibinfo{author}{\bibfnamefont{M.}~\bibnamefont{{Irwin}}},
  \bibinfo{author}{\bibfnamefont{P.}~\bibnamefont{{Parisi}}},
  \bibinfo{author}{\bibfnamefont{V.}~\bibnamefont{{Hill}}}, and
  \bibinfo{author}{\bibfnamefont{P.}~\bibnamefont{{Jablonka}}},
  \bibinfo{year}{2011}, \bibinfo{journal}{MNRAS}
  \textbf{\bibinfo{volume}{411}}, \bibinfo{pages}{1013}.

\bibitem[{\citenamefont{{Beers}} \emph{et~al.}(2012)\citenamefont{{Beers},
  {Carollo}, {Ivezi{\'c}}, {An}, {Chiba}, {Norris}, {Freeman}, {Lee}, {Munn},
  {Re Fiorentin}, {Sivarani}, {Wilhelm}} \emph{et~al.}}]{beer12}
\bibinfo{author}{\bibnamefont{{Beers}}, \bibfnamefont{T.~C.}},
  \bibinfo{author}{\bibfnamefont{D.}~\bibnamefont{{Carollo}}},
  \bibinfo{author}{\bibfnamefont{{\v Z}.}~\bibnamefont{{Ivezi{\'c}}}},
  \bibinfo{author}{\bibfnamefont{D.}~\bibnamefont{{An}}},
  \bibinfo{author}{\bibfnamefont{M.}~\bibnamefont{{Chiba}}},
  \bibinfo{author}{\bibfnamefont{J.~E.} \bibnamefont{{Norris}}},
  \bibinfo{author}{\bibfnamefont{K.~C.} \bibnamefont{{Freeman}}},
  \bibinfo{author}{\bibfnamefont{Y.~S.} \bibnamefont{{Lee}}},
  \bibinfo{author}{\bibfnamefont{J.~A.} \bibnamefont{{Munn}}},
  \bibinfo{author}{\bibfnamefont{P.}~\bibnamefont{{Re Fiorentin}}},
  \bibinfo{author}{\bibfnamefont{T.}~\bibnamefont{{Sivarani}}},
  \bibinfo{author}{\bibfnamefont{R.}~\bibnamefont{{Wilhelm}}}, \emph{et~al.},
  \bibinfo{year}{2012}, \bibinfo{journal}{ApJ} \textbf{\bibinfo{volume}{746}},
  \bibinfo{eid}{34}.

\bibitem[{\citenamefont{{Bland-Hawthorn}}
  \emph{et~al.}(2010{\natexlab{a}})\citenamefont{{Bland-Hawthorn}, {Karlsson},
  {Sharma}, {Krumholz}, and {Silk}}}]{blan10a}
\bibinfo{author}{\bibnamefont{{Bland-Hawthorn}}, \bibfnamefont{J.}},
  \bibinfo{author}{\bibfnamefont{T.}~\bibnamefont{{Karlsson}}},
  \bibinfo{author}{\bibfnamefont{S.}~\bibnamefont{{Sharma}}},
  \bibinfo{author}{\bibfnamefont{M.}~\bibnamefont{{Krumholz}}}, and
  \bibinfo{author}{\bibfnamefont{J.}~\bibnamefont{{Silk}}},
  \bibinfo{year}{2010}{\natexlab{a}}, \bibinfo{journal}{ApJ}
  \textbf{\bibinfo{volume}{721}}, \bibinfo{pages}{582}.

\bibitem[{\citenamefont{{Bland-Hawthorn}}
  \emph{et~al.}(2010{\natexlab{b}})\citenamefont{{Bland-Hawthorn}, {Krumholz},
  and {Freeman}}}]{blan10b}
\bibinfo{author}{\bibnamefont{{Bland-Hawthorn}}, \bibfnamefont{J.}},
  \bibinfo{author}{\bibfnamefont{M.~R.} \bibnamefont{{Krumholz}}}, and
  \bibinfo{author}{\bibfnamefont{K.}~\bibnamefont{{Freeman}}},
  \bibinfo{year}{2010}{\natexlab{b}}, \bibinfo{journal}{ApJ}
  \textbf{\bibinfo{volume}{713}}, \bibinfo{pages}{166}.

\bibitem[{\citenamefont{{Bovill} and {Ricotti}}(2009)}]{bovi09}
\bibinfo{author}{\bibnamefont{{Bovill}}, \bibfnamefont{M.~S.}}, and
  \bibinfo{author}{\bibfnamefont{M.}~\bibnamefont{{Ricotti}}},
  \bibinfo{year}{2009}, \bibinfo{journal}{ApJ} \textbf{\bibinfo{volume}{693}},
  \bibinfo{pages}{1859}.

\bibitem[{\citenamefont{{Bovill} and {Ricotti}}(2011{\natexlab{a}})}]{bovi11a}
\bibinfo{author}{\bibnamefont{{Bovill}}, \bibfnamefont{M.~S.}}, and
  \bibinfo{author}{\bibfnamefont{M.}~\bibnamefont{{Ricotti}}},
  \bibinfo{year}{2011}{\natexlab{a}}, \bibinfo{journal}{ApJ}
  \textbf{\bibinfo{volume}{741}}, \bibinfo{eid}{17}.

\bibitem[{\citenamefont{{Bovill} and {Ricotti}}(2011{\natexlab{b}})}]{bovi11b}
\bibinfo{author}{\bibnamefont{{Bovill}}, \bibfnamefont{M.~S.}}, and
  \bibinfo{author}{\bibfnamefont{M.}~\bibnamefont{{Ricotti}}},
  \bibinfo{year}{2011}{\natexlab{b}}, \bibinfo{journal}{ApJ}
  \textbf{\bibinfo{volume}{741}}, \bibinfo{eid}{18}.

\bibitem[{\citenamefont{{Bromm}} \emph{et~al.}(2009)\citenamefont{{Bromm},
  {Yoshida}, {Hernquist}, and {McKee}}}]{brom09}
\bibinfo{author}{\bibnamefont{{Bromm}}, \bibfnamefont{V.}},
  \bibinfo{author}{\bibfnamefont{N.}~\bibnamefont{{Yoshida}}},
  \bibinfo{author}{\bibfnamefont{L.}~\bibnamefont{{Hernquist}}}, and
  \bibinfo{author}{\bibfnamefont{C.~F.} \bibnamefont{{McKee}}},
  \bibinfo{year}{2009}, \bibinfo{journal}{Nature}
  \textbf{\bibinfo{volume}{459}}, \bibinfo{pages}{49}.

\bibitem[{\citenamefont{{Carollo}} \emph{et~al.}(2010)\citenamefont{{Carollo},
  {Beers}, {Chiba}, {Norris}, {Freeman}, {Lee}, {Ivezi{\'c}}, {Rockosi}, and
  {Yanny}}}]{caro10}
\bibinfo{author}{\bibnamefont{{Carollo}}, \bibfnamefont{D.}},
  \bibinfo{author}{\bibfnamefont{T.~C.} \bibnamefont{{Beers}}},
  \bibinfo{author}{\bibfnamefont{M.}~\bibnamefont{{Chiba}}},
  \bibinfo{author}{\bibfnamefont{J.~E.} \bibnamefont{{Norris}}},
  \bibinfo{author}{\bibfnamefont{K.~C.} \bibnamefont{{Freeman}}},
  \bibinfo{author}{\bibfnamefont{Y.~S.} \bibnamefont{{Lee}}},
  \bibinfo{author}{\bibfnamefont{{\v Z}.}~\bibnamefont{{Ivezi{\'c}}}},
  \bibinfo{author}{\bibfnamefont{C.~M.} \bibnamefont{{Rockosi}}}, and
  \bibinfo{author}{\bibfnamefont{B.}~\bibnamefont{{Yanny}}},
  \bibinfo{year}{2010}, \bibinfo{journal}{ApJ} \textbf{\bibinfo{volume}{712}},
  \bibinfo{pages}{692}.

\bibitem[{\citenamefont{{Carollo}} \emph{et~al.}(2007)\citenamefont{{Carollo},
  {Beers}, {Lee}, {Chiba}, {Norris}, {Wilhelm}, {Sivarani}, {Marsteller},
  {Munn}, {Bailer-Jones}, {Fiorentin}, and {York}}}]{caro07}
\bibinfo{author}{\bibnamefont{{Carollo}}, \bibfnamefont{D.}},
  \bibinfo{author}{\bibfnamefont{T.~C.} \bibnamefont{{Beers}}},
  \bibinfo{author}{\bibfnamefont{Y.~S.} \bibnamefont{{Lee}}},
  \bibinfo{author}{\bibfnamefont{M.}~\bibnamefont{{Chiba}}},
  \bibinfo{author}{\bibfnamefont{J.~E.} \bibnamefont{{Norris}}},
  \bibinfo{author}{\bibfnamefont{R.}~\bibnamefont{{Wilhelm}}},
  \bibinfo{author}{\bibfnamefont{T.}~\bibnamefont{{Sivarani}}},
  \bibinfo{author}{\bibfnamefont{B.}~\bibnamefont{{Marsteller}}},
  \bibinfo{author}{\bibfnamefont{J.~A.} \bibnamefont{{Munn}}},
  \bibinfo{author}{\bibfnamefont{C.~A.~L.} \bibnamefont{{Bailer-Jones}}},
  \bibinfo{author}{\bibfnamefont{P.~R.} \bibnamefont{{Fiorentin}}}, and
  \bibinfo{author}{\bibfnamefont{D.~G.} \bibnamefont{{York}}},
  \bibinfo{year}{2007}, \bibinfo{journal}{Nature}
  \textbf{\bibinfo{volume}{450}}, \bibinfo{pages}{1020}.

\bibitem[{\citenamefont{{Carretta}}
  \emph{et~al.}(2009)\citenamefont{{Carretta}, {Bragaglia}, {Gratton},
  {Lucatello}, {Catanzaro}, {Leone}, {Bellazzini}, {Claudi}, {D'Orazi},
  {Momany}, {Ortolani}, {Pancino}} \emph{et~al.}}]{carr09}
\bibinfo{author}{\bibnamefont{{Carretta}}, \bibfnamefont{E.}},
  \bibinfo{author}{\bibfnamefont{A.}~\bibnamefont{{Bragaglia}}},
  \bibinfo{author}{\bibfnamefont{R.~G.} \bibnamefont{{Gratton}}},
  \bibinfo{author}{\bibfnamefont{S.}~\bibnamefont{{Lucatello}}},
  \bibinfo{author}{\bibfnamefont{G.}~\bibnamefont{{Catanzaro}}},
  \bibinfo{author}{\bibfnamefont{F.}~\bibnamefont{{Leone}}},
  \bibinfo{author}{\bibfnamefont{M.}~\bibnamefont{{Bellazzini}}},
  \bibinfo{author}{\bibfnamefont{R.}~\bibnamefont{{Claudi}}},
  \bibinfo{author}{\bibfnamefont{V.}~\bibnamefont{{D'Orazi}}},
  \bibinfo{author}{\bibfnamefont{Y.}~\bibnamefont{{Momany}}},
  \bibinfo{author}{\bibfnamefont{S.}~\bibnamefont{{Ortolani}}},
  \bibinfo{author}{\bibfnamefont{E.}~\bibnamefont{{Pancino}}}, \emph{et~al.},
  \bibinfo{year}{2009}, \bibinfo{journal}{A\&A} \textbf{\bibinfo{volume}{505}},
  \bibinfo{pages}{117}.

\bibitem[{\citenamefont{{Cayrel}} \emph{et~al.}(2004)\citenamefont{{Cayrel},
  {Depagne}, {Spite}, {Hill}, {Spite}, {Fran{\c c}ois}, {Plez}, {Beers},
  {Primas}, {Andersen}, {Barbuy}, {Bonifacio}} \emph{et~al.}}]{cayr04}
\bibinfo{author}{\bibnamefont{{Cayrel}}, \bibfnamefont{R.}},
  \bibinfo{author}{\bibfnamefont{E.}~\bibnamefont{{Depagne}}},
  \bibinfo{author}{\bibfnamefont{M.}~\bibnamefont{{Spite}}},
  \bibinfo{author}{\bibfnamefont{V.}~\bibnamefont{{Hill}}},
  \bibinfo{author}{\bibfnamefont{F.}~\bibnamefont{{Spite}}},
  \bibinfo{author}{\bibfnamefont{P.}~\bibnamefont{{Fran{\c c}ois}}},
  \bibinfo{author}{\bibfnamefont{B.}~\bibnamefont{{Plez}}},
  \bibinfo{author}{\bibfnamefont{T.}~\bibnamefont{{Beers}}},
  \bibinfo{author}{\bibfnamefont{F.}~\bibnamefont{{Primas}}},
  \bibinfo{author}{\bibfnamefont{J.}~\bibnamefont{{Andersen}}},
  \bibinfo{author}{\bibfnamefont{B.}~\bibnamefont{{Barbuy}}},
  \bibinfo{author}{\bibfnamefont{P.}~\bibnamefont{{Bonifacio}}}, \emph{et~al.},
  \bibinfo{year}{2004}, \bibinfo{journal}{A\&A} \textbf{\bibinfo{volume}{416}},
  \bibinfo{pages}{1117}.

\bibitem[{\citenamefont{{Cescutti}}(2008)}]{cesc08}
\bibinfo{author}{\bibnamefont{{Cescutti}}, \bibfnamefont{G.}},
  \bibinfo{year}{2008}, \bibinfo{journal}{A\&A} \textbf{\bibinfo{volume}{481}},
  \bibinfo{pages}{691}.

\bibitem[{\citenamefont{{Clark}} \emph{et~al.}(2011)\citenamefont{{Clark},
  {Glover}, {Klessen}, and {Bromm}}}]{clar11}
\bibinfo{author}{\bibnamefont{{Clark}}, \bibfnamefont{P.~C.}},
  \bibinfo{author}{\bibfnamefont{S.~C.~O.} \bibnamefont{{Glover}}},
  \bibinfo{author}{\bibfnamefont{R.~S.} \bibnamefont{{Klessen}}}, and
  \bibinfo{author}{\bibfnamefont{V.}~\bibnamefont{{Bromm}}},
  \bibinfo{year}{2011}, \bibinfo{journal}{ApJ} \textbf{\bibinfo{volume}{727}},
  \bibinfo{pages}{110}.

\bibitem[{\citenamefont{{Cohen} and {Huang}}(2009)}]{cohe09}
\bibinfo{author}{\bibnamefont{{Cohen}}, \bibfnamefont{J.~G.}}, and
  \bibinfo{author}{\bibfnamefont{W.}~\bibnamefont{{Huang}}},
  \bibinfo{year}{2009}, \bibinfo{journal}{ApJ} \textbf{\bibinfo{volume}{701}},
  \bibinfo{pages}{1053}.

\bibitem[{\citenamefont{{Cohen} and {Huang}}(2010)}]{cohe10}
\bibinfo{author}{\bibnamefont{{Cohen}}, \bibfnamefont{J.~G.}}, and
  \bibinfo{author}{\bibfnamefont{W.}~\bibnamefont{{Huang}}},
  \bibinfo{year}{2010}, \bibinfo{journal}{ApJ} \textbf{\bibinfo{volume}{719}},
  \bibinfo{pages}{931}.

\bibitem[{\citenamefont{{Davies}} \emph{et~al.}(2011)\citenamefont{{Davies},
  {Bastian}, {Gieles}, {Seth}, {Mengel}, and {Konstantopoulos}}}]{davi11}
\bibinfo{author}{\bibnamefont{{Davies}}, \bibfnamefont{B.}},
  \bibinfo{author}{\bibfnamefont{N.}~\bibnamefont{{Bastian}}},
  \bibinfo{author}{\bibfnamefont{M.}~\bibnamefont{{Gieles}}},
  \bibinfo{author}{\bibfnamefont{A.~C.} \bibnamefont{{Seth}}},
  \bibinfo{author}{\bibfnamefont{S.}~\bibnamefont{{Mengel}}}, and
  \bibinfo{author}{\bibfnamefont{I.~S.} \bibnamefont{{Konstantopoulos}}},
  \bibinfo{year}{2011}, \bibinfo{journal}{MNRAS}
  \textbf{\bibinfo{volume}{411}}, \bibinfo{pages}{1386}.

\bibitem[{\citenamefont{{de Boer}} \emph{et~al.}(2011)\citenamefont{{de Boer},
  {Tolstoy}, {Saha}, {Olsen}, {Irwin}, {Battaglia}, {Hill}, {Shetrone},
  {Fiorentino}, and {Cole}}}]{debo11}
\bibinfo{author}{\bibnamefont{{de Boer}}, \bibfnamefont{T.~J.~L.}},
  \bibinfo{author}{\bibfnamefont{E.}~\bibnamefont{{Tolstoy}}},
  \bibinfo{author}{\bibfnamefont{A.}~\bibnamefont{{Saha}}},
  \bibinfo{author}{\bibfnamefont{K.}~\bibnamefont{{Olsen}}},
  \bibinfo{author}{\bibfnamefont{M.~J.} \bibnamefont{{Irwin}}},
  \bibinfo{author}{\bibfnamefont{G.}~\bibnamefont{{Battaglia}}},
  \bibinfo{author}{\bibfnamefont{V.}~\bibnamefont{{Hill}}},
  \bibinfo{author}{\bibfnamefont{M.~D.} \bibnamefont{{Shetrone}}},
  \bibinfo{author}{\bibfnamefont{G.}~\bibnamefont{{Fiorentino}}}, and
  \bibinfo{author}{\bibfnamefont{A.}~\bibnamefont{{Cole}}},
  \bibinfo{year}{2011}, \bibinfo{journal}{A\&A} \textbf{\bibinfo{volume}{528}},
  \bibinfo{pages}{A119}.

\bibitem[{\citenamefont{{De Silva}} \emph{et~al.}(2007)\citenamefont{{De
  Silva}, {Freeman}, {Asplund}, {Bland-Hawthorn}, {Bessell}, and
  {Collet}}}]{desi07}
\bibinfo{author}{\bibnamefont{{De Silva}}, \bibfnamefont{G.~M.}},
  \bibinfo{author}{\bibfnamefont{K.~C.} \bibnamefont{{Freeman}}},
  \bibinfo{author}{\bibfnamefont{M.}~\bibnamefont{{Asplund}}},
  \bibinfo{author}{\bibfnamefont{J.}~\bibnamefont{{Bland-Hawthorn}}},
  \bibinfo{author}{\bibfnamefont{M.~S.} \bibnamefont{{Bessell}}}, and
  \bibinfo{author}{\bibfnamefont{R.}~\bibnamefont{{Collet}}},
  \bibinfo{year}{2007}, \bibinfo{journal}{AJ} \textbf{\bibinfo{volume}{133}},
  \bibinfo{pages}{1161}.

\bibitem[{\citenamefont{{De Silva}} \emph{et~al.}(2006)\citenamefont{{De
  Silva}, {Sneden}, {Paulson}, {Asplund}, {Bland-Hawthorn}, {Bessell}, and
  {Freeman}}}]{desi06}
\bibinfo{author}{\bibnamefont{{De Silva}}, \bibfnamefont{G.~M.}},
  \bibinfo{author}{\bibfnamefont{C.}~\bibnamefont{{Sneden}}},
  \bibinfo{author}{\bibfnamefont{D.~B.} \bibnamefont{{Paulson}}},
  \bibinfo{author}{\bibfnamefont{M.}~\bibnamefont{{Asplund}}},
  \bibinfo{author}{\bibfnamefont{J.}~\bibnamefont{{Bland-Hawthorn}}},
  \bibinfo{author}{\bibfnamefont{M.~S.} \bibnamefont{{Bessell}}}, and
  \bibinfo{author}{\bibfnamefont{K.~C.} \bibnamefont{{Freeman}}},
  \bibinfo{year}{2006}, \bibinfo{journal}{AJ} \textbf{\bibinfo{volume}{131}},
  \bibinfo{pages}{455}.

\bibitem[{\citenamefont{{Elmegreen}}(2010)}]{elme10}
\bibinfo{author}{\bibnamefont{{Elmegreen}}, \bibfnamefont{B.~G.}},
  \bibinfo{year}{2010}, \bibinfo{journal}{ApJ} \textbf{\bibinfo{volume}{712}},
  \bibinfo{pages}{L184}.

\bibitem[{\citenamefont{{Fall}} \emph{et~al.}(2005)\citenamefont{{Fall},
  {Chandar}, and {Whitmore}}}]{fall05}
\bibinfo{author}{\bibnamefont{{Fall}}, \bibfnamefont{S.~M.}},
  \bibinfo{author}{\bibfnamefont{R.}~\bibnamefont{{Chandar}}}, and
  \bibinfo{author}{\bibfnamefont{B.~C.} \bibnamefont{{Whitmore}}},
  \bibinfo{year}{2005}, \bibinfo{journal}{ApJ} \textbf{\bibinfo{volume}{631}},
  \bibinfo{pages}{L133}.

\bibitem[{\citenamefont{{Fall}} \emph{et~al.}(2009)\citenamefont{{Fall},
  {Chandar}, and {Whitmore}}}]{fall09}
\bibinfo{author}{\bibnamefont{{Fall}}, \bibfnamefont{S.~M.}},
  \bibinfo{author}{\bibfnamefont{R.}~\bibnamefont{{Chandar}}}, and
  \bibinfo{author}{\bibfnamefont{B.~C.} \bibnamefont{{Whitmore}}},
  \bibinfo{year}{2009}, \bibinfo{journal}{ApJ} \textbf{\bibinfo{volume}{704}},
  \bibinfo{pages}{453}.

\bibitem[{\citenamefont{{Feltzing}}
  \emph{et~al.}(2009)\citenamefont{{Feltzing}, {Eriksson}, {Kleyna}, and
  {Wilkinson}}}]{felt09}
\bibinfo{author}{\bibnamefont{{Feltzing}}, \bibfnamefont{S.}},
  \bibinfo{author}{\bibfnamefont{K.}~\bibnamefont{{Eriksson}}},
  \bibinfo{author}{\bibfnamefont{J.}~\bibnamefont{{Kleyna}}}, and
  \bibinfo{author}{\bibfnamefont{M.~I.} \bibnamefont{{Wilkinson}}},
  \bibinfo{year}{2009}, \bibinfo{journal}{A\&A} \textbf{\bibinfo{volume}{508}},
  \bibinfo{pages}{L1}.

\bibitem[{\citenamefont{{Frebel}} \emph{et~al.}(2007)\citenamefont{{Frebel},
  {Johnson}, and {Bromm}}}]{freb07}
\bibinfo{author}{\bibnamefont{{Frebel}}, \bibfnamefont{A.}},
  \bibinfo{author}{\bibfnamefont{J.~L.} \bibnamefont{{Johnson}}}, and
  \bibinfo{author}{\bibfnamefont{V.}~\bibnamefont{{Bromm}}},
  \bibinfo{year}{2007}, \bibinfo{journal}{MNRAS}
  \textbf{\bibinfo{volume}{380}}, \bibinfo{pages}{L40}.

\bibitem[{\citenamefont{{Frebel}}
  \emph{et~al.}(2010{\natexlab{a}})\citenamefont{{Frebel}, {Kirby}, and
  {Simon}}}]{freb10a}
\bibinfo{author}{\bibnamefont{{Frebel}}, \bibfnamefont{A.}},
  \bibinfo{author}{\bibfnamefont{E.~N.} \bibnamefont{{Kirby}}}, and
  \bibinfo{author}{\bibfnamefont{J.~D.} \bibnamefont{{Simon}}},
  \bibinfo{year}{2010}{\natexlab{a}}, \bibinfo{journal}{Nature}
  \textbf{\bibinfo{volume}{464}}, \bibinfo{pages}{72}.

\bibitem[{\citenamefont{{Frebel}}
  \emph{et~al.}(2010{\natexlab{b}})\citenamefont{{Frebel}, {Simon}, {Geha}, and
  {Willman}}}]{freb10b}
\bibinfo{author}{\bibnamefont{{Frebel}}, \bibfnamefont{A.}},
  \bibinfo{author}{\bibfnamefont{J.~D.} \bibnamefont{{Simon}}},
  \bibinfo{author}{\bibfnamefont{M.}~\bibnamefont{{Geha}}}, and
  \bibinfo{author}{\bibfnamefont{B.}~\bibnamefont{{Willman}}},
  \bibinfo{year}{2010}{\natexlab{b}}, \bibinfo{journal}{ApJ}
  \textbf{\bibinfo{volume}{708}}, \bibinfo{pages}{560}.

\bibitem[{\citenamefont{{Fulbright}}
  \emph{et~al.}(2004)\citenamefont{{Fulbright}, {Rich}, and {Castro}}}]{fulb04}
\bibinfo{author}{\bibnamefont{{Fulbright}}, \bibfnamefont{J.~P.}},
  \bibinfo{author}{\bibfnamefont{R.~M.} \bibnamefont{{Rich}}}, and
  \bibinfo{author}{\bibfnamefont{S.}~\bibnamefont{{Castro}}},
  \bibinfo{year}{2004}, \bibinfo{journal}{ApJ} \textbf{\bibinfo{volume}{612}},
  \bibinfo{pages}{447}.

\bibitem[{\citenamefont{{Geisler}} \emph{et~al.}(2005)\citenamefont{{Geisler},
  {Smith}, {Wallerstein}, {Gonzalez}, and {Charbonnel}}}]{geis05}
\bibinfo{author}{\bibnamefont{{Geisler}}, \bibfnamefont{D.}},
  \bibinfo{author}{\bibfnamefont{V.~V.} \bibnamefont{{Smith}}},
  \bibinfo{author}{\bibfnamefont{G.}~\bibnamefont{{Wallerstein}}},
  \bibinfo{author}{\bibfnamefont{G.}~\bibnamefont{{Gonzalez}}}, and
  \bibinfo{author}{\bibfnamefont{C.}~\bibnamefont{{Charbonnel}}},
  \bibinfo{year}{2005}, \bibinfo{journal}{AJ} \textbf{\bibinfo{volume}{129}},
  \bibinfo{pages}{1428}.

\bibitem[{\citenamefont{{Gilmore}} \emph{et~al.}(2007)\citenamefont{{Gilmore},
  {Wilkinson}, {Wyse}, {Kleyna}, {Koch}, {Evans}, and {Grebel}}}]{gilm07}
\bibinfo{author}{\bibnamefont{{Gilmore}}, \bibfnamefont{G.}},
  \bibinfo{author}{\bibfnamefont{M.~I.} \bibnamefont{{Wilkinson}}},
  \bibinfo{author}{\bibfnamefont{R.~F.~G.} \bibnamefont{{Wyse}}},
  \bibinfo{author}{\bibfnamefont{J.~T.} \bibnamefont{{Kleyna}}},
  \bibinfo{author}{\bibfnamefont{A.}~\bibnamefont{{Koch}}},
  \bibinfo{author}{\bibfnamefont{N.~W.} \bibnamefont{{Evans}}}, and
  \bibinfo{author}{\bibfnamefont{E.~K.} \bibnamefont{{Grebel}}},
  \bibinfo{year}{2007}, \bibinfo{journal}{ApJ} \textbf{\bibinfo{volume}{663}},
  \bibinfo{pages}{948}.

\bibitem[{\citenamefont{{Grebel} and {Gallagher}}(2004)}]{greb04}
\bibinfo{author}{\bibnamefont{{Grebel}}, \bibfnamefont{E.~K.}}, and
  \bibinfo{author}{\bibfnamefont{J.~S.} \bibnamefont{{Gallagher}},
  \bibfnamefont{III}}, \bibinfo{year}{2004}, \bibinfo{journal}{ApJ}
  \textbf{\bibinfo{volume}{610}}, \bibinfo{pages}{L89}.

\bibitem[{\citenamefont{{Greif}} \emph{et~al.}(2010)\citenamefont{{Greif},
  {Glover}, {Bromm}, and {Klessen}}}]{grei10}
\bibinfo{author}{\bibnamefont{{Greif}}, \bibfnamefont{T.~H.}},
  \bibinfo{author}{\bibfnamefont{S.~C.~O.} \bibnamefont{{Glover}}},
  \bibinfo{author}{\bibfnamefont{V.}~\bibnamefont{{Bromm}}}, and
  \bibinfo{author}{\bibfnamefont{R.~S.} \bibnamefont{{Klessen}}},
  \bibinfo{year}{2010}, \bibinfo{journal}{ApJ} \textbf{\bibinfo{volume}{716}},
  \bibinfo{pages}{510}.

\bibitem[{\citenamefont{{Heger} and {Woosley}}(2002)}]{hege02}
\bibinfo{author}{\bibnamefont{{Heger}}, \bibfnamefont{A.}}, and
  \bibinfo{author}{\bibfnamefont{S.~E.} \bibnamefont{{Woosley}}},
  \bibinfo{year}{2002}, \bibinfo{journal}{ApJ} \textbf{\bibinfo{volume}{567}},
  \bibinfo{pages}{532}.

\bibitem[{\citenamefont{{Irwin} and {Hatzidimitriou}}(1995)}]{irwi95}
\bibinfo{author}{\bibnamefont{{Irwin}}, \bibfnamefont{M.}}, and
  \bibinfo{author}{\bibfnamefont{D.}~\bibnamefont{{Hatzidimitriou}}},
  \bibinfo{year}{1995}, \bibinfo{journal}{\mnras}
  \textbf{\bibinfo{volume}{277}}, \bibinfo{pages}{1354}.

\bibitem[{\citenamefont{{Ivans}} \emph{et~al.}(2003)\citenamefont{{Ivans},
  {Sneden}, {James}, {Preston}, {Fulbright}, {H{\"o}flich}, {Carney}, and
  {Wheeler}}}]{ivan03}
\bibinfo{author}{\bibnamefont{{Ivans}}, \bibfnamefont{I.~I.}},
  \bibinfo{author}{\bibfnamefont{C.}~\bibnamefont{{Sneden}}},
  \bibinfo{author}{\bibfnamefont{C.~R.} \bibnamefont{{James}}},
  \bibinfo{author}{\bibfnamefont{G.~W.} \bibnamefont{{Preston}}},
  \bibinfo{author}{\bibfnamefont{J.~P.} \bibnamefont{{Fulbright}}},
  \bibinfo{author}{\bibfnamefont{P.~A.} \bibnamefont{{H{\"o}flich}}},
  \bibinfo{author}{\bibfnamefont{B.~W.} \bibnamefont{{Carney}}}, and
  \bibinfo{author}{\bibfnamefont{J.~C.} \bibnamefont{{Wheeler}}},
  \bibinfo{year}{2003}, \bibinfo{journal}{ApJ} \textbf{\bibinfo{volume}{592}},
  \bibinfo{pages}{906}.

\bibitem[{\citenamefont{{Karlsson}}(2005)}]{karl05}
\bibinfo{author}{\bibnamefont{{Karlsson}}, \bibfnamefont{T.}},
  \bibinfo{year}{2005}, \bibinfo{journal}{A\&A} \textbf{\bibinfo{volume}{439}},
  \bibinfo{pages}{93}.

\bibitem[{\citenamefont{{Karlsson}}(2006)}]{karl06}
\bibinfo{author}{\bibnamefont{{Karlsson}}, \bibfnamefont{T.}},
  \bibinfo{year}{2006}, \bibinfo{journal}{ApJ} \textbf{\bibinfo{volume}{641}},
  \bibinfo{pages}{L41}.

\bibitem[{\citenamefont{{Karlsson}}
  \emph{et~al.}(2011)\citenamefont{{Karlsson}, {Bromm}, and
  {Bland-Hawthorn}}}]{karl11}
\bibinfo{author}{\bibnamefont{{Karlsson}}, \bibfnamefont{T.}},
  \bibinfo{author}{\bibfnamefont{V.}~\bibnamefont{{Bromm}}}, and
  \bibinfo{author}{\bibfnamefont{J.}~\bibnamefont{{Bland-Hawthorn}}},
  \bibinfo{year}{2011}, \bibinfo{note}{arXiv:1101.4024 [astro-ph.CO]}.

\bibitem[{\citenamefont{{Karlsson} and {Gustafsson}}(2005)}]{karl05b}
\bibinfo{author}{\bibnamefont{{Karlsson}}, \bibfnamefont{T.}}, and
  \bibinfo{author}{\bibfnamefont{B.}~\bibnamefont{{Gustafsson}}},
  \bibinfo{year}{2005}, \bibinfo{journal}{A\&A} \textbf{\bibinfo{volume}{436}},
  \bibinfo{pages}{879}.

\bibitem[{\citenamefont{{Karlsson}}
  \emph{et~al.}(2008)\citenamefont{{Karlsson}, {Johnson}, and
  {Bromm}}}]{karl08}
\bibinfo{author}{\bibnamefont{{Karlsson}}, \bibfnamefont{T.}},
  \bibinfo{author}{\bibfnamefont{J.~L.} \bibnamefont{{Johnson}}}, and
  \bibinfo{author}{\bibfnamefont{V.}~\bibnamefont{{Bromm}}},
  \bibinfo{year}{2008}, \bibinfo{journal}{ApJ} \textbf{\bibinfo{volume}{679}},
  \bibinfo{pages}{6}.

\bibitem[{\citenamefont{{Kirby}}
  \emph{et~al.}(2011{\natexlab{a}})\citenamefont{{Kirby}, {Cohen}, {Smith},
  {Majewski}, {Sohn}, and {Guhathakurta}}}]{kirb11a}
\bibinfo{author}{\bibnamefont{{Kirby}}, \bibfnamefont{E.~N.}},
  \bibinfo{author}{\bibfnamefont{J.~G.} \bibnamefont{{Cohen}}},
  \bibinfo{author}{\bibfnamefont{G.~H.} \bibnamefont{{Smith}}},
  \bibinfo{author}{\bibfnamefont{S.~R.} \bibnamefont{{Majewski}}},
  \bibinfo{author}{\bibfnamefont{S.~T.} \bibnamefont{{Sohn}}}, and
  \bibinfo{author}{\bibfnamefont{P.}~\bibnamefont{{Guhathakurta}}},
  \bibinfo{year}{2011}{\natexlab{a}}, \bibinfo{journal}{ApJ}
  \textbf{\bibinfo{volume}{727}}, \bibinfo{pages}{79}.

\bibitem[{\citenamefont{{Kirby}} \emph{et~al.}(2010)\citenamefont{{Kirby},
  {Guhathakurta}, {Simon}, {Geha}, {Rockosi}, {Sneden}, {Cohen}, {Sohn},
  {Majewski}, and {Siegel}}}]{kirb10}
\bibinfo{author}{\bibnamefont{{Kirby}}, \bibfnamefont{E.~N.}},
  \bibinfo{author}{\bibfnamefont{P.}~\bibnamefont{{Guhathakurta}}},
  \bibinfo{author}{\bibfnamefont{J.~D.} \bibnamefont{{Simon}}},
  \bibinfo{author}{\bibfnamefont{M.~C.} \bibnamefont{{Geha}}},
  \bibinfo{author}{\bibfnamefont{C.~M.} \bibnamefont{{Rockosi}}},
  \bibinfo{author}{\bibfnamefont{C.}~\bibnamefont{{Sneden}}},
  \bibinfo{author}{\bibfnamefont{J.~G.} \bibnamefont{{Cohen}}},
  \bibinfo{author}{\bibfnamefont{S.~T.} \bibnamefont{{Sohn}}},
  \bibinfo{author}{\bibfnamefont{S.~R.} \bibnamefont{{Majewski}}}, and
  \bibinfo{author}{\bibfnamefont{M.}~\bibnamefont{{Siegel}}},
  \bibinfo{year}{2010}, \bibinfo{journal}{ApJS} \textbf{\bibinfo{volume}{191}},
  \bibinfo{pages}{352}.

\bibitem[{\citenamefont{{Kirby}}
  \emph{et~al.}(2011{\natexlab{b}})\citenamefont{{Kirby}, {Lanfranchi},
  {Simon}, {Cohen}, and {Guhathakurta}}}]{kirb11b}
\bibinfo{author}{\bibnamefont{{Kirby}}, \bibfnamefont{E.~N.}},
  \bibinfo{author}{\bibfnamefont{G.~A.} \bibnamefont{{Lanfranchi}}},
  \bibinfo{author}{\bibfnamefont{J.~D.} \bibnamefont{{Simon}}},
  \bibinfo{author}{\bibfnamefont{J.~G.} \bibnamefont{{Cohen}}}, and
  \bibinfo{author}{\bibfnamefont{P.}~\bibnamefont{{Guhathakurta}}},
  \bibinfo{year}{2011}{\natexlab{b}}, \bibinfo{journal}{ApJ}
  \textbf{\bibinfo{volume}{727}}, \bibinfo{pages}{78}.

\bibitem[{\citenamefont{{Kirby}} \emph{et~al.}(2008)\citenamefont{{Kirby},
  {Simon}, {Geha}, {Guhathakurta}, and {Frebel}}}]{kirb08}
\bibinfo{author}{\bibnamefont{{Kirby}}, \bibfnamefont{E.~N.}},
  \bibinfo{author}{\bibfnamefont{J.~D.} \bibnamefont{{Simon}}},
  \bibinfo{author}{\bibfnamefont{M.}~\bibnamefont{{Geha}}},
  \bibinfo{author}{\bibfnamefont{P.}~\bibnamefont{{Guhathakurta}}}, and
  \bibinfo{author}{\bibfnamefont{A.}~\bibnamefont{{Frebel}}},
  \bibinfo{year}{2008}, \bibinfo{journal}{ApJ} \textbf{\bibinfo{volume}{685}},
  \bibinfo{pages}{L43}.

\bibitem[{\citenamefont{{Kleyna}} \emph{et~al.}(2004)\citenamefont{{Kleyna},
  {Wilkinson}, {Evans}, and {Gilmore}}}]{kley04}
\bibinfo{author}{\bibnamefont{{Kleyna}}, \bibfnamefont{J.~T.}},
  \bibinfo{author}{\bibfnamefont{M.~I.} \bibnamefont{{Wilkinson}}},
  \bibinfo{author}{\bibfnamefont{N.~W.} \bibnamefont{{Evans}}}, and
  \bibinfo{author}{\bibfnamefont{G.}~\bibnamefont{{Gilmore}}},
  \bibinfo{year}{2004}, \bibinfo{journal}{MNRAS}
  \textbf{\bibinfo{volume}{354}}, \bibinfo{pages}{L66}.

\bibitem[{\citenamefont{{Kleyna}} \emph{et~al.}(2003)\citenamefont{{Kleyna},
  {Wilkinson}, {Gilmore}, and {Evans}}}]{kley03}
\bibinfo{author}{\bibnamefont{{Kleyna}}, \bibfnamefont{J.~T.}},
  \bibinfo{author}{\bibfnamefont{M.~I.} \bibnamefont{{Wilkinson}}},
  \bibinfo{author}{\bibfnamefont{G.}~\bibnamefont{{Gilmore}}}, and
  \bibinfo{author}{\bibfnamefont{N.~W.} \bibnamefont{{Evans}}},
  \bibinfo{year}{2003}, \bibinfo{journal}{ApJ} \textbf{\bibinfo{volume}{588}},
  \bibinfo{pages}{L21}.

\bibitem[{\citenamefont{{Koch}} \emph{et~al.}(2008)\citenamefont{{Koch},
  {Grebel}, {Gilmore}, {Wyse}, {Kleyna}, {Harbeck}, {Wilkinson}, and {Wyn
  Evans}}}]{koch08}
\bibinfo{author}{\bibnamefont{{Koch}}, \bibfnamefont{A.}},
  \bibinfo{author}{\bibfnamefont{E.~K.} \bibnamefont{{Grebel}}},
  \bibinfo{author}{\bibfnamefont{G.~F.} \bibnamefont{{Gilmore}}},
  \bibinfo{author}{\bibfnamefont{R.~F.~G.} \bibnamefont{{Wyse}}},
  \bibinfo{author}{\bibfnamefont{J.~T.} \bibnamefont{{Kleyna}}},
  \bibinfo{author}{\bibfnamefont{D.~R.} \bibnamefont{{Harbeck}}},
  \bibinfo{author}{\bibfnamefont{M.~I.} \bibnamefont{{Wilkinson}}}, and
  \bibinfo{author}{\bibfnamefont{N.}~\bibnamefont{{Wyn Evans}}},
  \bibinfo{year}{2008}, \bibinfo{journal}{AJ} \textbf{\bibinfo{volume}{135}},
  \bibinfo{pages}{1580}.

\bibitem[{\citenamefont{{Kroupa}}(2001)}]{krou01}
\bibinfo{author}{\bibnamefont{{Kroupa}}, \bibfnamefont{P.}},
  \bibinfo{year}{2001}, \bibinfo{journal}{MNRAS}
  \textbf{\bibinfo{volume}{322}}, \bibinfo{pages}{231}.

\bibitem[{\citenamefont{{Lada} and {Lada}}(2003)}]{lada03}
\bibinfo{author}{\bibnamefont{{Lada}}, \bibfnamefont{C.~J.}}, and
  \bibinfo{author}{\bibfnamefont{E.~A.} \bibnamefont{{Lada}}},
  \bibinfo{year}{2003}, \bibinfo{journal}{ARA\&A}
  \textbf{\bibinfo{volume}{41}}, \bibinfo{pages}{57}.

\bibitem[{\citenamefont{{Lai}} \emph{et~al.}(2008)\citenamefont{{Lai}, {Bolte},
  {Johnson}, {Lucatello}, {Heger}, and {Woosley}}}]{lai08}
\bibinfo{author}{\bibnamefont{{Lai}}, \bibfnamefont{D.~K.}},
  \bibinfo{author}{\bibfnamefont{M.}~\bibnamefont{{Bolte}}},
  \bibinfo{author}{\bibfnamefont{J.~A.} \bibnamefont{{Johnson}}},
  \bibinfo{author}{\bibfnamefont{S.}~\bibnamefont{{Lucatello}}},
  \bibinfo{author}{\bibfnamefont{A.}~\bibnamefont{{Heger}}}, and
  \bibinfo{author}{\bibfnamefont{S.~E.} \bibnamefont{{Woosley}}},
  \bibinfo{year}{2008}, \bibinfo{journal}{ApJ} \textbf{\bibinfo{volume}{681}},
  \bibinfo{pages}{1524}.

\bibitem[{\citenamefont{{Lai}} \emph{et~al.}(2011)\citenamefont{{Lai}, {Lee},
  {Bolte}, {Lucatello}, {Beers}, {Johnson}, {Sivarani}, and {Rockosi}}}]{lai11}
\bibinfo{author}{\bibnamefont{{Lai}}, \bibfnamefont{D.~K.}},
  \bibinfo{author}{\bibfnamefont{Y.~S.} \bibnamefont{{Lee}}},
  \bibinfo{author}{\bibfnamefont{M.}~\bibnamefont{{Bolte}}},
  \bibinfo{author}{\bibfnamefont{S.}~\bibnamefont{{Lucatello}}},
  \bibinfo{author}{\bibfnamefont{T.~C.} \bibnamefont{{Beers}}},
  \bibinfo{author}{\bibfnamefont{J.~A.} \bibnamefont{{Johnson}}},
  \bibinfo{author}{\bibfnamefont{T.}~\bibnamefont{{Sivarani}}}, and
  \bibinfo{author}{\bibfnamefont{C.~M.} \bibnamefont{{Rockosi}}},
  \bibinfo{year}{2011}, \bibinfo{journal}{ApJ} \textbf{\bibinfo{volume}{738}},
  \bibinfo{eid}{51}.

\bibitem[{\citenamefont{{Larsen}}(2009)}]{lars09}
\bibinfo{author}{\bibnamefont{{Larsen}}, \bibfnamefont{S.~S.}},
  \bibinfo{year}{2009}, \bibinfo{journal}{A\&A} \textbf{\bibinfo{volume}{503}},
  \bibinfo{pages}{467}.

\bibitem[{\citenamefont{{Lee}} \emph{et~al.}(2009)\citenamefont{{Lee}, {Yuk},
  {Park}, {Harris}, and {Zaritsky}}}]{lee09}
\bibinfo{author}{\bibnamefont{{Lee}}, \bibfnamefont{M.~G.}},
  \bibinfo{author}{\bibfnamefont{I.-S.} \bibnamefont{{Yuk}}},
  \bibinfo{author}{\bibfnamefont{H.~S.} \bibnamefont{{Park}}},
  \bibinfo{author}{\bibfnamefont{J.}~\bibnamefont{{Harris}}}, and
  \bibinfo{author}{\bibfnamefont{D.}~\bibnamefont{{Zaritsky}}},
  \bibinfo{year}{2009}, \bibinfo{journal}{ApJ} \textbf{\bibinfo{volume}{703}},
  \bibinfo{pages}{692}.

\bibitem[{\citenamefont{{Letarte}} \emph{et~al.}(2006)\citenamefont{{Letarte},
  {Hill}, {Jablonka}, {Tolstoy}, {Fran{\c c}ois}, and {Meylan}}}]{leta06}
\bibinfo{author}{\bibnamefont{{Letarte}}, \bibfnamefont{B.}},
  \bibinfo{author}{\bibfnamefont{V.}~\bibnamefont{{Hill}}},
  \bibinfo{author}{\bibfnamefont{P.}~\bibnamefont{{Jablonka}}},
  \bibinfo{author}{\bibfnamefont{E.}~\bibnamefont{{Tolstoy}}},
  \bibinfo{author}{\bibfnamefont{P.}~\bibnamefont{{Fran{\c c}ois}}}, and
  \bibinfo{author}{\bibfnamefont{G.}~\bibnamefont{{Meylan}}},
  \bibinfo{year}{2006}, \bibinfo{journal}{A\&A} \textbf{\bibinfo{volume}{453}},
  \bibinfo{pages}{547}.

\bibitem[{\citenamefont{{Letarte}} \emph{et~al.}(2010)\citenamefont{{Letarte},
  {Hill}, {Tolstoy}, {Jablonka}, {Shetrone}, {Venn}, {Spite}, {Irwin},
  {Battaglia}, {Helmi}, {Primas}, {Fran{\c c}ois}} \emph{et~al.}}]{leta10}
\bibinfo{author}{\bibnamefont{{Letarte}}, \bibfnamefont{B.}},
  \bibinfo{author}{\bibfnamefont{V.}~\bibnamefont{{Hill}}},
  \bibinfo{author}{\bibfnamefont{E.}~\bibnamefont{{Tolstoy}}},
  \bibinfo{author}{\bibfnamefont{P.}~\bibnamefont{{Jablonka}}},
  \bibinfo{author}{\bibfnamefont{M.}~\bibnamefont{{Shetrone}}},
  \bibinfo{author}{\bibfnamefont{K.~A.} \bibnamefont{{Venn}}},
  \bibinfo{author}{\bibfnamefont{M.}~\bibnamefont{{Spite}}},
  \bibinfo{author}{\bibfnamefont{M.~J.} \bibnamefont{{Irwin}}},
  \bibinfo{author}{\bibfnamefont{G.}~\bibnamefont{{Battaglia}}},
  \bibinfo{author}{\bibfnamefont{A.}~\bibnamefont{{Helmi}}},
  \bibinfo{author}{\bibfnamefont{F.}~\bibnamefont{{Primas}}},
  \bibinfo{author}{\bibfnamefont{P.}~\bibnamefont{{Fran{\c c}ois}}},
  \emph{et~al.}, \bibinfo{year}{2010}, \bibinfo{journal}{A\&A}
  \textbf{\bibinfo{volume}{523}}, \bibinfo{eid}{A17}.

\bibitem[{\citenamefont{{Marigo}} \emph{et~al.}(2008)\citenamefont{{Marigo},
  {Girardi}, {Bressan}, {Groenewegen}, {Silva}, and {Granato}}}]{mari08}
\bibinfo{author}{\bibnamefont{{Marigo}}, \bibfnamefont{P.}},
  \bibinfo{author}{\bibfnamefont{L.}~\bibnamefont{{Girardi}}},
  \bibinfo{author}{\bibfnamefont{A.}~\bibnamefont{{Bressan}}},
  \bibinfo{author}{\bibfnamefont{M.~A.~T.} \bibnamefont{{Groenewegen}}},
  \bibinfo{author}{\bibfnamefont{L.}~\bibnamefont{{Silva}}}, and
  \bibinfo{author}{\bibfnamefont{G.~L.} \bibnamefont{{Granato}}},
  \bibinfo{year}{2008}, \bibinfo{journal}{A\&A} \textbf{\bibinfo{volume}{482}},
  \bibinfo{pages}{883}.

\bibitem[{\citenamefont{{Matteucci} and {Greggio}}(1986)}]{matt86}
\bibinfo{author}{\bibnamefont{{Matteucci}}, \bibfnamefont{F.}}, and
  \bibinfo{author}{\bibfnamefont{L.}~\bibnamefont{{Greggio}}},
  \bibinfo{year}{1986}, \bibinfo{journal}{A\&A} \textbf{\bibinfo{volume}{154}},
  \bibinfo{pages}{279}.

\bibitem[{\citenamefont{{Mengel} and {Tacconi-Garman}}(2009)}]{meng09}
\bibinfo{author}{\bibnamefont{{Mengel}}, \bibfnamefont{S.}}, and
  \bibinfo{author}{\bibfnamefont{L.~E.} \bibnamefont{{Tacconi-Garman}}},
  \bibinfo{year}{2009}, \bibinfo{journal}{Ap\&SS}
  \textbf{\bibinfo{volume}{324}}, \bibinfo{pages}{321}.

\bibitem[{\citenamefont{{Nakamura} and {Umemura}}(2001)}]{naka01}
\bibinfo{author}{\bibnamefont{{Nakamura}}, \bibfnamefont{F.}}, and
  \bibinfo{author}{\bibfnamefont{M.}~\bibnamefont{{Umemura}}},
  \bibinfo{year}{2001}, \bibinfo{journal}{ApJ} \textbf{\bibinfo{volume}{548}},
  \bibinfo{pages}{19}.

\bibitem[{\citenamefont{{Nomoto}} \emph{et~al.}(2006)\citenamefont{{Nomoto},
  {Tominaga}, {Umeda}, {Kobayashi}, and {Maeda}}}]{nomo06}
\bibinfo{author}{\bibnamefont{{Nomoto}}, \bibfnamefont{K.}},
  \bibinfo{author}{\bibfnamefont{N.}~\bibnamefont{{Tominaga}}},
  \bibinfo{author}{\bibfnamefont{H.}~\bibnamefont{{Umeda}}},
  \bibinfo{author}{\bibfnamefont{C.}~\bibnamefont{{Kobayashi}}}, and
  \bibinfo{author}{\bibfnamefont{K.}~\bibnamefont{{Maeda}}},
  \bibinfo{year}{2006}, \bibinfo{journal}{Nucl. Phys. A}
  \textbf{\bibinfo{volume}{777}}, \bibinfo{pages}{424}.

\bibitem[{\citenamefont{{Norris}}
  \emph{et~al.}(2010{\natexlab{a}})\citenamefont{{Norris}, {Gilmore}, {Wyse},
  {Yong}, and {Frebel}}}]{norr10a}
\bibinfo{author}{\bibnamefont{{Norris}}, \bibfnamefont{J.~E.}},
  \bibinfo{author}{\bibfnamefont{G.}~\bibnamefont{{Gilmore}}},
  \bibinfo{author}{\bibfnamefont{R.~F.~G.} \bibnamefont{{Wyse}}},
  \bibinfo{author}{\bibfnamefont{D.}~\bibnamefont{{Yong}}}, and
  \bibinfo{author}{\bibfnamefont{A.}~\bibnamefont{{Frebel}}},
  \bibinfo{year}{2010}{\natexlab{a}}, \bibinfo{journal}{ApJ}
  \textbf{\bibinfo{volume}{722}}, \bibinfo{pages}{L104}.

\bibitem[{\citenamefont{{Norris}}
  \emph{et~al.}(2010{\natexlab{b}})\citenamefont{{Norris}, {Wyse}, {Gilmore},
  {Yong}, {Frebel}, {Wilkinson}, {Belokurov}, and {Zucker}}}]{norr10b}
\bibinfo{author}{\bibnamefont{{Norris}}, \bibfnamefont{J.~E.}},
  \bibinfo{author}{\bibfnamefont{R.~F.~G.} \bibnamefont{{Wyse}}},
  \bibinfo{author}{\bibfnamefont{G.}~\bibnamefont{{Gilmore}}},
  \bibinfo{author}{\bibfnamefont{D.}~\bibnamefont{{Yong}}},
  \bibinfo{author}{\bibfnamefont{A.}~\bibnamefont{{Frebel}}},
  \bibinfo{author}{\bibfnamefont{M.~I.} \bibnamefont{{Wilkinson}}},
  \bibinfo{author}{\bibfnamefont{V.}~\bibnamefont{{Belokurov}}}, and
  \bibinfo{author}{\bibfnamefont{D.~B.} \bibnamefont{{Zucker}}},
  \bibinfo{year}{2010}{\natexlab{b}}, \bibinfo{journal}{ApJ}
  \textbf{\bibinfo{volume}{723}}, \bibinfo{pages}{1632}.

\bibitem[{\citenamefont{{Norris}}
  \emph{et~al.}(2010{\natexlab{c}})\citenamefont{{Norris}, {Yong}, {Gilmore},
  and {Wyse}}}]{norr10c}
\bibinfo{author}{\bibnamefont{{Norris}}, \bibfnamefont{J.~E.}},
  \bibinfo{author}{\bibfnamefont{D.}~\bibnamefont{{Yong}}},
  \bibinfo{author}{\bibfnamefont{G.}~\bibnamefont{{Gilmore}}}, and
  \bibinfo{author}{\bibfnamefont{R.~F.~G.} \bibnamefont{{Wyse}}},
  \bibinfo{year}{2010}{\natexlab{c}}, \bibinfo{journal}{ApJ}
  \textbf{\bibinfo{volume}{711}}, \bibinfo{pages}{350}.

\bibitem[{\citenamefont{{Pe{\~n}arrubia}}
  \emph{et~al.}(2008)\citenamefont{{Pe{\~n}arrubia}, {Navarro}, and
  {McConnachie}}}]{pena08}
\bibinfo{author}{\bibnamefont{{Pe{\~n}arrubia}}, \bibfnamefont{J.}},
  \bibinfo{author}{\bibfnamefont{J.~F.} \bibnamefont{{Navarro}}}, and
  \bibinfo{author}{\bibfnamefont{A.~W.} \bibnamefont{{McConnachie}}},
  \bibinfo{year}{2008}, \bibinfo{journal}{ApJ} \textbf{\bibinfo{volume}{673}},
  \bibinfo{pages}{226}.

\bibitem[{\citenamefont{{Pe{\~n}arrubia}}
  \emph{et~al.}(2009)\citenamefont{{Pe{\~n}arrubia}, {Walker}, and
  {Gilmore}}}]{pena09}
\bibinfo{author}{\bibnamefont{{Pe{\~n}arrubia}}, \bibfnamefont{J.}},
  \bibinfo{author}{\bibfnamefont{M.~G.} \bibnamefont{{Walker}}}, and
  \bibinfo{author}{\bibfnamefont{G.}~\bibnamefont{{Gilmore}}},
  \bibinfo{year}{2009}, \bibinfo{journal}{MNRAS}
  \textbf{\bibinfo{volume}{399}}, \bibinfo{pages}{1275}.

\bibitem[{\citenamefont{{Pflamm-Altenburg}}
  \emph{et~al.}(2007)\citenamefont{{Pflamm-Altenburg}, {Weidner}, and
  {Kroupa}}}]{pfla07}
\bibinfo{author}{\bibnamefont{{Pflamm-Altenburg}}, \bibfnamefont{J.}},
  \bibinfo{author}{\bibfnamefont{C.}~\bibnamefont{{Weidner}}}, and
  \bibinfo{author}{\bibfnamefont{P.}~\bibnamefont{{Kroupa}}},
  \bibinfo{year}{2007}, \bibinfo{journal}{ApJ} \textbf{\bibinfo{volume}{671}},
  \bibinfo{pages}{1550}.

\bibitem[{\citenamefont{{Piskunov}}
  \emph{et~al.}(2008)\citenamefont{{Piskunov}, {Kharchenko}, {Schilbach},
  {R{\"o}ser}, {Scholz}, and {Zinnecker}}}]{pisk08}
\bibinfo{author}{\bibnamefont{{Piskunov}}, \bibfnamefont{A.~E.}},
  \bibinfo{author}{\bibfnamefont{N.~V.} \bibnamefont{{Kharchenko}}},
  \bibinfo{author}{\bibfnamefont{E.}~\bibnamefont{{Schilbach}}},
  \bibinfo{author}{\bibfnamefont{S.}~\bibnamefont{{R{\"o}ser}}},
  \bibinfo{author}{\bibfnamefont{R.-D.} \bibnamefont{{Scholz}}}, and
  \bibinfo{author}{\bibfnamefont{H.}~\bibnamefont{{Zinnecker}}},
  \bibinfo{year}{2008}, \bibinfo{journal}{A\&A} \textbf{\bibinfo{volume}{487}},
  \bibinfo{pages}{557}.

\bibitem[{\citenamefont{{Pritzl}} \emph{et~al.}(2005)\citenamefont{{Pritzl},
  {Venn}, and {Irwin}}}]{prit05}
\bibinfo{author}{\bibnamefont{{Pritzl}}, \bibfnamefont{B.~J.}},
  \bibinfo{author}{\bibfnamefont{K.~A.} \bibnamefont{{Venn}}}, and
  \bibinfo{author}{\bibfnamefont{M.}~\bibnamefont{{Irwin}}},
  \bibinfo{year}{2005}, \bibinfo{journal}{AJ} \textbf{\bibinfo{volume}{130}},
  \bibinfo{pages}{2140}.

\bibitem[{\citenamefont{{Revaz}} \emph{et~al.}(2009)\citenamefont{{Revaz},
  {Jablonka}, {Sawala}, {Hill}, {Letarte}, {Irwin}, {Battaglia}, {Helmi},
  {Shetrone}, {Tolstoy}, and {Venn}}}]{reva09}
\bibinfo{author}{\bibnamefont{{Revaz}}, \bibfnamefont{Y.}},
  \bibinfo{author}{\bibfnamefont{P.}~\bibnamefont{{Jablonka}}},
  \bibinfo{author}{\bibfnamefont{T.}~\bibnamefont{{Sawala}}},
  \bibinfo{author}{\bibfnamefont{V.}~\bibnamefont{{Hill}}},
  \bibinfo{author}{\bibfnamefont{B.}~\bibnamefont{{Letarte}}},
  \bibinfo{author}{\bibfnamefont{M.}~\bibnamefont{{Irwin}}},
  \bibinfo{author}{\bibfnamefont{G.}~\bibnamefont{{Battaglia}}},
  \bibinfo{author}{\bibfnamefont{A.}~\bibnamefont{{Helmi}}},
  \bibinfo{author}{\bibfnamefont{M.~D.} \bibnamefont{{Shetrone}}},
  \bibinfo{author}{\bibfnamefont{E.}~\bibnamefont{{Tolstoy}}}, and
  \bibinfo{author}{\bibfnamefont{K.~A.} \bibnamefont{{Venn}}},
  \bibinfo{year}{2009}, \bibinfo{journal}{A\&A} \textbf{\bibinfo{volume}{501}},
  \bibinfo{pages}{189}.

\bibitem[{\citenamefont{{Ricotti}}(2010)}]{rico10}
\bibinfo{author}{\bibnamefont{{Ricotti}}, \bibfnamefont{M.}},
  \bibinfo{year}{2010}, \bibinfo{journal}{Adv. Astron.}
  \textbf{\bibinfo{volume}{2010}}, \bibinfo{eid}{271592}.

\bibitem[{\citenamefont{{Roederer} and {Sneden}}(2011)}]{roed11}
\bibinfo{author}{\bibnamefont{{Roederer}}, \bibfnamefont{I.~U.}}, and
  \bibinfo{author}{\bibfnamefont{C.}~\bibnamefont{{Sneden}}},
  \bibinfo{year}{2011}, \bibinfo{journal}{AJ} \textbf{\bibinfo{volume}{142}},
  \bibinfo{eid}{22}.

\bibitem[{\citenamefont{{Ryan}} \emph{et~al.}(1996)\citenamefont{{Ryan},
  {Norris}, and {Beers}}}]{ryan96}
\bibinfo{author}{\bibnamefont{{Ryan}}, \bibfnamefont{S.~G.}},
  \bibinfo{author}{\bibfnamefont{J.~E.} \bibnamefont{{Norris}}}, and
  \bibinfo{author}{\bibfnamefont{T.~C.} \bibnamefont{{Beers}}},
  \bibinfo{year}{1996}, \bibinfo{journal}{ApJ} \textbf{\bibinfo{volume}{471}},
  \bibinfo{pages}{254}.

\bibitem[{\citenamefont{{Sadakane}}
  \emph{et~al.}(2004)\citenamefont{{Sadakane}, {Arimoto}, {Ikuta}, {Aoki},
  {Jablonka}, and {Tajitsu}}}]{sada04}
\bibinfo{author}{\bibnamefont{{Sadakane}}, \bibfnamefont{K.}},
  \bibinfo{author}{\bibfnamefont{N.}~\bibnamefont{{Arimoto}}},
  \bibinfo{author}{\bibfnamefont{C.}~\bibnamefont{{Ikuta}}},
  \bibinfo{author}{\bibfnamefont{W.}~\bibnamefont{{Aoki}}},
  \bibinfo{author}{\bibfnamefont{P.}~\bibnamefont{{Jablonka}}}, and
  \bibinfo{author}{\bibfnamefont{A.}~\bibnamefont{{Tajitsu}}},
  \bibinfo{year}{2004}, \bibinfo{journal}{PASJ} \textbf{\bibinfo{volume}{56}},
  \bibinfo{pages}{1041}.

\bibitem[{\citenamefont{{Salvadori} and {Ferrara}}(2009)}]{salv09}
\bibinfo{author}{\bibnamefont{{Salvadori}}, \bibfnamefont{S.}}, and
  \bibinfo{author}{\bibfnamefont{A.}~\bibnamefont{{Ferrara}}},
  \bibinfo{year}{2009}, \bibinfo{journal}{MNRAS}
  \textbf{\bibinfo{volume}{395}}, \bibinfo{pages}{L6}.

\bibitem[{\citenamefont{{Saunders}}
  \emph{et~al.}(2010)\citenamefont{{Saunders}, {Colless}, {Saunders},
  {Hopkins}, {Goodwin}, {Heijmans}, {Brzeski}, and {Farrell}}}]{saun10}
\bibinfo{author}{\bibnamefont{{Saunders}}, \bibfnamefont{W.}},
  \bibinfo{author}{\bibfnamefont{M.}~\bibnamefont{{Colless}}},
  \bibinfo{author}{\bibfnamefont{I.}~\bibnamefont{{Saunders}}},
  \bibinfo{author}{\bibfnamefont{A.}~\bibnamefont{{Hopkins}}},
  \bibinfo{author}{\bibfnamefont{M.}~\bibnamefont{{Goodwin}}},
  \bibinfo{author}{\bibfnamefont{J.}~\bibnamefont{{Heijmans}}},
  \bibinfo{author}{\bibfnamefont{J.}~\bibnamefont{{Brzeski}}}, and
  \bibinfo{author}{\bibfnamefont{T.}~\bibnamefont{{Farrell}}},
  \bibinfo{year}{2010}, in \emph{\bibinfo{booktitle}{Society of Photo-Optical
  Instrumentation Engineers (SPIE) Conference Series}}, volume
  \bibinfo{volume}{7735} of \emph{\bibinfo{series}{Society of Photo-Optical
  Instrumentation Engineers (SPIE) Conference Series}}.

\bibitem[{\citenamefont{{Sch{\"o}nrich}}
  \emph{et~al.}(2011)\citenamefont{{Sch{\"o}nrich}, {Asplund}, and
  {Casagrande}}}]{scho11}
\bibinfo{author}{\bibnamefont{{Sch{\"o}nrich}}, \bibfnamefont{R.}},
  \bibinfo{author}{\bibfnamefont{M.}~\bibnamefont{{Asplund}}}, and
  \bibinfo{author}{\bibfnamefont{L.}~\bibnamefont{{Casagrande}}},
  \bibinfo{year}{2011}, \bibinfo{journal}{MNRAS}
  \textbf{\bibinfo{volume}{415}}, \bibinfo{pages}{3807}.

\bibitem[{\citenamefont{{Sch{\"o}rck}}
  \emph{et~al.}(2009)\citenamefont{{Sch{\"o}rck}, {Christlieb}, {Cohen},
  {Beers}, {Shectman}, {Thompson}, {McWilliam}, {Bessell}, {Norris},
  {Mel{\'e}ndez}, {Ram{\'{\i}}rez}, {Haynes}} \emph{et~al.}}]{scho09}
\bibinfo{author}{\bibnamefont{{Sch{\"o}rck}}, \bibfnamefont{T.}},
  \bibinfo{author}{\bibfnamefont{N.}~\bibnamefont{{Christlieb}}},
  \bibinfo{author}{\bibfnamefont{J.~G.} \bibnamefont{{Cohen}}},
  \bibinfo{author}{\bibfnamefont{T.~C.} \bibnamefont{{Beers}}},
  \bibinfo{author}{\bibfnamefont{S.}~\bibnamefont{{Shectman}}},
  \bibinfo{author}{\bibfnamefont{I.}~\bibnamefont{{Thompson}}},
  \bibinfo{author}{\bibfnamefont{A.}~\bibnamefont{{McWilliam}}},
  \bibinfo{author}{\bibfnamefont{M.~S.} \bibnamefont{{Bessell}}},
  \bibinfo{author}{\bibfnamefont{J.~E.} \bibnamefont{{Norris}}},
  \bibinfo{author}{\bibfnamefont{J.}~\bibnamefont{{Mel{\'e}ndez}}},
  \bibinfo{author}{\bibfnamefont{S.}~\bibnamefont{{Ram{\'{\i}}rez}}},
  \bibinfo{author}{\bibfnamefont{D.}~\bibnamefont{{Haynes}}}, \emph{et~al.},
  \bibinfo{year}{2009}, \bibinfo{journal}{A\&A} \textbf{\bibinfo{volume}{507}},
  \bibinfo{pages}{817}.

\bibitem[{\citenamefont{{Schuster}}
  \emph{et~al.}(2012)\citenamefont{{Schuster}, {Moreno}, {Nissen}, and
  {Pichardo}}}]{schu12}
\bibinfo{author}{\bibnamefont{{Schuster}}, \bibfnamefont{W.~J.}},
  \bibinfo{author}{\bibfnamefont{E.}~\bibnamefont{{Moreno}}},
  \bibinfo{author}{\bibfnamefont{P.~E.} \bibnamefont{{Nissen}}}, and
  \bibinfo{author}{\bibfnamefont{B.}~\bibnamefont{{Pichardo}}},
  \bibinfo{year}{2012}, \bibinfo{journal}{A\&A} \textbf{\bibinfo{volume}{538}},
  \bibinfo{eid}{A21}.

\bibitem[{\citenamefont{{Sharma} and {Johnston}}(2009)}]{shar09}
\bibinfo{author}{\bibnamefont{{Sharma}}, \bibfnamefont{S.}}, and
  \bibinfo{author}{\bibfnamefont{K.~V.} \bibnamefont{{Johnston}}},
  \bibinfo{year}{2009}, \bibinfo{journal}{ApJ} \textbf{\bibinfo{volume}{703}},
  \bibinfo{pages}{1061}.

\bibitem[{\citenamefont{{Shetrone}}
  \emph{et~al.}(2003)\citenamefont{{Shetrone}, {Venn}, {Tolstoy}, {Primas},
  {Hill}, and {Kaufer}}}]{shet03}
\bibinfo{author}{\bibnamefont{{Shetrone}}, \bibfnamefont{M.}},
  \bibinfo{author}{\bibfnamefont{K.~A.} \bibnamefont{{Venn}}},
  \bibinfo{author}{\bibfnamefont{E.}~\bibnamefont{{Tolstoy}}},
  \bibinfo{author}{\bibfnamefont{F.}~\bibnamefont{{Primas}}},
  \bibinfo{author}{\bibfnamefont{V.}~\bibnamefont{{Hill}}}, and
  \bibinfo{author}{\bibfnamefont{A.}~\bibnamefont{{Kaufer}}},
  \bibinfo{year}{2003}, \bibinfo{journal}{AJ} \textbf{\bibinfo{volume}{125}},
  \bibinfo{pages}{684}.

\bibitem[{\citenamefont{{Shetrone}}
  \emph{et~al.}(2001)\citenamefont{{Shetrone}, {C{\^o}t{\'e}}, and
  {Sargent}}}]{shet01}
\bibinfo{author}{\bibnamefont{{Shetrone}}, \bibfnamefont{M.~D.}},
  \bibinfo{author}{\bibfnamefont{P.}~\bibnamefont{{C{\^o}t{\'e}}}}, and
  \bibinfo{author}{\bibfnamefont{W.~L.~W.} \bibnamefont{{Sargent}}},
  \bibinfo{year}{2001}, \bibinfo{journal}{ApJ} \textbf{\bibinfo{volume}{548}},
  \bibinfo{pages}{592}.

\bibitem[{\citenamefont{{Simon}} \emph{et~al.}(2010)\citenamefont{{Simon},
  {Frebel}, {McWilliam}, {Kirby}, and {Thompson}}}]{simo10}
\bibinfo{author}{\bibnamefont{{Simon}}, \bibfnamefont{J.~D.}},
  \bibinfo{author}{\bibfnamefont{A.}~\bibnamefont{{Frebel}}},
  \bibinfo{author}{\bibfnamefont{A.}~\bibnamefont{{McWilliam}}},
  \bibinfo{author}{\bibfnamefont{E.~N.} \bibnamefont{{Kirby}}}, and
  \bibinfo{author}{\bibfnamefont{I.~B.} \bibnamefont{{Thompson}}},
  \bibinfo{year}{2010}, \bibinfo{journal}{ApJ} \textbf{\bibinfo{volume}{716}},
  \bibinfo{pages}{446}.

\bibitem[{\citenamefont{{Sneden}} \emph{et~al.}(2008)\citenamefont{{Sneden},
  {Cowan}, and {Gallino}}}]{sned08}
\bibinfo{author}{\bibnamefont{{Sneden}}, \bibfnamefont{C.}},
  \bibinfo{author}{\bibfnamefont{J.~J.} \bibnamefont{{Cowan}}}, and
  \bibinfo{author}{\bibfnamefont{R.}~\bibnamefont{{Gallino}}},
  \bibinfo{year}{2008}, \bibinfo{journal}{ARA\&A}
  \textbf{\bibinfo{volume}{46}}, \bibinfo{pages}{241}.

\bibitem[{\citenamefont{{Sobeck}} \emph{et~al.}(2011)\citenamefont{{Sobeck},
  {Kraft}, {Sneden}, {Preston}, {Cowan}, {Smith}, {Thompson}, {Shectman}, and
  {Burley}}}]{sobe11}
\bibinfo{author}{\bibnamefont{{Sobeck}}, \bibfnamefont{J.~S.}},
  \bibinfo{author}{\bibfnamefont{R.~P.} \bibnamefont{{Kraft}}},
  \bibinfo{author}{\bibfnamefont{C.}~\bibnamefont{{Sneden}}},
  \bibinfo{author}{\bibfnamefont{G.~W.} \bibnamefont{{Preston}}},
  \bibinfo{author}{\bibfnamefont{J.~J.} \bibnamefont{{Cowan}}},
  \bibinfo{author}{\bibfnamefont{G.~H.} \bibnamefont{{Smith}}},
  \bibinfo{author}{\bibfnamefont{I.~B.} \bibnamefont{{Thompson}}},
  \bibinfo{author}{\bibfnamefont{S.~A.} \bibnamefont{{Shectman}}}, and
  \bibinfo{author}{\bibfnamefont{G.~S.} \bibnamefont{{Burley}}},
  \bibinfo{year}{2011}, \bibinfo{journal}{AJ} \textbf{\bibinfo{volume}{141}},
  \bibinfo{eid}{175}.

\bibitem[{\citenamefont{{Stacy}} \emph{et~al.}(2010)\citenamefont{{Stacy},
  {Greif}, and {Bromm}}}]{stac10}
\bibinfo{author}{\bibnamefont{{Stacy}}, \bibfnamefont{A.}},
  \bibinfo{author}{\bibfnamefont{T.~H.} \bibnamefont{{Greif}}}, and
  \bibinfo{author}{\bibfnamefont{V.}~\bibnamefont{{Bromm}}},
  \bibinfo{year}{2010}, \bibinfo{journal}{MNRAS}
  \textbf{\bibinfo{volume}{403}}, \bibinfo{pages}{45}.

\bibitem[{\citenamefont{{Starkenburg}}
  \emph{et~al.}(2010)\citenamefont{{Starkenburg}, {Hill}, {Tolstoy},
  {Gonz{\'a}lez Hern{\'a}ndez}, {Irwin}, {Helmi}, {Battaglia}, {Jablonka},
  {Tafelmeyer}, {Shetrone}, {Venn}, and {de Boer}}}]{star10}
\bibinfo{author}{\bibnamefont{{Starkenburg}}, \bibfnamefont{E.}},
  \bibinfo{author}{\bibfnamefont{V.}~\bibnamefont{{Hill}}},
  \bibinfo{author}{\bibfnamefont{E.}~\bibnamefont{{Tolstoy}}},
  \bibinfo{author}{\bibfnamefont{J.~I.} \bibnamefont{{Gonz{\'a}lez
  Hern{\'a}ndez}}}, \bibinfo{author}{\bibfnamefont{M.}~\bibnamefont{{Irwin}}},
  \bibinfo{author}{\bibfnamefont{A.}~\bibnamefont{{Helmi}}},
  \bibinfo{author}{\bibfnamefont{G.}~\bibnamefont{{Battaglia}}},
  \bibinfo{author}{\bibfnamefont{P.}~\bibnamefont{{Jablonka}}},
  \bibinfo{author}{\bibfnamefont{M.}~\bibnamefont{{Tafelmeyer}}},
  \bibinfo{author}{\bibfnamefont{M.}~\bibnamefont{{Shetrone}}},
  \bibinfo{author}{\bibfnamefont{K.}~\bibnamefont{{Venn}}}, and
  \bibinfo{author}{\bibfnamefont{T.}~\bibnamefont{{de Boer}}},
  \bibinfo{year}{2010}, \bibinfo{journal}{A\&A} \textbf{\bibinfo{volume}{513}},
  \bibinfo{pages}{A34}.

\bibitem[{\citenamefont{{Tafelmeyer}}
  \emph{et~al.}(2010)\citenamefont{{Tafelmeyer}, {Jablonka}, {Hill},
  {Shetrone}, {Tolstoy}, {Irwin}, {Battaglia}, {Helmi}, {Starkenburg}, {Venn},
  {Abel}, {Francois}} \emph{et~al.}}]{tafe10}
\bibinfo{author}{\bibnamefont{{Tafelmeyer}}, \bibfnamefont{M.}},
  \bibinfo{author}{\bibfnamefont{P.}~\bibnamefont{{Jablonka}}},
  \bibinfo{author}{\bibfnamefont{V.}~\bibnamefont{{Hill}}},
  \bibinfo{author}{\bibfnamefont{M.}~\bibnamefont{{Shetrone}}},
  \bibinfo{author}{\bibfnamefont{E.}~\bibnamefont{{Tolstoy}}},
  \bibinfo{author}{\bibfnamefont{M.~J.} \bibnamefont{{Irwin}}},
  \bibinfo{author}{\bibfnamefont{G.}~\bibnamefont{{Battaglia}}},
  \bibinfo{author}{\bibfnamefont{A.}~\bibnamefont{{Helmi}}},
  \bibinfo{author}{\bibfnamefont{E.}~\bibnamefont{{Starkenburg}}},
  \bibinfo{author}{\bibfnamefont{K.~A.} \bibnamefont{{Venn}}},
  \bibinfo{author}{\bibfnamefont{T.}~\bibnamefont{{Abel}}},
  \bibinfo{author}{\bibfnamefont{P.}~\bibnamefont{{Francois}}}, \emph{et~al.},
  \bibinfo{year}{2010}, \bibinfo{journal}{A\&A} \textbf{\bibinfo{volume}{524}},
  \bibinfo{eid}{A58}.

\bibitem[{\citenamefont{{Tinsley}}(1980)}]{tins80}
\bibinfo{author}{\bibnamefont{{Tinsley}}, \bibfnamefont{B.~M.}},
  \bibinfo{year}{1980}, \bibinfo{journal}{Fund. Cosmic Phys.}
  \textbf{\bibinfo{volume}{5}}, \bibinfo{pages}{287}.

\bibitem[{\citenamefont{{Tolstoy}} \emph{et~al.}(2009)\citenamefont{{Tolstoy},
  {Hill}, and {Tosi}}}]{tols09}
\bibinfo{author}{\bibnamefont{{Tolstoy}}, \bibfnamefont{E.}},
  \bibinfo{author}{\bibfnamefont{V.}~\bibnamefont{{Hill}}}, and
  \bibinfo{author}{\bibfnamefont{M.}~\bibnamefont{{Tosi}}},
  \bibinfo{year}{2009}, \bibinfo{journal}{ARA\&A}
  \textbf{\bibinfo{volume}{47}}, \bibinfo{pages}{371}.

\bibitem[{\citenamefont{{Turk}} \emph{et~al.}(2009)\citenamefont{{Turk},
  {Abel}, and {O'Shea}}}]{turk09}
\bibinfo{author}{\bibnamefont{{Turk}}, \bibfnamefont{M.~J.}},
  \bibinfo{author}{\bibfnamefont{T.}~\bibnamefont{{Abel}}}, and
  \bibinfo{author}{\bibfnamefont{B.}~\bibnamefont{{O'Shea}}},
  \bibinfo{year}{2009}, \bibinfo{journal}{Science}
  \textbf{\bibinfo{volume}{325}}, \bibinfo{pages}{601}.

\bibitem[{\citenamefont{{Umeda} and {Nomoto}}(2002)}]{umed02}
\bibinfo{author}{\bibnamefont{{Umeda}}, \bibfnamefont{H.}}, and
  \bibinfo{author}{\bibfnamefont{K.}~\bibnamefont{{Nomoto}}},
  \bibinfo{year}{2002}, \bibinfo{journal}{ApJ} \textbf{\bibinfo{volume}{565}},
  \bibinfo{pages}{385}.

\bibitem[{\citenamefont{{van den Bergh}}(1962)}]{berg62}
\bibinfo{author}{\bibnamefont{{van den Bergh}}, \bibfnamefont{S.}},
  \bibinfo{year}{1962}, \bibinfo{journal}{AJ} \textbf{\bibinfo{volume}{67}},
  \bibinfo{pages}{486}.

\bibitem[{\citenamefont{{Walker}} \emph{et~al.}(2006)\citenamefont{{Walker},
  {Mateo}, {Olszewski}, {Pal}, {Sen}, and {Woodroofe}}}]{walk06}
\bibinfo{author}{\bibnamefont{{Walker}}, \bibfnamefont{M.~G.}},
  \bibinfo{author}{\bibfnamefont{M.}~\bibnamefont{{Mateo}}},
  \bibinfo{author}{\bibfnamefont{E.~W.} \bibnamefont{{Olszewski}}},
  \bibinfo{author}{\bibfnamefont{J.~K.} \bibnamefont{{Pal}}},
  \bibinfo{author}{\bibfnamefont{B.}~\bibnamefont{{Sen}}}, and
  \bibinfo{author}{\bibfnamefont{M.}~\bibnamefont{{Woodroofe}}},
  \bibinfo{year}{2006}, \bibinfo{journal}{ApJ} \textbf{\bibinfo{volume}{642}},
  \bibinfo{pages}{L41}.

\bibitem[{\citenamefont{{Walker} and {Pe{\~n}arrubia}}(2011)}]{pena11}
\bibinfo{author}{\bibnamefont{{Walker}}, \bibfnamefont{M.~G.}}, and
  \bibinfo{author}{\bibfnamefont{J.}~\bibnamefont{{Pe{\~n}arrubia}}},
  \bibinfo{year}{2011}, \bibinfo{journal}{ApJ} \textbf{\bibinfo{volume}{742}},
  \bibinfo{eid}{20}.

\bibitem[{\citenamefont{{Weidner} and {Kroupa}}(2005)}]{weid05}
\bibinfo{author}{\bibnamefont{{Weidner}}, \bibfnamefont{C.}}, and
  \bibinfo{author}{\bibfnamefont{P.}~\bibnamefont{{Kroupa}}},
  \bibinfo{year}{2005}, \bibinfo{journal}{ApJ} \textbf{\bibinfo{volume}{625}},
  \bibinfo{pages}{754}.

\bibitem[{\citenamefont{{Wise} and {Abel}}(2008)}]{wise08}
\bibinfo{author}{\bibnamefont{{Wise}}, \bibfnamefont{J.~H.}}, and
  \bibinfo{author}{\bibfnamefont{T.}~\bibnamefont{{Abel}}},
  \bibinfo{year}{2008}, \bibinfo{journal}{ApJ} \textbf{\bibinfo{volume}{685}},
  \bibinfo{pages}{40}.

\bibitem[{\citenamefont{{Wise}}
  \emph{et~al.}(2012{\natexlab{a}})\citenamefont{{Wise}, {Abel}, {Turk},
  {Norman}, and {Smith}}}]{wise12a}
\bibinfo{author}{\bibnamefont{{Wise}}, \bibfnamefont{J.~H.}},
  \bibinfo{author}{\bibfnamefont{T.}~\bibnamefont{{Abel}}},
  \bibinfo{author}{\bibfnamefont{M.~J.} \bibnamefont{{Turk}}},
  \bibinfo{author}{\bibfnamefont{M.~L.} \bibnamefont{{Norman}}}, and
  \bibinfo{author}{\bibfnamefont{B.~D.} \bibnamefont{{Smith}}},
  \bibinfo{year}{2012}{\natexlab{a}}, \bibinfo{note}{submitted, arXiv:1206.1043
  [astro-ph.CO]}.

\bibitem[{\citenamefont{{Wise}}
  \emph{et~al.}(2012{\natexlab{b}})\citenamefont{{Wise}, {Turk}, {Norman}, and
  {Abel}}}]{wise12b}
\bibinfo{author}{\bibnamefont{{Wise}}, \bibfnamefont{J.~H.}},
  \bibinfo{author}{\bibfnamefont{M.~J.} \bibnamefont{{Turk}}},
  \bibinfo{author}{\bibfnamefont{M.~L.} \bibnamefont{{Norman}}}, and
  \bibinfo{author}{\bibfnamefont{T.}~\bibnamefont{{Abel}}},
  \bibinfo{year}{2012}{\natexlab{b}}, \bibinfo{journal}{ApJ}
  \textbf{\bibinfo{volume}{745}}, \bibinfo{eid}{50}.

\bibitem[{\citenamefont{{Woo}} \emph{et~al.}(2008)\citenamefont{{Woo},
  {Courteau}, and {Dekel}}}]{woo08}
\bibinfo{author}{\bibnamefont{{Woo}}, \bibfnamefont{J.}},
  \bibinfo{author}{\bibfnamefont{S.}~\bibnamefont{{Courteau}}}, and
  \bibinfo{author}{\bibfnamefont{A.}~\bibnamefont{{Dekel}}},
  \bibinfo{year}{2008}, \bibinfo{journal}{MNRAS}
  \textbf{\bibinfo{volume}{390}}, \bibinfo{pages}{1453}.

\bibitem[{\citenamefont{{Yoshida}} \emph{et~al.}(2008)\citenamefont{{Yoshida},
  {Omukai}, and {Hernquist}}}]{yosh08}
\bibinfo{author}{\bibnamefont{{Yoshida}}, \bibfnamefont{N.}},
  \bibinfo{author}{\bibfnamefont{K.}~\bibnamefont{{Omukai}}}, and
  \bibinfo{author}{\bibfnamefont{L.}~\bibnamefont{{Hernquist}}},
  \bibinfo{year}{2008}, \bibinfo{journal}{Science}
  \textbf{\bibinfo{volume}{321}}, \bibinfo{pages}{669}.

\end{thebibliography}

\end{document}